\documentclass[nohyper,12pt,letterpaper]{JHEP3} 
\usepackage{epsfig}
\usepackage{amsmath}
\usepackage{amssymb}

\author{Neil R. Constable${}^\dagger$,                           
Daniel Z. Freedman${}^\dagger{}^\ddagger$,                       
Matthew Headrick${}^\star$, \newline
Shiraz Minwalla${}^\star$,
Lubo\v{s} Motl${}^\star$,
Alexander Postnikov${}^\ddagger$,                                    
Witold Skiba${}^\dagger$\\                                    
\qquad\\
  $\!\!{}^\ddagger$Department of Mathematics \quad and \quad   
${}^\dagger$Center for Theoretical Physics\\                    
  Massachusetts Institute of Technology\\          
  Cambridge, MA 02139\\
\qquad {\scriptsize and} \\
  $\!\!\!{}^\star$Jefferson Physical Laboratory\\
  Harvard University\\
  Cambridge, MA 02138\\ 
\qquad\\
E-mail: \email{ constabl@lns.mit.edu, dzf@math.mit.edu,\\      
\qquad\qquad headrick@riemann.harvard.edu, minwalla@born.harvard.edu,\\
\qquad\qquad motl@feynman.harvard.edu, apost@math.mit.edu, skiba@mit.edu}}

\abstract{Recently, Berenstein et al.\ have proposed a duality between a
sector of $\N=4$ super-Yang-Mills theory with large R-charge $J$, and
string theory in a pp-wave background. In the limit considered, the
effective 't Hooft coupling has been argued to be $\lambda'=g_{\rm YM}^2
N/J^2=1/(\mu p^+ \apm)^2$. We study Yang-Mills theory at small $\lambda'$
(large $\mu$) with a view to reproducing string interactions. We
demonstrate that the effective genus counting parameter of the Yang-Mills
theory is $g^2_2= J^4/N^2 =(4\pi g_{\rm s})^2 (\mu p^+ \apm)^4$, the
effective two-dimensional Newton constant for strings propagating on the
pp-wave background.  We identify $g_2 \sqrt{\lambda'}$ as the effective
coupling between a wide class of excited string states on the pp-wave
background.  We compute the anomalous dimensions of BMN operators at 
first
order in $g_2^2$ and $\lambda'$ and interpret our result as the genus one
mass renormalization of the corresponding string state. We postulate a
relation between the three-string vertex function and the gauge theory
three-point function and compare our proposal to string field theory.  We
utilize this proposal, together with quantum mechanical perturbation
theory, to recompute the genus one energy shift of string states, and 
find precise agreement with our gauge theory computation.}

\preprint{{\tt hep-th/0205089 v3}\\   
  MIT-CTP-3271, HUTP-02/A013}
\title{PP-wave string interactions\\
from perturbative Yang-Mills theory}

\def\endOFproof{  {\bf QED~$\square$}\par\vspace{3mm}}
\newenvironment{proof}[1][\proofname]{\par \noindent
{\bf Proof. }}{\endOFproof}

\def\eqn#1#2{\begin{equation}#2\label{#1}\end{equation}}
\def\ket#1{\vert #1\rangle}

\def\abs#1{\left\vert#1\right\vert}

\renewcommand{\thanks}[1]{\footnote{#1}} 

\newcommand{\te}[1]{{\text{#1}}}

\makeatletter
\newcounter{fig}
\renewcommand\thefig{\arabic{fig}}

\def\fps@fig{tbp}
\def\ftype@fig{1}
\def\ext@fig{lof}
\def\fnum@fig{\figurename~\thefig}

\newenvironment{fig*}
               {\@dblfloat{fig}}
               {\end@dblfloat}

\makeatother

\usepackage{epsfig}

\newcommand{\bea}{\begin{eqnarray}}
\newcommand{\eea}{\end{eqnarray}}
\newcommand{\<}{\langle}
\renewcommand{\>}{\rangle}
\def\ba{\begin{eqnarray}}
\def\ea{\end{eqnarray}}

\DeclareMathOperator{\I}{Im}
\DeclareMathOperator{\Tr}{Tr}

\def\ap{{\alpha'}}

\def\p{\partial}

\def\unit{1 \hskip-.3em \raise2pt\hbox{$ \scriptstyle |$ } }

\newtheorem{theorem}{Theorem}[section]
\newtheorem{proposition}[theorem]{Proposition}

\newtheorem{lemma}[theorem]{Lemma}

\def\I{\sqrt{-1}}

\def\Z{\mathbb{Z}}
\def\C{\mathbb{C}}
\def\pa{\partial}
\def\O{\mathcal{O}}
\def\cotg{\cot}

\def\BR{\mathrm{BR}}
\def\ll{\mathrm{ll}}
\def\lr{\mathrm{lr}}
\def\rl{\mathrm{rl}}
\def\rr{\mathrm{rr}}
\def\ds{\displaystyle}

\def\d{\delta}

\def\apm{{\alpha^\prime}}    

\def\N{{\cal N}}
\def\O{{\cal O}}

\def\CO{{\cal O}}

\def\gYM{g_\te{YM}^2}

\def\bop#1{\setbox0=\hbox{$#1M$}\mkern1.5mu
        \vbox{\hrule height0pt depth.04\ht0
        \hbox{\vrule width.04\ht0 height.9\ht0 \kern.9\ht0
        \vrule width.04\ht0}\hrule height.04\ht0}\mkern1.5mu}
\def\rarrow{\rightarrow}

\def\leftrighthookfill#1{$\mathsurround=0pt \mathord\hook#1
       \hrulefill\mathord\hook#1$}
\def\underhook#1{\vtop{\ialign{##\crcr                 
       $\hfil\displaystyle{#1}\hfil$\crcr
       \noalign{\kern-1pt\nointerlineskip\vskip2pt}
       \leftrighthookfill5\crcr}}}

\makeatletter

\begin{document}

\section{Introduction}

Many years ago 't Hooft \cite{'tHooft:1973jz} demonstrated the existence of 
a nontrivial large $N$ limit of $SU(N)$ gauge theories
\begin{equation}\label{hooft}
N \to \infty,~~~ g^2_{\rm YM} \to 0,~~~ \lambda = \gYM N ~~{\rm fixed}.
\end{equation}   
In the 't Hooft limit \eqref{hooft}, Yang-Mills interactions  
are controlled by the 't Hooft coupling 
$\lambda = \gYM N$. Away from the strict $N \to \infty$
limit, Yang-Mills perturbation theory may be 
organized as a double expansion. Feynman graphs are summed over their
genus (controlled by the genus counting parameter
$1/N^2$) and over Feynman loops (controlled by the
effective coupling $\lambda$). 
These observations led 't~Hooft to conjecture a duality between 
large $N$ gauge theories and weakly interacting string theories. 't~Hooft 
proposed that the genus expansion on the two sides of this duality could be
identified, leading to the identification of $1/N$ as the effective string
coupling. The $AdS$/CFT conjecture and its generalizations have generated 
dramatic evidence for these proposals by supplying several 
concrete examples of such dualities. The study of these special examples
has also led to the identification of $\lambda^{\frac{1}{4}}$ as the 
effective string scale of the dual string theory, in units appropriate for 
comparison with the gauge theory. This implies, in particular, that 
as $\lambda \to \infty$, all string oscillator states have infinite 
mass and all unprotected single trace gauge theory operators have 
infinite dimension.

Recently, Berenstein, Maldacena, and Nastase \cite{ias} have
drawn attention to a different $N \to \infty$ limit of the
$N=4$, $d=4$ Super Yang-Mills theory. The $N=4$ theory has an $SO(6)$
R symmetry group under which its six scalar fields $X^1 \ldots X^6$
transform in the vector representation. Consider an arbitrarily chosen 
$U(1)$ subgroup of this R-symmetry group; for definiteness let this $U(1)$  
represent rotations in the $X^5$ and $X^6$ plane. BMN study the sector 
of this theory with $R$ charge $J$, and let $J$ scale with $N$ 
according to 
\begin{equation}\label{bmnlimit}
N\to\infty\te{, with }
\frac{J}{\sqrt{N}}\te{ and }\gYM~~\te{and}~~ \Delta -J \te{ fixed.}
\end{equation}

Note that $\lambda \to \infty$ and $1/N \to 0$ in the BMN limit.
Consequently, according to the 't~Hooftian lore reviewed above, 
SYM theory in the limit \eqref{bmnlimit} is infinitely strongly coupled. 
Furthermore its string dual appears to be a {\it free} string theory 
with {\it infinite} effective string mass. None of these expectations is true;
usual 't~Hooftian reasoning fails as a consequence of the fact that 
observables in BMN limit are not held fixed, but scale to infinite 
charge as $N \to \infty$. We will explain these remarks further below. 
However, it is useful to first review the string dual 
of Super Yang-Mills theory in the BMN limit. 

BMN were led to the large $N$ scaling \eqref{bmnlimit} by the
consideration of a limit of the $AdS$/CFT duality. 
Super Yang-Mills in the seemingly singular regime \eqref{bmnlimit} is 
actually dual to
a well behaved closed string theory:  IIB theory on the Ramond-Ramond 
pp-wave \cite{Blau:2002dy}:
\begin{equation}\label{ppback}
ds^2 = -4dx^+dx^- - \mu^2 z^2dx^{+2}+dz^2, \qquad
F_{+1234} = F_{+5678} = \frac{\mu}{4\pi^3g_{\rm s}\alpha^{\prime2}},\qquad
e^\Phi = g_{\rm s}.
\end{equation} 
According to this duality\footnote{This duality and its generalizations 
have been studied further by many authors, see 
\cite{Metsaev:1999gz}-\cite{Oh:2002sv}.}, the R charge $J$ 
of a Yang-Mills operator is
proportional to the light-cone momentum $p^+$ of the corresponding string 
state, while $\Delta -J$ of the Yang-Mills operator is proportional to 
the light-cone energy $p^-$ of the same state. 
The detailed dictionary between charges of the string theory and 
the gauge theory is given by 
\begin{equation}\label{bmnmap}
\mu p^+\alpha' = \frac{J}{\sqrt\lambda}, \qquad
\frac{2p^-}{\mu} = \Delta-J, \qquad \gYM=4\pi g_{\rm s}.
\end{equation}
Consequently, the $AdS$/CFT duality predicts that Super Yang-Mills theory 
in the limit \eqref{bmnlimit} is dual to an {\it interacting} string 
theory with {\it finite} effective scale. This prediction is 
in conflict with the 't~Hooftian expectations of the previous paragraph.

We first address the puzzle of the effective string 
mass \cite{ias}. It is certainly true that all {\it fixed} unprotected 
 single trace operators scale to infinite anomalous dimension 
(consequently the corresponding modes in the dual string theory scale to 
infinite mass) as $\lambda$ is taken 
to infinity. However, as we have emphasized above, observables are not held 
fixed, but scale with $N$ in the BMN limit. While most such operators
leave the spectrum in the $N\to \infty$, $\lambda \to \infty$ limit
\eqref{bmnlimit}, BMN have identified a special set of operators whose 
anomalous dimension remains finite in this limit. These operators are dual to 
stringy oscillator states on the background \eqref{ppback}. These 
operators
are special; though they are not BPS, in the large $N$ limit they are  `locally' chiral 
(see section two for more details), and so are nearly BPS. 
Scaling dimensions of these special operators do receive loop corrections;
 however the supersymmetric cancellations responsible for the non 
renormalization of exactly chiral operators also ensure that the 
anomalous dimensions of these almost BPS operators are much smaller 
than the power series in $\gYM N$ that naive perturbative estimates suggest. 
Indeed BMN have argued that the anomalous dimensions of these special 
operators are not just finite, but actually 
{\it computable perturbatively}, even though the 't~Hooft coupling 
$\lambda$ diverges in the limit \eqref{bmnlimit}. Supersymmetric
cancellations produce a new coupling constant
\begin{equation}
\lambda' = \frac{\gYM N}{J^2} = \frac1{(\mu p^+\alpha')^2},
\end{equation}
which appears to play the role of the loop counting parameter in the 
computation of two point functions of these operators. 

Like their scaling dimensions, three point functions of chiral
operators are not renormalized \cite{Lee:1998bx,9907098}. 
Consequently we expect analogous 
supersymmetric cancellations to permit the perturbative computation of
three point couplings of BMN operators (hence interactions of the
corresponding string modes) at small $\lambda'$. In the rest of this paper 
(which is devoted to the study of PP-wave string interactions from
perturbative Yang-Mills theory) we proceed on this assumption. 
The coherence and consistency of the picture that emerges provide 
some justification for this assumption. 

We now turn to the puzzle of the effective string coupling. 
String loops certainly contribute to scattering of modes of IIB theory  at
nonzero $g_s$ on the background \eqref{ppback} (see 
\cite{Spradlin:2002ar}),
consequently generic correlation functions in Yang-Mills
must also receive contributions from higher genus graphs even though 
$N=\infty$, as in the limit \eqref{bmnlimit}. As we will demonstrate
in section 3 of this paper, this puzzle has a simple resolution. 
It is certainly true that each graph at genus $h$ is suppressed relative 
to a planar graph by the factor $1/N^{2h}$. However we will demonstrate 
below that the {\it number} of diagrams at genus $h$ is proportional 
to $J^{4h}$, so that the effective genus-counting parameter is actually
\begin{equation}\label{gt}
g_2^2 = \left(\frac{J^2}N\right)^2 = 16\pi^2 g_{\rm s}^2(\mu p^+\alpha')^4.
\end{equation}
$g_2^2$, the effective genus counting parameter,  
must also control the mixing between single and 
multi trace operators; this is easy to see directly. 
The two point function between single trace and double trace 
operator is of order $g_2/\sqrt{J}$ (see section 3). A single trace operator 
of size $J$ mixes with $J$ different double trace operators; 
consequently this mixing contributes to two point functions at order 
$J \times ( g_2/ \sqrt{J})^2=g_2^2$, in agreement with \eqref{gt} for a 
genus one process.  

The identification of $g_2^2$ with the Yang-Mills genus counting parameter
 fits naturally into duality between Yang-Mills and String theory, 
as $g_2$ has a rather natural interpretation in IIB
theory on \eqref{ppback}. In the pp-wave background, the worldsheet 
fields for the eight transverse directions are massive, so low energy 
excitations are confined to a distance $1/\sqrt{\mu p^+}$ from the origin. 
$g_2^2=g^2\sqrt{\apm \mu p^+}^8$ is simply the effective two dimensional 
Newton's constant, obtained after a `dimensional reduction' on the 
8 transverse dimensions.

In summary, despite first appearances, Yang-Mills theory in the limit 
\eqref{bmnlimit} appears to develop a new perturbative parameter 
$\lambda'$. 
In particular the theory is weakly coupled at small $\lambda'$ or 
large $\mu$. Further, the genus expansion and mixing between single and 
multi trace operators---effects related to interactions in the string 
dual---are controlled by $g_2$, the effective two dimensional Newton's
constant of the string theory. With this framework in place we
proceed, in the rest of this introduction, to describe the precise
relationship between string interactions and Yang-Mills correlators. 
As Yang-Mills correlators are perturbatively computable only 
at small $\lambda'$ or large $\mu$, some of the discussion that follows 
applies only to this limit. 

The first and most important qualitative issue concerns the identification 
of the effective string coupling in the background \eqref{ppback}. 
Following 
our discussion of the genus expansion in gauge theory, it is tempting to 
identify the effective string coupling with $g_2$. This guess is incorrect. 
In  section 5 we will argue that the effective string coupling 
between states with the same $\Delta_0-J$
(at small $\lambda'$) in the pp-wave background is $g_2 \sqrt{\lambda'}$,
where $\Delta_0$ is the scaling dimension at $\lambda'=0$. 
Note that the genus expansion of Yang-Mills theory (governed by the
parameter $g_2^2$) survives even in the free limit ($\lambda'=0$)
 when the effective string coupling is
zero. This genus expansion appears to be rather unphysical; 
we believe it contains information about the map between string states and 
Yang-Mills operators, but does not appear to directly encode interesting 
stringy dynamics. Physical 
effects (like anomalous dimensions) from higher genus graphs are obtained
only upon adding some Yang-Mills interaction vertices to 
these graphs; this addition leads to the re identification of the string
coupling as $g_2 \sqrt{\lambda'}$. Note that string states in the 
pp-wave background blow up into giant gravitons \cite{giant} when 
$J^2/N \gg 1/g_{\rm s}$, i.e. precisely when $g_2 \sqrt{\lambda'}$, the 
effective string coupling, is large.

\EPSFIGURE[r]{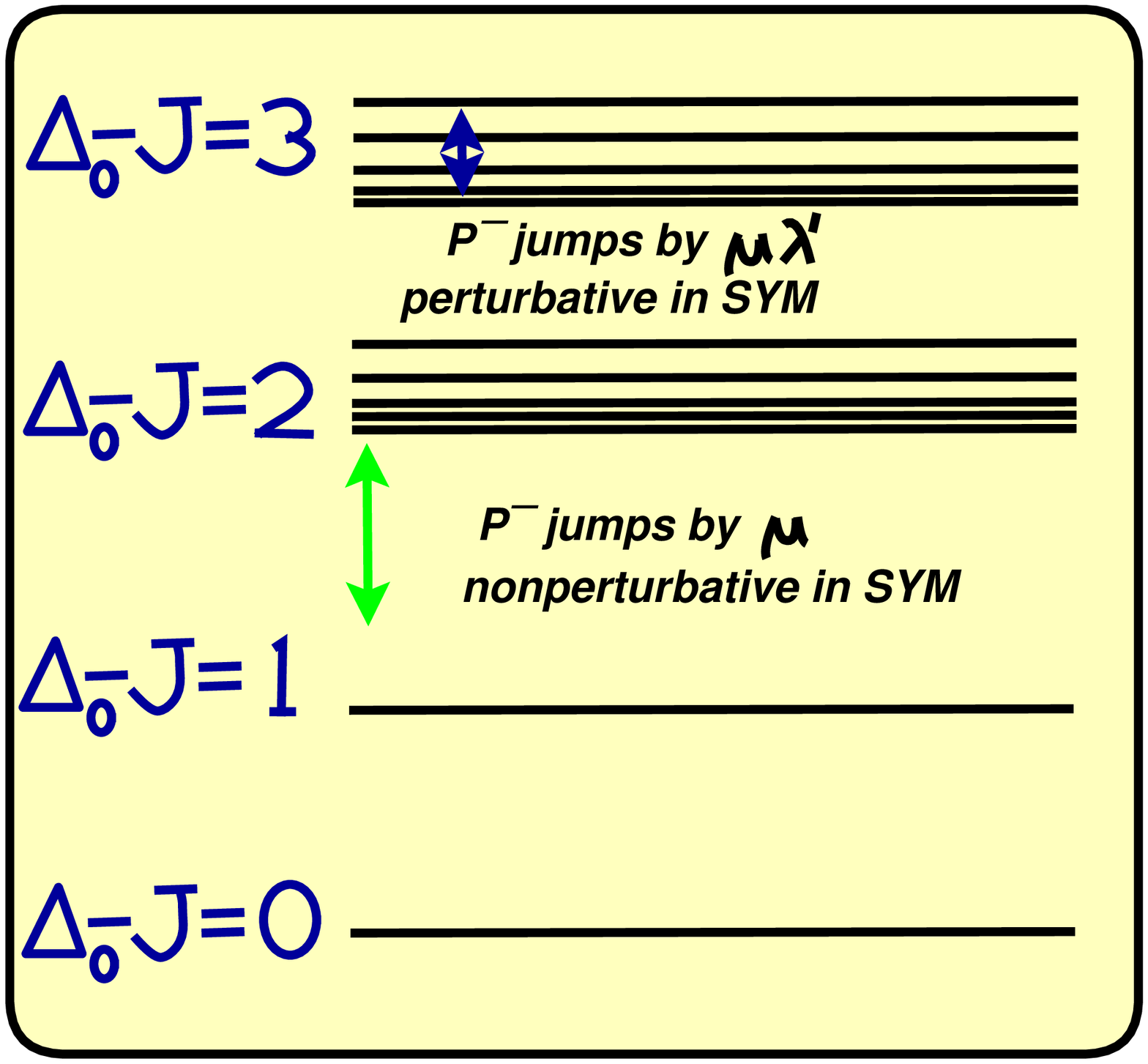,width=8cm}{Hierarchy of energy scales
for large $\mu p^+\alpha'$.
Only the transitions between states with the same $\Delta_0-J$, i.e.\
with the same number of impurities, can be calculated perturbatively.
The big jumps change the energy by an amount $1/\lambda'$ times bigger 
and therefore they result from non-perturbative effects in Yang-Mills
theory.\label{hierarchy}}

Let us explain our proposal for string interactions in more detail. 
The spectrum of string states in the pp-wave background clumps into 
almost degenerate multiplets at large $\mu$. 
The splittings between states of the same multiplet 
are of order $\mu \lambda'$, while the energy gap between distinct multiplets 
is of order $\mu$.  In section 5 we propose that the matrix element 
of the light-cone Hamiltonian between single and double string states  
within  the {\it same} multiplet is the three-point
coefficient of the suitably normalized operator product coefficient of the 
corresponding operators (this quantity is $\CO(g_2)$), 
multiplied by the difference between their unperturbed light-cone
energies. Since energy splittings within a multiplet are of order
$\mu\lambda'$, Hamiltonian matrix elements between such states are of order 
$\mu g_2 \lambda'$, corresponding to an invariant string coupling of order
$g_2 \sqrt{\lambda'}$. 
For a class of BMN states we compute these matrix elements  perturbatively 
in Yang-Mills theory. Note that transitions between states with 
different $\Delta_0-J$ involve  large changes in energy; such transitions 
appear to be non-perturbative in the gauge theory (see figure
\ref{hierarchy}).

In the paragraphs above we have presented a specific proposal 
relating interaction amplitudes in string theory with correlation functions 
of the dual gauge theory. In the next two paragraphs we describe the
evidence in support of our proposal. As we describe below, our proposal
passes a rather nontrivial consistency check. Further we have also 
partially verified our proposal  by direct comparison of three point functions
(computed in Yang-Mills perturbation theory) with three string light-cone
matrix elements (computed using light-cone string field theory). 
 
We first describe the consistency check on our proposal. 
In section 5 we compute the shift in dimension of a class of BMN
operators to first order in $\lambda'$ and first order in $g_2^2$,
i.e.\ on the torus. We find that the anomalous dimensions 
receive non-zero corrections from the torus diagrams with a quartic 
interaction between ``non-nearest neighbor'' fields (see figure
\ref{nonnears}).
The anomalous dimensions are proportional to $g_2^2 \lambda'$, 
the square of the effective string coupling,  and are interpreted as  
mass renormalizations of excited  string states. We then proceed to 
recompute the mass renormalization of excited string states 
using second-order quantum mechanical perturbation theory.
We obtain the light-cone matrix elements needed for this computation from 
correlators computed in perturbative gauge theory, utilizing 
our prescription described above. These two independent computations agree
exactly, constituting a highly nontrivial ``unitarity'' check on the 
consistency of our proposals.  

In section five we also compare our proposal for string interactions with 
matrix elements of the light-cone Hamiltonian of string field theory.
The light-cone Hamiltonian is generated by a two-derivative prefactor 
acting on a delta functional overlap. In section 5 we demonstrate that 
the three-point function of three BMN operators, computed in free 
Yang-Mills theory, reproduces the delta functional overlap between 
three string states at large $\mu$. We conjecture that, in the same limit, 
 the prefactor of this delta functional overlap reproduces the second 
element of our formula for matrix elements (the difference between the 
unperturbed energies of the corresponding states). We sketch how string 
field theory predicts a modification of our prescription for the case of
the operators involving $D_\mu Z$ and fermions.

We conclude this introduction with a digression that may help to put our 
work in perspective. Yang-Mills/String theory dualities have hitherto 
been understood, even qualitatively, only in regimes of strong
gauge theory coupling.  
For instance, it has long been suspected that confining gauge theories may be
reformulated as string theories, with tubes of gauge theory flux 
constituting the dual string. However, as flux tubes emerge at distance 
scales larger than $1/\Lambda_{QCD}$, their dynamics is nonperturbative in 
the gauge theory.  More recently the Maldacena conjecture has established 
a duality between a conformal gauge theory (with a fixed line of couplings) 
and string theories on an $AdS$ background. However these dualities are 
well understood only at large values of the gauge coupling. In this paper, 
utilizing the BMN duality, we have taken the first
steps in explicitly reformulating an effectively {\it weakly coupled} 
gauge theory as an {\it interacting} sting theory (IIB theory on the 
pp-wave background at large $\mu$). As perturbative gauge theories are 
under complete control, a detailed understanding of this
extremely explicit duality holds the promise of significantly enhancing our
understanding of gauge-string dualities in general. 

The rest of this paper is organized as follows. Section 2 
contains a review of subtle aspects of the BMN paper of importance to us. 
In section 3 we explain the counting that identifies $g_2=J^2/N$ as the 
genus counting parameter in free Yang-Mills. We also present the
computation of planar three-point functions and torus two-point
functions of BMN operators in free Yang-Mills theory. In section 4 we
compute the torus contribution to the anomalous dimensions of BMN
operators. In section 5 we present our proposals relating Yang-Mills 
computations to amplitudes of the string Hamiltonian. We also present 
a nontrivial unitarity check of our proposals, and compare our proposals to
string field theory. In section 6 we conclude with a discussion of our 
results and directions for future work. The reader who is uninterested in 
the details of perturbative computations of Yang-Mills correlators 
can skip from section 3.1 to section 5. In Appendix A we present a precise 
definition of a class of BMN operators. In Appendix B we prove that D-terms 
interactions do not contribute to the correlation function computations
presented in this paper. In Appendix C we present a rigorous and
self-contained derivation of two point functions of BMN operators. In
Appendix D we present an alternative method for Yang-Mills computations. 

\vspace{3mm}

{\bf Note:} As we were completing our manuscript, related papers appeared
on the internet archive \cite{germanpaper,berencompete,grosscompete}. 
\cite{germanpaper} overlaps with parts of sections three and four of our
paper, while \cite{berencompete} overlaps with parts of section 3 and
section 5.3 of this paper. Our results disagree
with those of \cite{germanpaper} and \cite{berencompete} in certain 
important respects. Unlike both of these papers we find non vanishing anomalous
dimensions for BMN operators on the torus at first order in Yang-Mills
coupling. We identify $g_2\sqrt{\lambda'}$ rather 
than $g_2$ as the effective string coupling at large $\mu$. 
As noted above, we have presented a rather non-trivial
unitarity check of our proposals. We have also compared our proposal 
for the three-string vertex with the Green-Schwarz string field theory
\cite{Spradlin:2002ar}.

\section{Preliminaries\label{prelims}}

\subsection{The BMN operators\label{prelimsbmn}}

The simplest single-trace operator with R-charge $J$ is
\begin{equation}\label{cpdef}
O^J = \frac1{\sqrt{JN^J}}\Tr Z^J,
\end{equation}
where
\begin{equation}
Z = \frac{X^5+iX^6}{\sqrt2}.
\end{equation}
This is a chiral primary operator, with scaling dimension exactly equal to
$J$ at all $\lambda'$. According to the BMN proposal it corresponds
to the light-cone ground state $\ket{0,p^+}$, where the map between
parameters is given by \eqref{bmnmap}.

Other protected operators may be generated from $O^J$ by acting on it 
with $SO(6)$, conformal, or supersymmetry lowering operators. For example,
by acting on $O^{J+1}$ with a particular $SO(6)$ lowering operator yields
\begin{equation}\label{oneimpurity}
O^J_0 = \frac1{\sqrt{N^{J+1}}}\Tr\left(\phi Z^J\right),
\end{equation}
where we have defined the complex combinations of the scalars:
\begin{equation}\label{fielddef}
\phi = \frac{X^1+iX^2}{\sqrt{2}}, \qquad
\psi = \frac{X^3+iX^4}{\sqrt{2}}. \qquad
\end{equation}
$O^J_0$ is chiral with scaling dimension $\Delta=\Delta_0=J+1$; 
it corresponds to the string state $a_0^{\phi\dag}\ket{0,p^+}$, where
$a_0^{\phi\dag} = (a_0^{1\dag}+ia_0^{2\dag})/\sqrt{2}$. To take another 
example, $O^{J+2}$ acted on by two distinct $SO(6)$ lowering operators 
yields the protected operator 
\begin{equation}\label{bnnopf}
O^J_{n,-n} =
\frac1{\sqrt{JN^{J+2}}}
\sum_{l=0}^J\Tr\left(\phi Z^l\psi Z^{J-l}\right)
\end{equation}
which corresponds to the BPS string state $a_0^{\psi \dag}a_0^{\phi\dag}
\ket{0,p^+}$. Proceeding in this manner, all protected operators (operators 
dual to supergravity modes) of the Yang-Mills theory may be obtained by 
acting on $O^J$, for some $J$, with the appropriate number of 
lowering operators of various sorts.

As noted in the introduction,  
only protected operators remain in the spectrum as $N$ is taken to infinity 
with $\gYM$ held fixed, in any sector of fixed charge $J$. 
However when $J$ is taken to infinity together with $N$, it is possible to 
construct operators that are locally BPS. These operators consist of
finite strings of fields (all of which are BPS) that are sewn together (in
the trace) with varying phases into an operator of length $J \to \infty$
that is not precisely BPS.
%
An example of such a near BPS operator is 
\begin{equation}\label{bnnop}
O^J_{n,-n} =
\frac1{\sqrt{JN^{J+2}}}
\sum_{l=0}^Je^{2\pi inl/J}\Tr\left(\phi Z^l\psi Z^{J-l}\right)
\end{equation}
We will usually abbreviate this as $O_n^J$; however we must be careful to
distinguish between the two chiral operators $O_0^J$ and $O^J_{0,0}$.
In an inspired guess, BMN conjectured that the operator $O_n^J$ corresponds
to the string state $a^{\phi\dag}_na^{\psi\dag}_{-n}\ket{0,p^+}$.
As we have emphasized above, for $n\neq0$ this operator is 
weakly non-chiral and its scaling dimension is corrected. However these 
corrections are finite, and may be expanded in a power series in $\lambda'$
(this result follows to low orders from direct computation, but
independently,  to all orders by comparison 
 with the exactly known string spectrum). Operators corresponding 
to more than two string oscillators acting on the vacuum are 
discussed in appendix A.

$O_n^J$ was obtained from $O^{J+2}$ by replacing two $Z$'s by the 
`impurities' $\phi$ and $\psi$, and sprinkling in position dependent
phases. The impurities $\phi$ and $\psi$ were obtained by the action of 
$SO(6)$ lowering operators on $Z$. In an analogous manner the impurity  
$D_\mu Z$ may be obtained by acting on $Z$ with the generators of conformal
invariance. Similarly, supersymmetry operators acting on $Z$ produce
gauginos. General BMN operators consist of these impurities sprinkled in a
trace of $Z$'s, together with phases. For the purposes of this paper it 
will be
sufficient to consider only scalar impurities, but we will explain in section 
\ref{secfive} how our ideas can be extended and checked with 
the other types of impurities.

As we have stressed in the introduction, 
the dimensions of operators such as $O_n^J$ remain finite (and
perturbatively computable at small $\lambda'$) in the limit of infinite 
't~Hooft coupling only because these operators differ very slightly from 
protected chiral operators. It is very important that the operator
$O_n^J$ is defined to reduce precisely to the chiral operator $O_{0,0}^J$
when $n$
is set to zero. Even a small modification in the definition of this
operator (such as a modification of the range of summation of the variable $l$ 
to $1,\dots,J$, as originally written in \cite{ias}) introduces a
small---$\CO(1/\sqrt{J})$---projection onto operators that are far from
chiral, resulting in 
perturbative contributions to scaling dimensions like
$\gYM N/J$, which diverges in the BMN limit, and hence a breakdown of 
perturbation theory.\footnote{The importance of the summation
range $0,\dots,J$ has been also recognized by the authors of
\cite{germanpaper}.}

\subsection{On the applicability of perturbation theory in the BMN limit}

Consider the perturbative computation of, say, the planar 
scaling dimension $\Delta$ of a BMN operator such as $O_n^J$ 
in \eqref{bnnop} above. 
Suppressing all dependence on $n$, the results of a
perturbative computation may be organized (under mild assumptions) as  
\begin{equation} \label{ddim}
\Delta=\sum_{m=0}^\infty \left(\frac{\gYM N}{J^2}\right)^m f_m(\gYM N),
\end{equation}
where $f_m$ are unknown functions of the 't~Hooft coupling. BMN computed
the planar part of
$f_1(0)$ using Yang-Mills perturbation theory, and deduced $f_1(\infty)$ 
using the duality to string theory on the pp-wave background. 
 Quite remarkably they found that $f_1(0)=f_1(\infty)$. This result 
suggests that $f_1(x)$ is a constant function at the planar level. 
Recently, the authors of 
\cite{grosscompete} have demonstrated that $f_2(0)=f_2(\infty)$, and 
have presented arguments which suggest that the planar components of 
$f_m(x)$ are constant functions for all 
$m$. Note that a term proportional to $x^m$ in $f_m(x)$  would result 
in the breakdown of perturbation theory, in the BMN limit, at order 
$(\gYM N)^{m+n}$. Consequently, the conjecture that $f_m(x)$ are 
constant functions for all $m$ is identical to the conjecture that 
$\lambda'=\gYM N/J^2$ is the true perturbation parameter, 
for the computation under consideration, in the BMN limit. 

In this paper we will proceed on the assumption that $\lambda'$ is indeed
the perturbative parameter for the computations we perform, namely low
order calculations of non-planar anomalous dimensions and three-point
functions of BMN operators. We will see that \eqref{ddim} acquires
extra non-planar contributions proportional to
$(J^4/N^2)^h$ from genus $h$ diagrams. These contributions
are finite in the BMN limit but they
can be expanded in $\lambda'$ just like \eqref{ddim}. All our results are
consistent with this conjecture, and lend it further support;  however, it
would certainly be interesting to understand this issue better.

\vspace{-1.7mm}

\section{Correlators in free Yang-Mills theory\label{kore}}

\vspace{-1.8mm}

\subsection{Correlators of chiral operators at arbitrary 
genus\label{korech}}

\vspace{-0.6mm}

\EPSFIGURE[l]{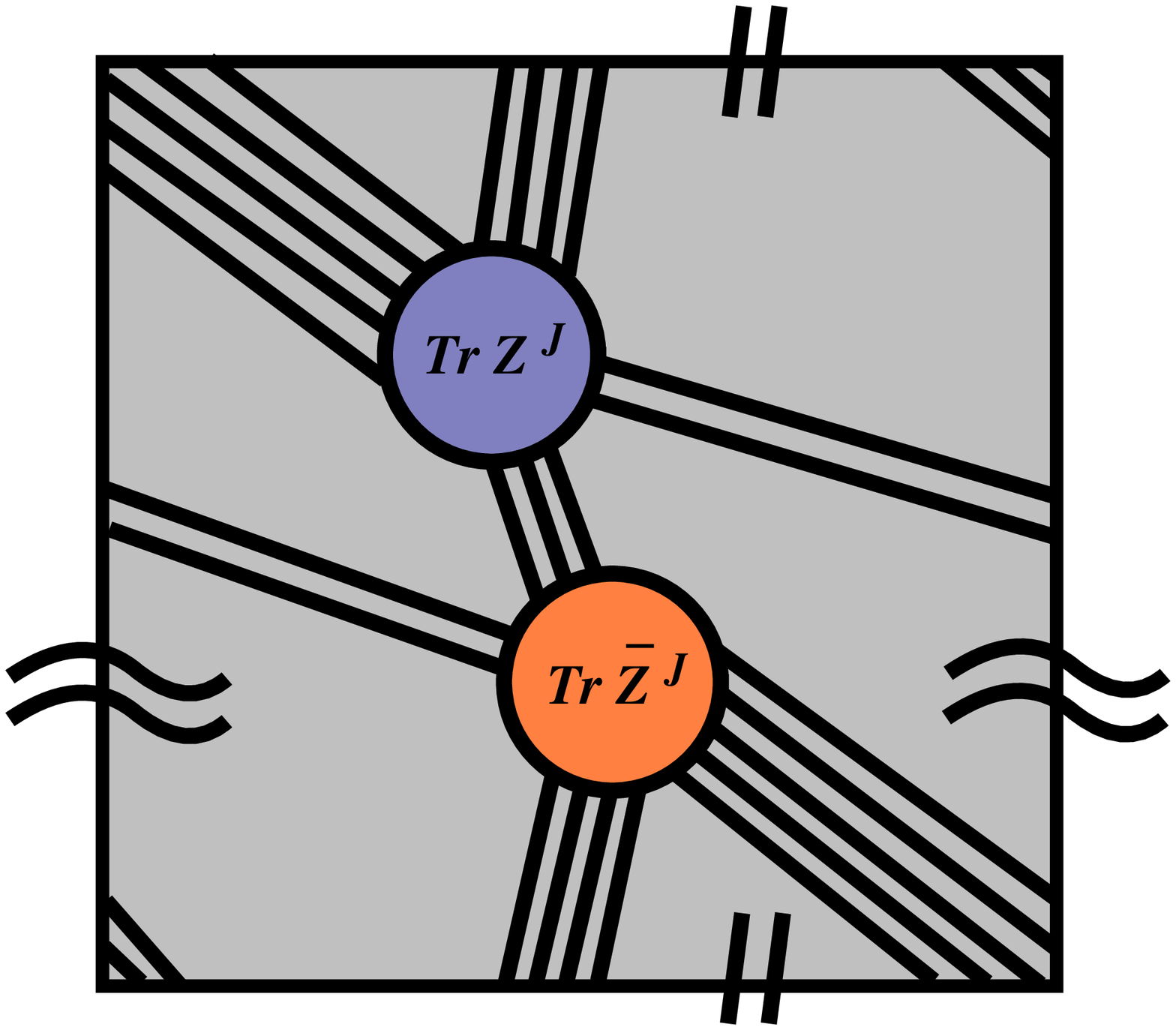,width=7.0cm}{Genus one diagram
drawn on a square.\label{shiraztorus}}

Consider a correlation function 
involving operators of typical size (R-charge) $J$ in free $U(N)$ 
Yang-Mills 
theory.\footnote{Most formulae simplify for $U(N)$
as compared to $SU(N)$ and the relative difference is of 
order $1/N$. We therefore choose to work with the gauge group $U(N)$.}
Following BMN, we study this correlator in the large $N$ limit; 
$J$ is simultaneously scaled to infinity with $J^2/N$ held fixed.
In this section we will demonstrate that the 
number of graphs that contribute to this correlation function at genus $h$ 
scales with $J$ like $J^{4h}$. Since any particular genus $h$ graph is 
suppressed by a factor of $1/N^{2h}$ compared to a planar graph, 
we conclude that the net contribution of all genus $h$ graphs remains
finite in the BMN limit, scaling like $g_2^{2h}$ where 
$g_2=J^2/N$. Consequently $g_2^2$ is a genus counting parameter; 
it determines the relative importance of higher genus graphs in free
Yang-Mills theory.\footnote{It was previously observed in 
\cite{Balasubramanian:2001nh, Corley:2001zk} that operator mixing and 
higher genus contributions to correlation functions are important, even  
as $N \to \infty$ for operators whose size scales with $N$. 
\cite{Corley:2001zk} has also presented detailed formulae for free field 
correlators at all $N$ in a basis different from that employed in this paper.}

The free Yang-Mills genus expansion encodes a modification in the dictionary 
between string states and Yang-Mills operators, but does not in itself
appear to 
contain information about string interactions. We will return to the
question of true string interactions in sections 4 and 5 below. 

Consider the two-point function $\langle \bar O^J(0)O^J(x)\rangle$
in free Yang-Mills theory, where the operators $O^J$ are defined in
\eqref{cpdef}. The planar contribution to this two-point function is
\begin{equation}\label{planartpp}
\langle \bar O^J(0)O^J(x)\rangle_\te{planar}
=\frac{1}{(4\pi^2x^{2})^J}.
\end{equation}

\FIGURE[ht]{\quad\epsfig{file=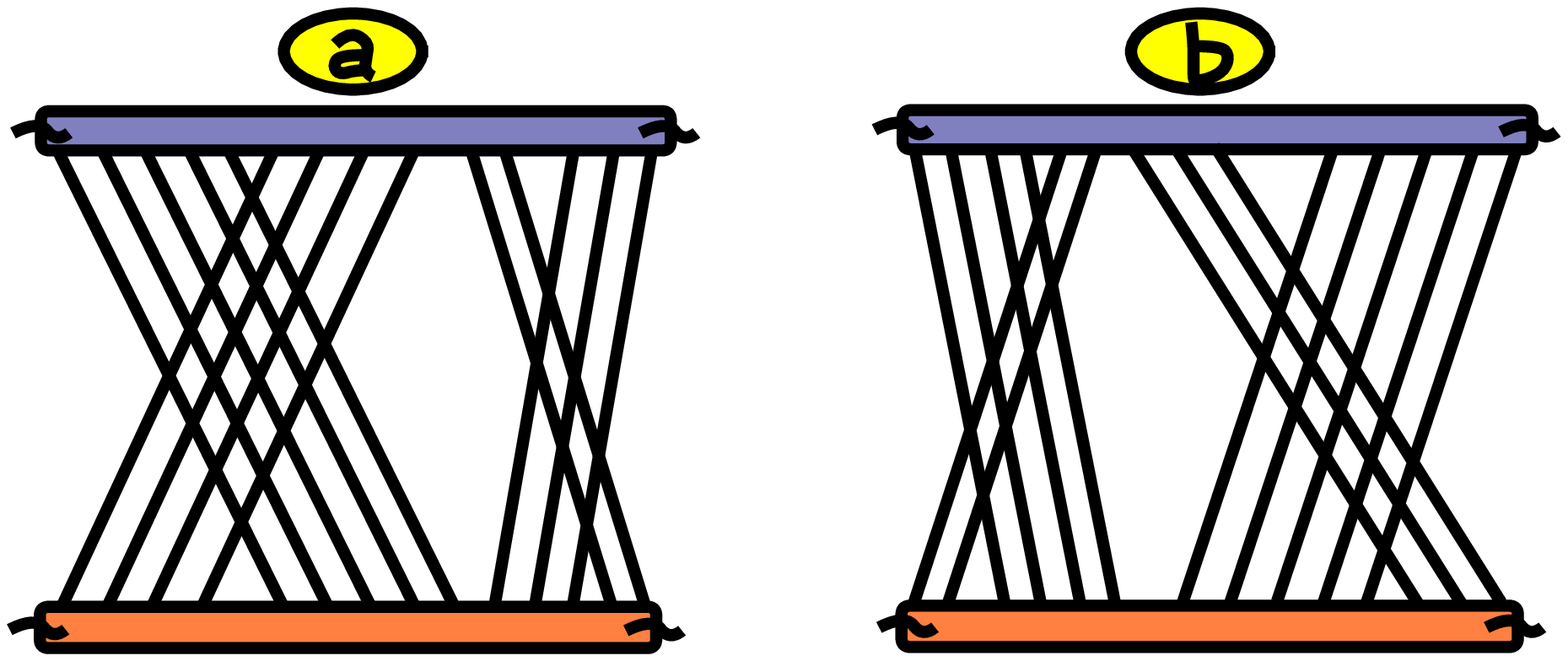,width=12cm}
\caption{The same genus one diagram in an alternative representation.
The diagram can be interpreted so that (a) the blue string
$\Tr Z^{5+4+2+3}$ splits into $\Tr Z^{5+4}$ and $\Tr Z^{2+3}$;
these two strings rotate by 5 or 2 units, respectively, to get
the final state contracted with $\Tr \bar Z^{4+5+3+2}$. The same free
theory diagram however also counts (b) a similar split of the string into
$\Tr Z^{4+2}$ and $\Tr Z^{3+5}$.}
\label{torusvertical}}

To find the genus 1 contribution to the correlator, we must find all the
free diagrams that can be drawn on the torus but not on the sphere. To do
this the $J$ propagators must be divided into either 3 or 4 groups (see
figure \ref{shiraztorus}). The number of ways to do this is
\begin{equation}
\binom{J}{4} + \binom{J}{3} = \binom{J+1}{4} \approx 
\frac{J^4}{4!}.
\end{equation}
This must be multiplied by $J$ for overall cyclic permutations, but then 
divided by $J$ again due to the normalization of the operator, and
also by $N^2$ due to the genus. The resulting quantity is finite in the BMN
limit, and proportional to $g_2^2$:
\begin{equation}\label{torustp}
\langle \bar O^J(0)O^J(x)\rangle_\te{torus}
=\frac{g_2^2}{4!(4\pi^2x^{2})^J}.
\end{equation}

\FIGURE[ht]{\qquad\epsfig{file=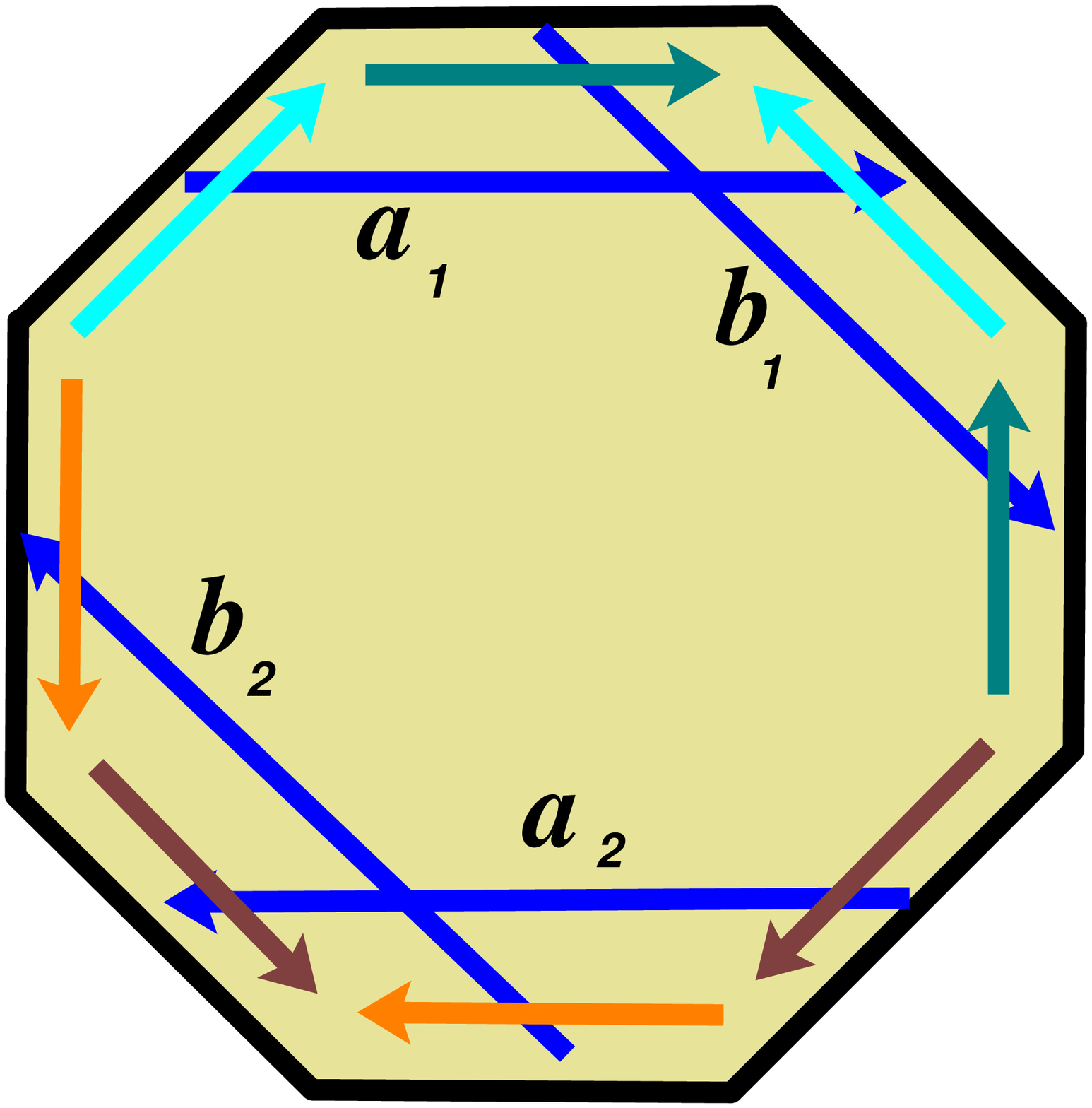,width=4.5cm}\qquad\qquad
\epsfig{file=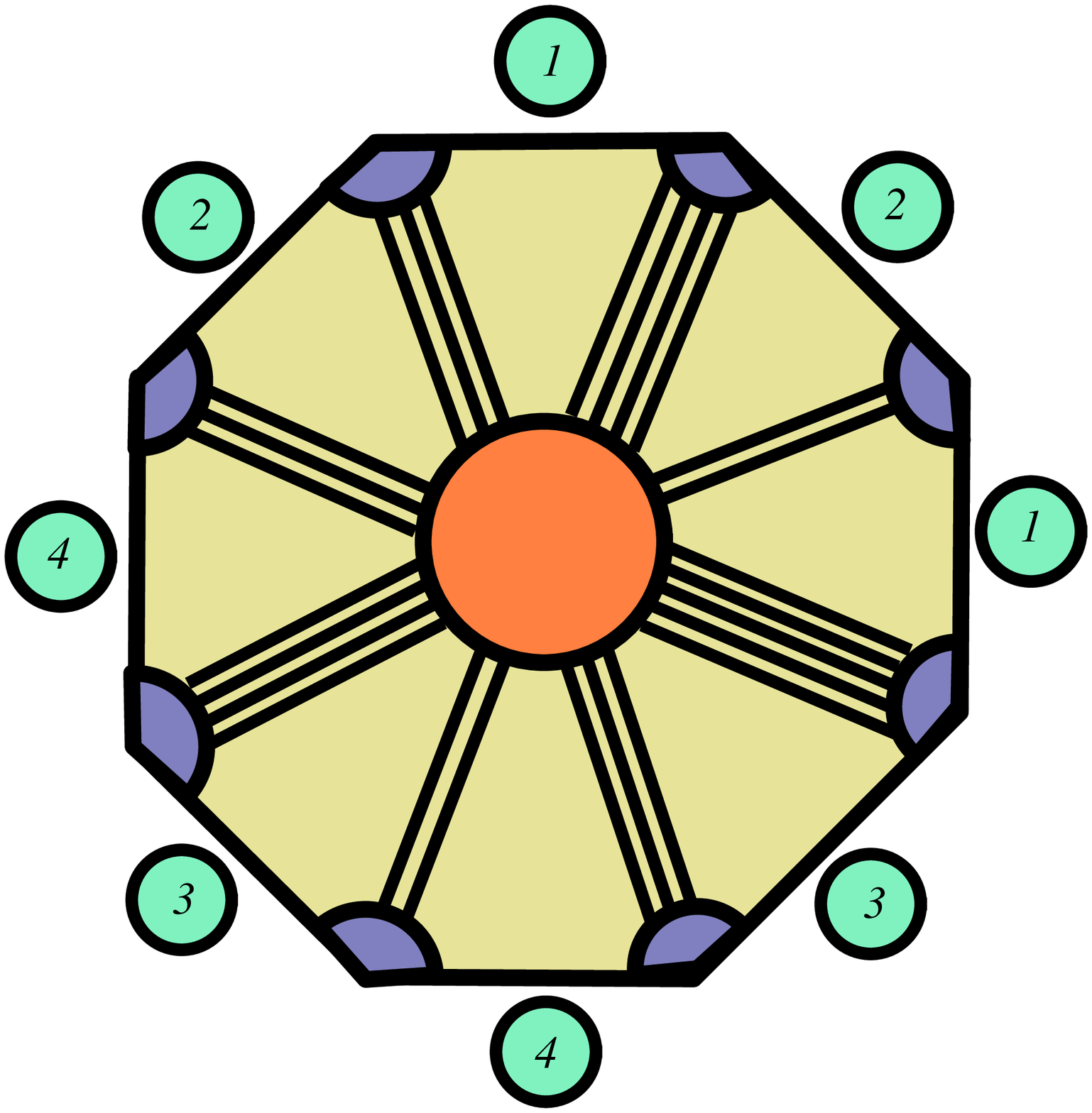,width=4.5cm}
\caption{(a) Genus two surface represented as an octagon.
The vertices of the octagon are all identified,
while the edges are identified pairwise. The usual homology
one-cycles with intersection numbers $\#(a_i,b_j)=\delta_{ij}$ are 
depicted.
\newline 
(b) Genus two diagram. One operator is located at the center, and the 
other at the vertices. This way of drawing the diagram immediately 
generalizes to any genus.\vspace{-1mm}}
\label{octagons}}

This counting is easily extended to arbitrary genus. A genus $h$ Feynman
graph can be drawn on a $4h$-gon with sides identified pairwise. As we see
from figure \ref{octagons}, the number of graphs that can be drawn on a
$4h$-gon is the number of ways of dividing $J$ lines into $4h$ groups,
which is $J^{4h}/(4h)!$. (The lines may also be divided into $4h-1$ groups,
but this gives a vanishing contribution in the BMN limit.) We must
multiply this by the number of inequivalent ways of gluing the sides of a 
$4h$-gon into a genus $h$ surface. This number has been computed \cite{HZ};
the result is
\begin{equation}\label{combin}
{\frac{1\cdot 3 \cdots (4h-1)}{2h+1}}.
\end{equation}
Consequently a total of $2^{-2h}J^{4h}/(2h+1)!$ graphs contribute to 
this
correlator at genus $h$.  Summing over genera we find
\begin{equation}\label{sumall}
\langle \bar O^J(0)O^J(x)\rangle =
\frac{1}{(4\pi^2x^2)^J}\sum_{h=0}^\infty
\frac1{(2h+1)!}\left(\frac{g_2}2\right)^{2h}
= \frac{2\sinh(g_2/2)}{g_2(4\pi^2x^2)^J}.
\end{equation}

This method can easily be generalized to show that the two-point function
for an arbitrary chiral BMN operator
such as $O^J_0$ or $O^J_{0,0}$ (defined in \eqref{oneimpurity} and
\eqref{bnnop} respectively) has the same coefficient as in
\eqref{sumall}. Thus for example,
\begin{equation}\label{po}
\langle \bar O^J_{0,0}(0)O^J_{0,0}(x) \rangle
= \frac{2\sinh(g_2/2)}{g_2(4\pi^2x^2)^{J+2}}.
\end{equation}

The easiest way to generalize to higher-point functions of chiral operators
is probably via a Gaussian matrix model.  For example, it is not 
difficult to compute 
\begin{equation}\label{act}
\int {\cal D}Z\, {\cal D} {\bar Z} \left[ 
\left( 
\prod_{i=1}^k \Tr Z^{J_i} \right) \Tr{\bar Z}^J
e^{-\Tr(Z \bar{Z})} \right];
\qquad
\sum_{i=1}^k J_i=J
\end{equation}
yielding simple explicit formulae
that generalize \eqref{sumall}.\footnote{These formulae have been 
obtained
in collaboration with M. van Raamsdonk. They have also been presented in
detail in the recent paper \cite{germanpaper}.}

\subsection{Planar three-point functions}

In this subsection we compute free planar three-point functions
for the BMN operators defined in subsection \ref{prelimsbmn}.
The results we obtain will be used in section \ref{secfive} when we discuss
the construction of  string interactions.

We will first compute 
\begin{equation}\label{tpf}
\langle \bar O_n^J(0)O_m^{J_1}(x_1)O^{J_2}(x_2) \rangle,
\end{equation}
where $J_1+J_2=J$. The planar, free field computation of this correlator is
summarized in figure \ref{threedelta}. The only complication is that we
must sum over all of the possible positions for the $\phi$ and $\psi$
fields and carefully keep track of combinatorial factors as well as
normalizations. The summation over the position of $\phi$ and $\psi$ in
$O_n^{J_1}$ may be converted into integrals in the large $N$ limit,
\begin{equation}\label{threeint} 
J_1^2 \int_{0}^1 da \int_{0}^1 db\,e^{2 \pi i a (m -ny)}
e^{-2 \pi i b(m-ny)} = 
J_1^2\frac{\sin^2\pi ny}{\pi^2(ny-m)^2}.
\end{equation}
where $y=J_1/J$. The final result for the correlator is obtained by
multiplying this integral by $J_2$ (from cyclic rotations of $O^{J_2}$)
and dividing by $\sqrt{J_1J_2J}$ (from the normalization of each operator)
and by $N$ (from $1/N$ counting). We find
\begin{equation}\label{tpff}
\langle {\bar O_n^J}(0)O_m^{J_1}(x_1)O^{J_2}(x_2) \rangle =
\frac{g_2y^{3/2}\sqrt{1-y}\sin^2(\pi ny)}
{\sqrt{J}\pi^2(ny-m)^2(4\pi^2x_1^2)^{J_1+2}(4\pi^2x_2^2)^{J_2}}.
\end{equation}

A similar calculation yields
\begin{equation}\label{threebps2}
\langle \bar  O_n^J(0) O_0^{J_1}(x_1) O_0^{J_2}(x_2) \rangle = 
\frac{g_2\sin^2(\pi ny)}
{\sqrt{J}\pi^2n^2(4\pi^2x_1^2)^{J_1+1}(4\pi^2x_2^2)^{J_2+1}},
\end{equation}
where $O_0^{J_1}$ and $O_0^{J_2}$ have $\phi$ and $\psi$ impurities
respectively.

These expressions for the three-point functions will play an important
role in our comparison between perturbative string theory and perturbative
Yang-Mills theory in section \ref{secfive}.

\subsection{Torus two-point functions of BMN operators}

\EPSFIGURE[r]{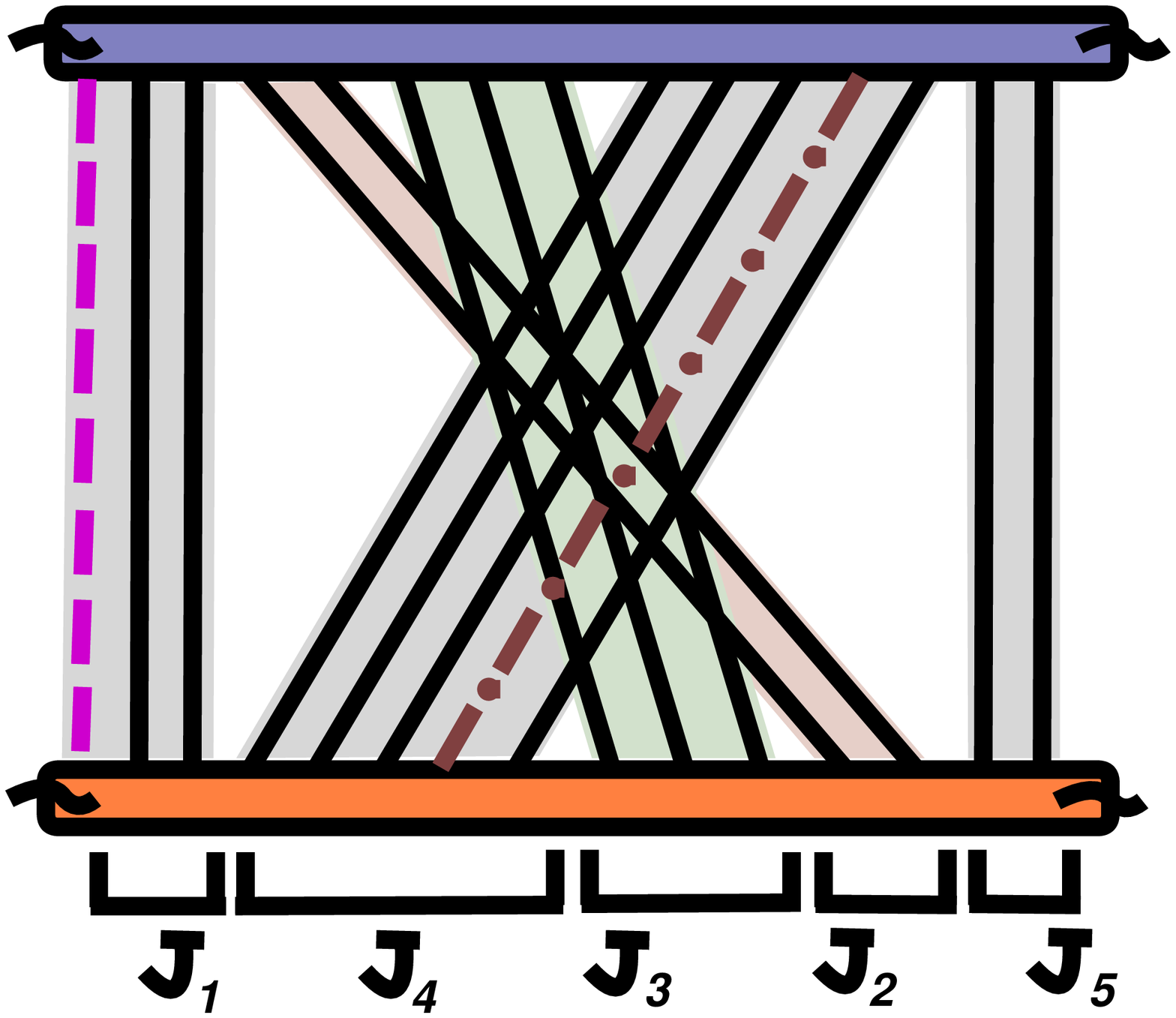,width=7cm}{A typical free   
genus one graph for the two-point function of an operator with two   
impurities.\label{torusc}}

In this subsection we present an 
explicit computation of the two-point functions for the BMN operators
\eqref{bnnop} at genus one in free Yang-Mills theory. The operator $O^J_n$
differs from the chiral operator $O^J_{0,0}$ only in the presence of 
phases. Torus (and, indeed all genus) two point functions of $O^J_{0,0}$ 
were computed rather easily in \eqref{po}. The additional phases 
complicates matters somewhat, as we will see below.  Nonetheless, it is not 
difficult to convince oneself that these additional complications 
affect only the details of the result, but not its scalings with $J$
and $N$. Indeed, $g_2^2$ is the correct 
genus counting parameter for all free Yang-Mills computations in the BMN
limit.
 
We consider~first~the~correlator $\langle \bar O_n^J(0) O_n^J(x)\rangle$. 
The
free genus one diagrams are given by the torus diagrams presented in the
last section (figure \ref{torusvertical}) with $J+2$ lines, summed over
all ways of replacing one line by a $\psi$ line and another by a $\phi$
line, with the rest becoming $Z$ lines. There are four groups of lines,
and if the $\psi$ and $\phi$ are
in different groups then their relative positions will be different in the
first and second operators, giving a non-trivial phase (unlike in the case of
planar diagrams where they are always the same distance apart in the first
and second operators). 

In fact it is convenient always to put the $\phi$
line at the beginning of both the first and second operator. With this
convention every torus diagram with one $\phi$ line and $J+1$ other lines can
be drawn as in figure \ref{torusc}, where each solid line
represents a group of $Z$
lines, and $J_1+\dots+J_5=J+1$. Now we must put in the $\psi$ line. If it
is in the first group ($J_1$ possibilities) or the last group ($J_5$
possibilities), then the phase associated with the diagram is 1, because it
doesn't move relative to the $\phi$ line. On the other hand if it's in the
second group ($J_2$ possibilities) then it moves to the right by $J_3+J_4$
steps, giving a phase $\exp(2\pi in(J_3+J_4)/J)$. Similarly for the
third and fourth groups, giving in all
$J_1 + J_2e^{2\pi in(J_3+J_4)/J} + J_3e^{2\pi in(J_4-J_2)/J}
+ J_4e^{-2\pi in(J_2+J_3)/J}+J_5$. We must now sum this over all ways of
dividing the $J+1$ lines into five groups:
\begin{align}
&\frac{(4\pi^2x^2)^J}{g_2^2}
\langle \bar O_n^J(0)O_n^J(x)\rangle_\te{free torus}
\notag \\
&\qquad= \frac1{J^5}
\sum_{\substack{J_1+\dots+J_5\\=J+1}}
\left(
J_1 + J_2e^{2\pi in(J_3+J_4)/J} + J_3e^{2\pi in(J_4-J_2)/J}
+ J_4e^{-2\pi in(J_2+J_3)/J} + J_5
\right) \notag \\
&\qquad \xrightarrow[N\to\infty]{}\int_0^1 \mathrm dj_1\cdots \mathrm dj_5\,
\delta(j_1+\dots+j_5-1) \notag \\
&\qquad\qquad\times\left(
j_1 + j_2e^{2\pi in(j_3+j_4)} + j_3e^{2\pi in(j_4-j_2)} +j_4e^{-2\pi
in(j_2+j_3)} +j_5
\right) \notag \\
&\qquad= \begin{cases}
\frac1{24}, & n=0, \\
\frac1{60}-\frac1{6(2\pi n)^2}+\frac7{(2\pi n)^4}, & n\neq0.
\end{cases}\label{freegenusone}
\end{align}
In taking the limit
$N\to\infty$ the fractions $j_i=J_i/J$ go over to continuous variables.

If we now consider a correlator of two different operators $O_n^J$ and 
$O_m^J$,
then the phase
associated with a diagram depends not just on which group the $\psi$ is
inserted into, but on where in the group it is inserted. The formulae are
therefore somewhat more complicated, but it's clear that again in the limit
$N\to\infty$ the two-point function will reduce to $g_2^2$ times a
finite integral:
\begin{align}
\frac{(4\pi^2x^2)^J}{g_2^2}&
\langle \bar O_n^J(0)O_m^J(x)\rangle_\te{free torus} \notag 
\xrightarrow[N\to\infty]{}
\int_0^1\frac{dj_1\cdots dj_5\,
\delta(j_1+\dots+j_5-1)}{i(U-V)}\left(
e^{ i(U-V)j_1}-1 \right. \notag \\
&\qquad\qquad\qquad\qquad\qquad\qquad\qquad\qquad
{}+e^{iUj_1- iV(j_1+j_3+j_4)}(e^{ i(U-V)j_2}-1) \notag \\
&\qquad\qquad\qquad\qquad\qquad\qquad\qquad\qquad{}
+e^{iU(j_1+j_2)- iV(j_1+j_4)}(e^{ i(U-V)j_3}-1)
\notag \\
& \qquad\qquad\qquad\qquad\qquad\qquad\qquad\qquad{}
+e^{iU(j_1+j_2+j_3)- iVj_1}(e^{ i(U-V)j_4}-1) \notag \\
&\qquad\qquad\qquad\qquad\qquad\qquad\qquad\qquad{}
+\left.e^{i(U-V)(j_1+j_2+j_3+j_4)}(e^{ i(U-V)j_5}-1)
\right) \notag \\
&= \begin{cases}
\frac1{24}, & m=n=0; \\
0, & m=0, n\neq0 \te{ or } n=0, m\neq0; \\
\frac1{60} - \frac1{6U^2} + \frac7{U^4}, & m=n\neq0; \\
\frac1{4U^2}\left(\frac13+\frac{35}{2U^2}\right), & m=-n\neq0; \\
\frac1{(U-V)^2}
\left(\frac13+\frac4{V^2}+\frac4{U^2}-\frac6{UV}-\frac2{(U-V)^2}\right), &
\te{all other cases}
\end{cases}\!\!\!\!\!\!\!\!\!\!\!\!\!\!\!\!\!\!\!\!\!\!\!\!\!\!\!\!\!\!\!\!\!
\end{align}

\EPSFIGURE[r]{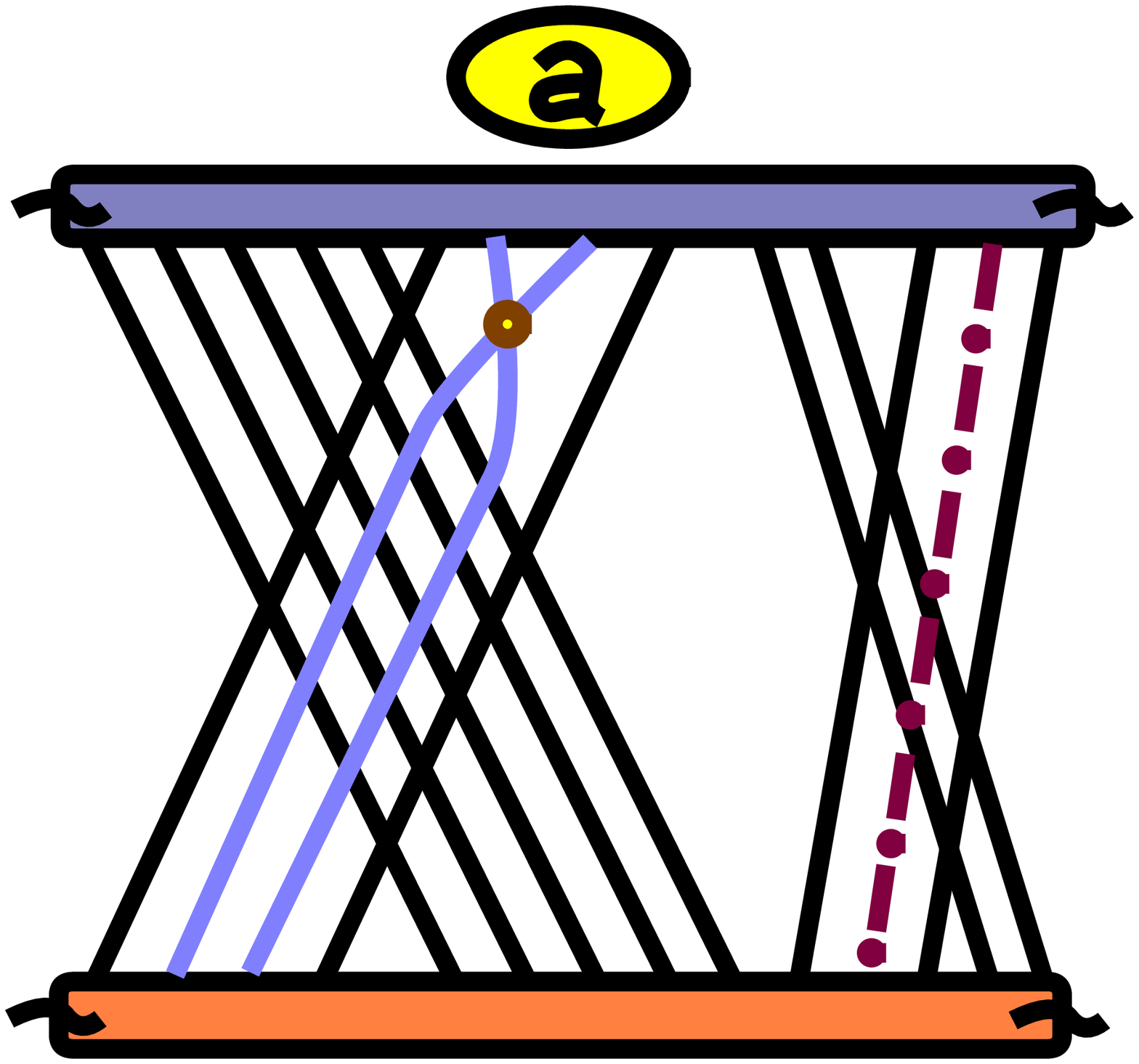,width=7cm}{A typical graph
with nearest neighbor interaction. Four (blue) lines coming from the
vertex should be replaced by all possible terms from figure
\ref{ftermfig}. The dashed line is a $\psi$ propagator.
\label{nearestnei}}

where $U=2\pi m, V=2\pi n$.
The result for the free two-point function including genus one
corrections can thus be summarized as
\begin{equation} \label{tfree}
\langle \bar O_n^J(0)O_m^J(x)\rangle_\te{free torus} = 
\frac{\delta_{nm} + g^2_2 A_{nm}}{(4\pi^2x^2)^J},
\end{equation}
where the entries for $A_{nm}$ are given above.
As $\langle O^J_n \bar{O}^J_n \rangle$ is non-zero for $n\neq m$ 
(unless either $n$ or $m$ is zero), we see that $O^J_n$ and $O^J_m$ mix
with each other, and that the mixing matrix elements are $\CO(g_2^2)$. 
 
It is clear that the above procedure generalizes to the higher genus free
diagrams described in section \ref{prelims}, in which the lines are 
divided into $4h$ groups. The genus $h$ contribution to the two point
function may be written as $g_2^{2h}$ times a finite  integral over
$4h+1$ parameters. See appendix C for a rigorous, general discussion.

\section{Anomalous dimensions from torus two-point 
functions\label{anosec}}


The planar anomalous dimension of the operator $O_n^J$ is related, 
via the duality with string theory, to the 
light-cone energy (or dispersion relation) of the
corresponding free string state. The planar anomalous dimension
was computed to first
order in $\gYM$ in \cite{ias}. Their result was of order 
$\CO(\lambda'$), in precise agreement with the free spectrum of strings in
the pp-wave background \eqref{ppback}. On the other hand, the contribution 
to the anomalous dimensions from genus one gauge theory diagrams is 
related to the string one loop corrected dispersion relation for the
corresponding state (see Section 5 for more details). In this section we 
compute the anomalous dimension of $O_n^J$ on the torus, to first order in 
$\gYM$. We find a result proportional to $g_2^2 \lambda'$, in accord with 
the identification of $g_2^2$ as the gauge theory genus counting parameter, 
and $\lambda'$ as the effective gauge coupling. This result is a prediction 
for the one string loop `mass renormalization' of the corresponding state.

Below we present a diagrammatic computation of this anomalous
dimension; in appendices C and D two independent rigorous calculations
confirm and generalize the results of this section.

We will find it convenient to think of the  $\N=4$ Lagrangian in $\N=1$
language; $Z,\phi,\psi$  are the lowest components of the three adjoint
chiral superfields of this theory. Most of the interactions of the theory, 
including scalar-gluon (and ghost) interactions and scalar-scalar 
interactions of the form $\Tr\left|[Z,\bar\phi]\right|^2$  
are `flavor blind' 
(see Appendix B). The contribution of these terms to this correlator is
identical to their contribution to $\langle \Tr O^J \bar{O}^J \rangle$;
consequently they vanish to order 
$\gYM$ by the theorem proved in  
\cite{9907098} (see Appendix B for more details). 
Consequently, only flavor sensitive terms in the Lagrangian, i.e. F-terms,
contribute to our calculation. The F-term
interactions between scalars are very simple 
\eqn{ftermm}{
V_F = -4\gYM\Tr\left(|[Z,\phi]|^2 + |[Z,\psi]|^2 + 
|[\phi,\psi]|^2\right).
}
Further, at the order under consideration, the last term does not
contribute, as it is the square of a term anti-symmetric under 
$\phi\leftrightarrow \psi$ and so has vanishing Wick contractions with
$O_n^J$ and its conjugate, as these operators are symmetric in $\phi$ and
$\psi$. In summary, to the order under consideration, the two impurities do
not talk to each other, and may be dealt with individually. Further, 
each impurity effectively only interacts quartically with the $Z$ fields
through the interactions in \eqref{ftermm}.

We now turn to the computation of all diagrams with a single F-term
interaction. Consider contributions to the two point function 
\begin{equation} \label{tpfd}
\langle O^J_n (0) \bar {O}^J_n(x)\rangle.
\end{equation}
from Feynman diagrams with a single $Z, \phi$ interaction vertex.
All such graphs (see figure \ref{nearestnei} for one example)
have identical spacetime dependence and their Feynman
integral is proportional to 
\eqn{lambdaint}{
\frac{1}{16\pi^4}
\int \frac{d^4 y}{y^4(y-x)^4}=\frac{\ln (\Lambda^2x^2)}{8\pi^2 x^4}.
}
We work in  position space in \eqref{lambdaint}; $y$ represents the 
position of the interaction point, which must be integrated over all 
space. 
The integrand in \eqref{lambdaint} consists of two propagators from 
$O^J_n (0)$ to the interaction point multiplied by two propagators from 
$\bar{O}^J_n(x)$ to the interaction point $y$, (see figures 
\ref{nearestnei} and \ref{cpdef}). 

\FIGURE[ht]{\epsfig{file=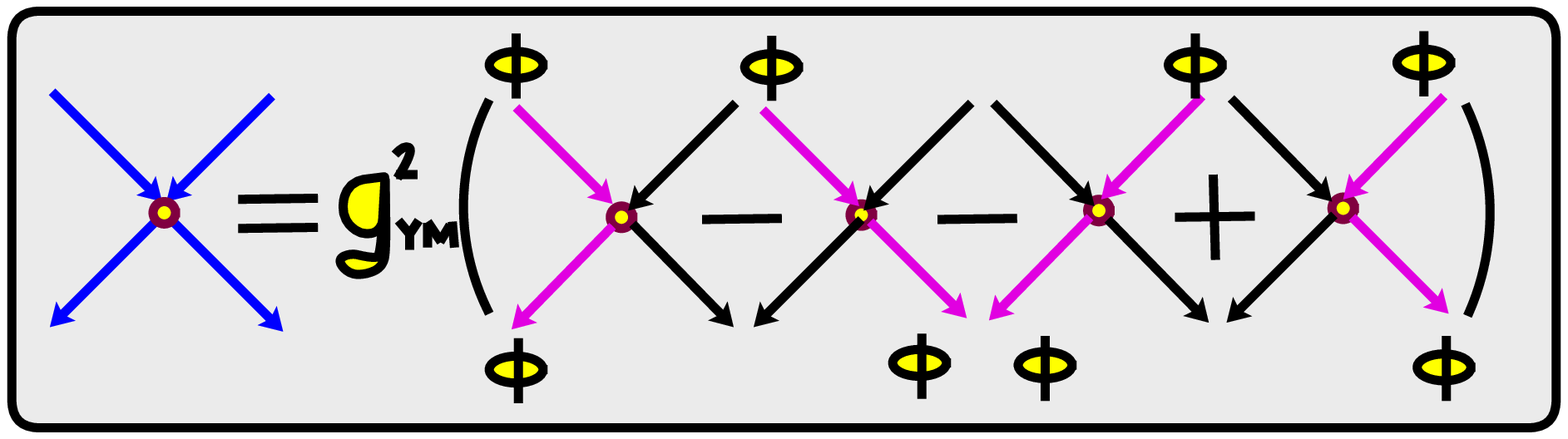,width=15cm}
\caption{The quartic $Z$-$\phi$ vertices coming from the F-term   
\eqref{ftermm}.
\label{ftermfig}}}

\vspace{2mm}

In order to complete the computation of the torus two point function 
we must 
\begin{itemize}
\item{\bf (a)} Enumerate all graphs that can be drawn with a single 
F-term
interaction on the torus. 
\item{\bf (b)} Evaluate each of these graphs ignoring the propagators from 
the 
two operators to the interaction point (this corresponds to evaluating the 
corresponding free graph) and then multiply the result by 
\eqref{lambdaint}.
\item{\bf (c)} Sum over the contribution from all these graphs. 
\end{itemize}

In the rest of this section we carefully carry through this process to 
compute $\langle O^J_n (0) 
\bar{O}^J_n(x)
\rangle$ on the torus, to first
order in the Yang-Mills coupling. It turns out that the  graphs that 
contribute may be categorized into three separate groups;
nearest neighbor, semi-nearest neighbor, and non-nearest
neighbor graphs, respectively. 

\vspace{2mm}

\subsection{Nearest neighbor interactions}

Consider for example the diagram shown in figure \ref{nearestnei},
in which
two adjacent lines in a free diagram such as figure \ref{torusvertical} are
brought together at an interaction vertex. We use the convention that 
diagrams at figure \ref{nearestnei} actually represent the sum of four 
different Feynman diagrams. In diagrams such as figure \ref{nearestnei}, 
one of the lines connecting each of the
operators to the interaction point is always  $\phi$ 
propagator (two choices for each operator) and the other line always 
represents a $Z$ propagator. The four Feynman diagrams correspond to the 
four possible choices. 
The dashed line on the right in figure \ref{nearestnei} represents a 
$\psi$ propagator. The four Feynman graphs that constitute the process 
depicted in figure \ref{nearestnei} each contributes with the same weight; 
but graphs in which a $\phi$ line crosses the $Z$ line contribute with a 
relative minus sign (this follows from the fact that the interaction 
is derived from  $\gYM\Tr\left|[Z,\phi]\right|^2$), as shown in figure
\ref{ftermfig}. The total contribution of these 
four diagrams is thus
\eqn{nnt}{
-\frac{\gYM}{N}(1-e^{2\pi in/J})(1-e^{-2\pi in/J}) \approx
-\frac{\lambda'}{N^2}(2\pi)^2n^2}
\EPSFIGURE[r]{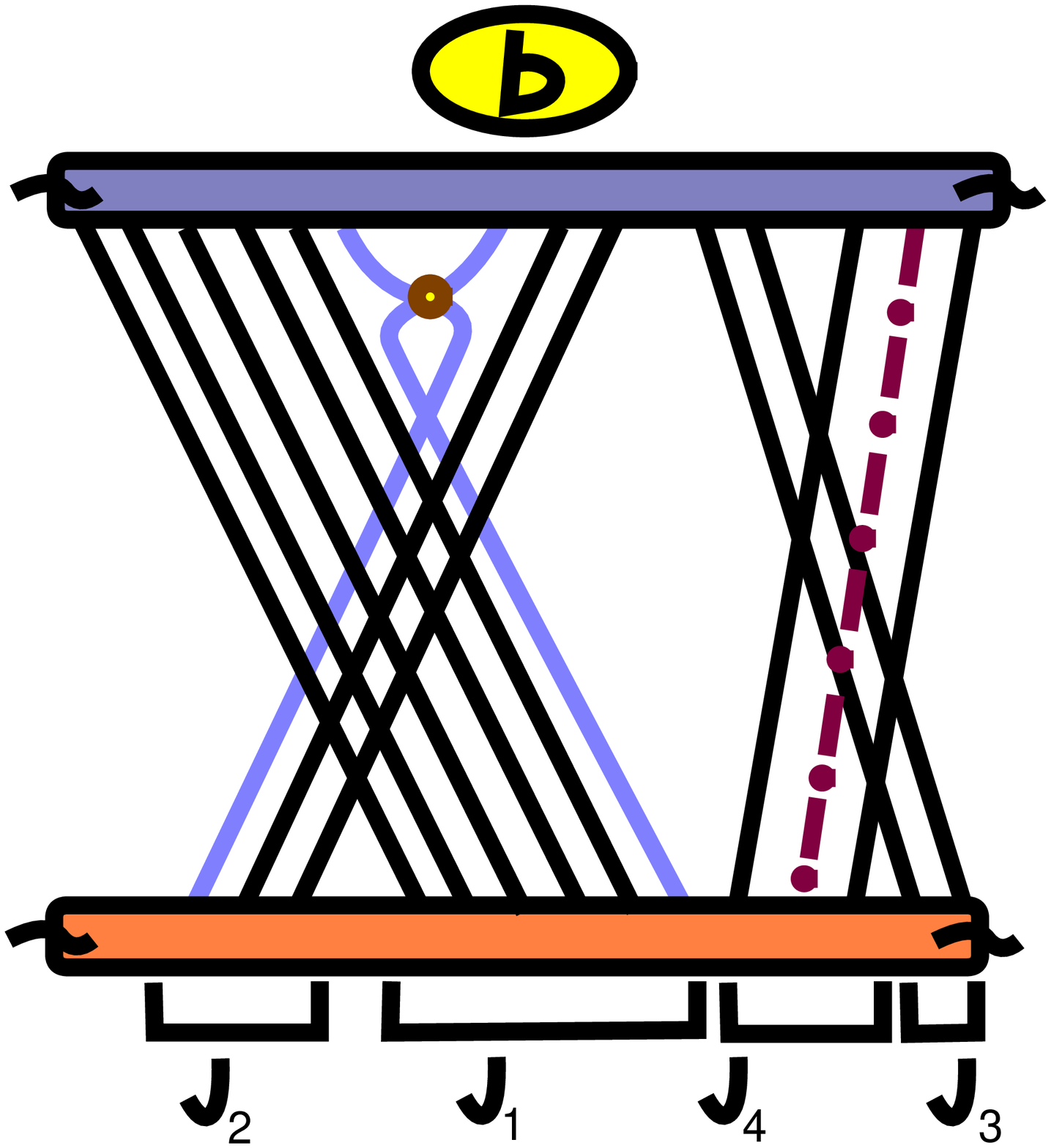,width=7cm}{A typical graph
with semi-nearest neighbor interaction.\label{seminear}}
\noindent times the phase associated to the corresponding free diagram.
\eqref{nnt} is independent both of which two lines in the free diagram 
figure
\ref{torusvertical} we are considering. It also does not depend on 
which particular free diagram is under consideration. Consequently,  
the sum of all such ``nearest-neighbor'' diagrams is simply
\eqref{nnt} multiplied by $A_{nn}$, the genus one contribution to the free
correlator \eqref{tfree} calculated in the previous section (with an
additional factor of two from diagrams in which the interaction involves 
the $\psi$ rather than the $\phi$ field).

Summing up all these diagrams, together with the free torus diagrams
computed in this section, and adding these contributions to the free and 
one loop planar results computed in BMN we obtain the following correlator:
\begin{equation}\label{incomplete}
\< \bar O_n^J(0) O_n^J(x)\> =\frac{1}{(4\pi^2x^2)^{J+2}}
\left(1 - \lambda^{\prime} n^2 \ln(\Lambda^2x^2)\right) 
\left(1+g_2^2 A_{nn}\right) + \cdots.
\end{equation}

Consequently, the diagrams studied in this subsection merely correct the
coefficient of the logarithm in the two point function to account for the 
changed normalization of the operator $O_n^J$, as computed in the previous 
section. If there were no further contributions to the coefficient of the 
logarithm, this result would imply that torus diagrams do not contribute to
the anomalous dimensions of the BMN operators.\footnote{The  
unphysical nature of these contributions to the two-point
function was also recognized in \cite{germanpaper}.}
In fact other diagrams we describe in the next two sections do modify the
scaling dimensions, as we describe below. 

\EPSFIGURE[r]{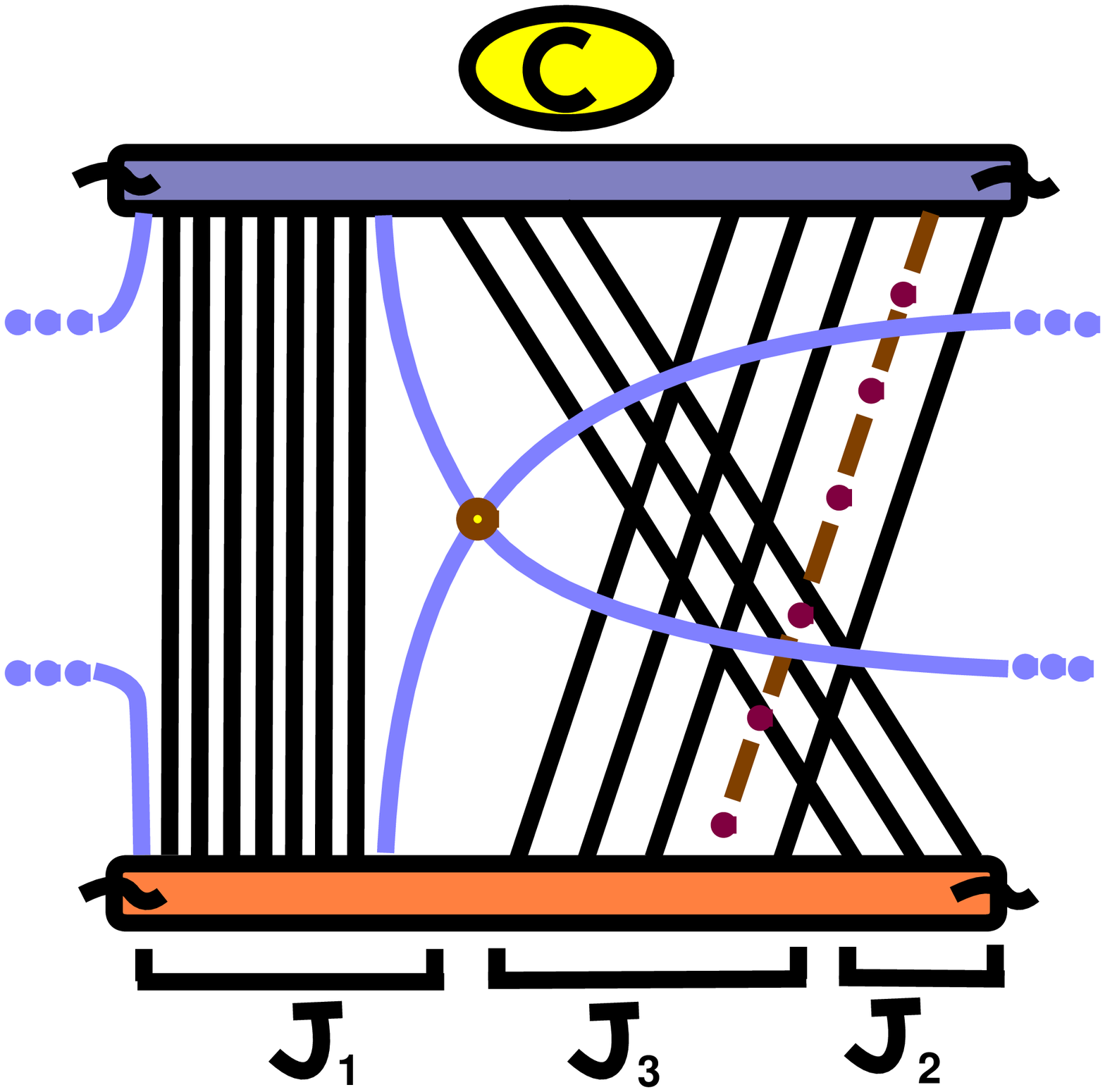,width=7.5cm}{A typical graph
with non-nearest neighbor interaction. Those graphs contribute to the
actual anomalous dimension.
\label{nonnearr}}

\subsection{Semi-nearest neighbor interactions$\!\!\!\!\!\!\!\!\!\!$}

There are two other classes of
diagrams, illustrated in figures \ref{seminear} and \ref{nonnearr} (see also
figure \ref{nonnears}) that could potentially contribute to the correlator at order
$g_2^2\lambda'$, and thus to the anomalous dimension.
As we will demonstrate below, the ``semi-nearest-neighbor''
diagrams of figure \ref{seminear}, in which the fields involved in the
interaction are adjacent in one but not the other operator, 
contribute to the two point function $\langle O^J_n(0)
\bar{O}^J_m(x)\rangle$ only when $m \neq n$. 
However, the ``non-nearest neighbor'' diagrams of figure
\ref{nonnearr} contribute to the logarithmic divergence of this 
correlator whether or not $n=m$. Consequently, these diagrams 
result in a genuine shift in the anomalous dimension of $O^J_n$. 

It is not difficult to argue that no other classes of diagrams contribute
to this process. To verify this claim, consider all diagrams with a single
quartic interaction, that can be drawn on a torus. Each such diagram
involves two propagator loops involving the interaction point. All diagrams
fall into four classes; diagrams in which each of these loops is
contractible, in which one loop is contractible and the other wraps a cycle
of the torus, in which both loops wrap the same cycle of the torus, and
finally those in which the two loops wrap different cycles of the torus. 
Further dressing these diagrams with all sets of propagators that leave it 
genus one, we find that the first class constitutes nearest neighbor
graphs of the form figure \ref{nearestnei}, the second set constitutes 
semi-nearest neighbor graphs of the form figure \ref{seminear}, the third 
set
constitutes nonnearest neighbor graphs of the form \ref{nonnearr} and the
last set cannot be implemented with F-term interactions. In Appendix C 
and D we verify this result using different techniques. 

In this subsection we discuss the semi-nearest-neighbor diagrams.
There are exactly eight diagrams of this type corresponding to
the number of ways one may choose the last member of a given
group in figure \ref{seminear} to
interact with the first member of the next group.
The number of semi-nearest neighbor diagrams is smaller than the number
of nearest neighbor diagrams by a factor of $\CO(1/J)$ 
as either $\phi$ or $\psi$ must be located at the edge of one of the 
four `groups' of lines in figure \ref{seminear}. Consequently such 
diagrams
are naively negligible in the $J\to\infty$ limit. However, each 
individual semi-nearest neighbor diagram is enhanced by $\CO(J)$ 
relative to a nearest neighbor diagram. In order to understand this, 
consider for example the case illustrated in figure \ref{seminear}. 
As in the previous case, this figure really represents four diagrams, 
which contribute a total
\eqn{snt}{
\frac{\gYM}{N}\frac{2\pi in}{J}
\left(e^{2\pi i n J_2/J}-e^{-2\pi i nJ_1/J}\right)
e^{2\pi i nJ_3/J}
}
(the last factor is due to the $\psi$ field, which in this particular
example happens to sit in the fourth block, but whose position
should be summed over). The fact that there is only one power of $J$ 
in the denominator, rather than two as in \eqref{nnt}, compensates the fact
that these diagrams are rarer by a factor of $1/J$ than the
nearest-neighbor ones. Consequently, such diagrams could make non-vanishing
contribution in the BMN limit $J\to\infty$, and they do contribute to 
$\langle O^J_n(0)\bar{O}^J_m(x)\rangle$ for $m \neq n$. However it turns
out that the full contribution from semi-nearest neighbor graphs to the
correlator above vanishes for the case  $m = n$ considered in this
section. One can see this either by
considering the other semi-nearest neighbor diagrams at fixed
$J_1,J_2,J_3,J_4$ (there are 32 such diagrams in total), and seeing the
cancellations explicitly, or
by considering a fixed diagram such as figure \ref{seminear} summed over
$J_1,J_2,J_3,J_4$---since \eqref{snt} is antisymmetric under exchange of
$J_1$ and $J_2$ it must vanish in the sum.


\subsection{Non-nearest neighbor interactions}

Finally, we turn to the non-nearest-neighbor graphs  
described in the figure \ref{nonnearr} (redrawn
differently in figure \ref{nonnears}). The external legs of our
operator are divided into three groups containing $J_1$ and   
$J_2$ or $J_3$ $Z$'s, respectively. Because we have divided the $Z$
propagators into three rather than four lines, these diagrams are rarer
still by a factor $1/J$ than the semi-nearest-neighbor diagrams, or 
by a factor of $\CO(1/J^2)$ compared to nearest neighbor diagrams. However,
this is compensated by the fact
that each non-nearest neighbor diagram is enhanced
by a factor $\CO(J)$ compared to semi-nearest neighbor diagrams, or 
$\CO(J^2)$ compared to non-nearest neighbor diagrams. 
In order to see this note that in the diagram of figure \ref{nonnearr}
both ends of the $\phi$ propagator jump by a macroscopic 
(i.e.\ order $J$) distance along the
string of $Z$'s. Let the impurity $\phi$ be located
on the left and/or right end of the group $J_1$ whose first and last
propagators are ``pinched''. The four diagrams represented by 
figure \ref{nonnearr} consequently contribute equally, but weighted by 
phase $1$, (for the two diagrams in which $\phi$ does not jump) 
or $-\exp(\pm 2\pi i n J_1/J)$ for the diagrams in which $\phi$ jumps
either to the left/ right; consequently the sum of these four
diagrams is proportional to 
$$(1-e^{2\pi i n J_1/J})(1-e^{-2\pi i n J_1/J}).$$

\EPSFIGURE[l]{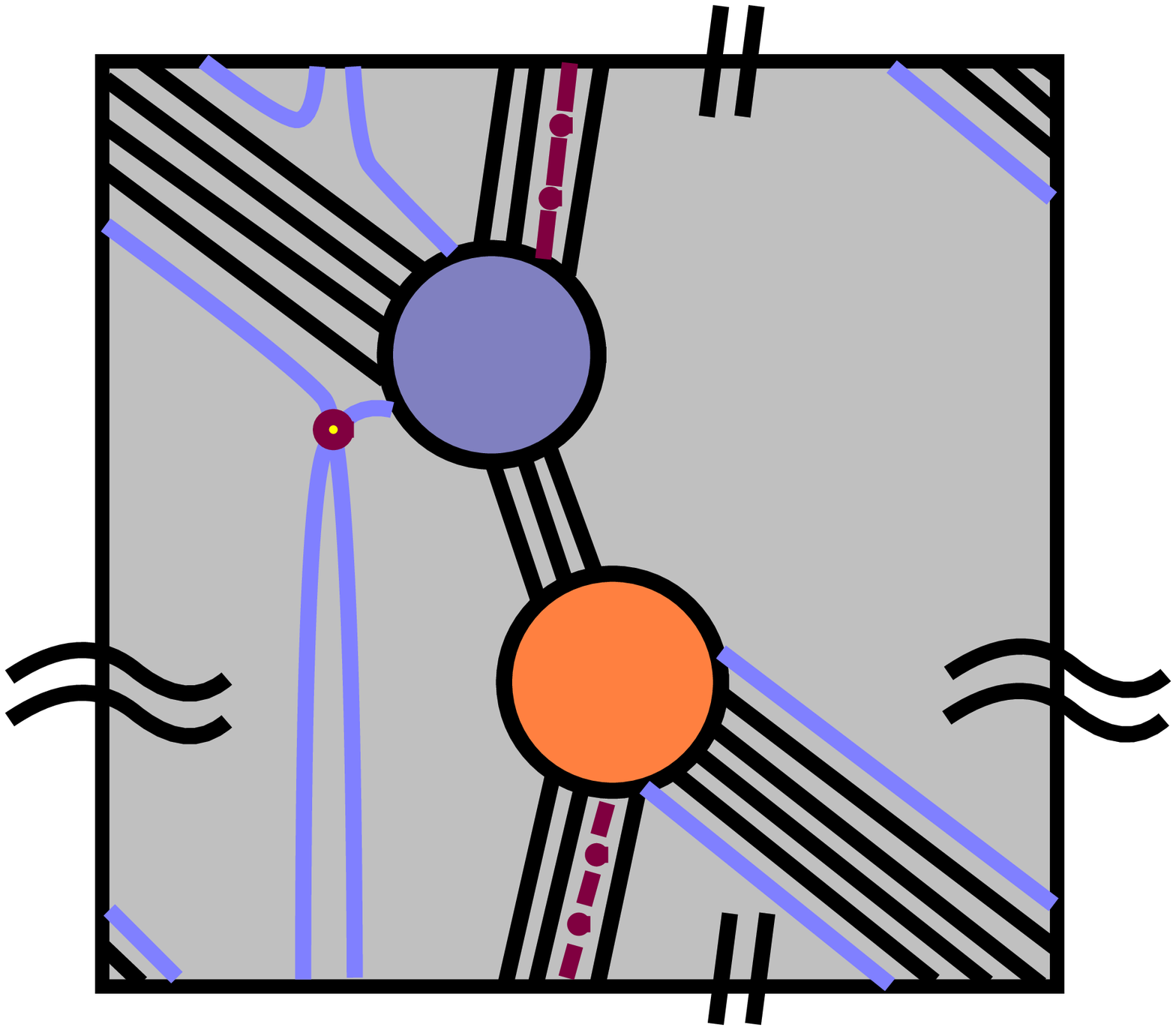,width=7cm}{Figure
\ref{nonnearr} represented as a periodic
square.\label{nonnears}}

We now turn to the contribution to the diagram from the phase associated 
with the second impurity $\psi$. If $\psi$ is in one of $J_1$ places 
inside the first vertical block its relative position on the two operators 
is the same, and so these $J_1$ diagrams contribute with no phase. 
On the other hand, if $\psi$ is in the block with $J_2$ propagators; 
its relative position on the two operators is different by $J_3$; 
the corresponding $J_2$ diagrams contribute with phase $\exp(2\pi i n
J_3/J)$. Finally, if $\psi$ can be located in the third block (with $J_3$ 
propagators) its relative position on the two operators slides
to the left by $J_2$ units. Consequently, these $J_3$ diagrams 
are each proportional to $\exp(-2\pi i n J_2/J)$. 
Replacing the sum over $J^i$ (with  $J_1+J_2+J_3=J$) by an integral 
 over $j_i = J_i/J$ (with $j_1+j_2+j_3=1$),
we arrive at the integral
\begin{equation}
\int_0^1 dj_1
dj_2 dj_3 \,\delta(j_1+j_2+j_3-1)   
(j_2e^{2\pi i n j_3} + j_3 e^{-2\pi i n j_2} + j_1)
|1-e^{2\pi i n j_1}|^2
= \frac 13+ \frac{5}{2\pi^2 n^2}
\label{integralj}
\end{equation}
(for $n\neq0$). The two-point function is thus, at first order in
$\lambda'$ and $g_2^2$, given by
\begin{multline}
(4\pi^2x^2)^{J+2}\langle\bar O_n^J(0) O_n^J(x)\rangle = \\
(1 + g_2^2 A_{nn})\left(1-n^2\lambda'\ln (\Lambda^2x^2)\right) +
\frac{\lambda^{\prime}g_2^2}{4\pi^2}
\left(\frac 13+ \frac{5}{2\pi^2 n^2}\right)\ln(\Lambda^2x^2),
\label{finalanom}
\end{multline}
and the anomalous dimension is given by
\eqn{toruanoresult}{
\Delta =
J + 2 + \lambda'n^2
-\frac{g_2^2\lambda'}{4\pi^2}
\left(
\frac 13+ \frac{5}{2\pi^2 n^2}
\right).}

The methods described in this section, or the ones described in appendices
C and D, may also be used to calculate the two-point function between
different operators. We summarize the result here; the reader will find the
details in the appendices:
\begin{multline}\label{generaltwopoint}
(4\pi^2x^2)^{J+2}\langle\bar O_n^J(0) O_m^J(x)\rangle = \\
(\delta_{nm} + g_2^2 A_{nm})
\left(1-
(n^2-nm+m^2)
\lambda'\ln(\Lambda^2x^2)\right) +
\frac{\lambda' g_2^2}{4\pi^2}B_{nm}\ln(\Lambda^2x^2).
\end{multline}
Here the first, factorized term contains the contributions of the
nearest-neighbor (proportional to $nm$) and semi-nearest-neighbor
(proportional to $(n-m)^2
$) diagrams. The second
term contains the contribution of the non-nearest-neighbor diagrams:
\begin{equation}
B_{nm} = 
\begin{cases}
0, & n=0\te{ or }m=0; \\
\frac13+\frac{10}{U^2}, & n=m\neq0; \\
-\frac{15}{2U^2}, & n=-m\neq0; \\
\frac6{UV}+\frac2{(U-V)^2}, & \te{all other cases,}
\end{cases}
\end{equation}
where $U=2\pi m$ and $V=2\pi n$.

The classification of
diagrams into nearest-, non-nearest-, and semi-nearest-neighbor continues
to be valid at higher genus (at first order in $\lambda'$). Interestingly,
the factorization of the first two contributions, as in
\eqref{generaltwopoint}, is true at all genera.

\section{String interactions from Yang-Mills correlators\label{secfive}}

In this section we finally turn to the relationship between correlation 
functions in Yang-Mills and dual string interactions. We make two specific
proposals at large $\mu$:
\begin{itemize}
\item
Three-point functions of suitably 
normalized BMN operators, multiplied by the difference in $p^-$ between the
ingoing and outgoing operators, may be identified 
with the matrix elements of the light-cone Hamiltonian between the
corresponding one string and two string states.
\item
The one string loop mass renormalization of a class of excited 
string states is reproduced by the $\CO(g_2^2)$ anomalous 
dimensions of the corresponding operators.
\end{itemize}
We believe that these proposals form part of a larger dictionary relating
the Yang-Mills theory and the string theory; however we leave the
determination of the rest of this dictionary to future work. 

This section is organized as follows. In subsection 5.1
we motivate and explain our proposals in detail, and elaborate on some of
their consequences. In the rest of this section we provide evidence for 
the validity of our proposals. In subsection 5.2 we demonstrate that our 
proposals pass a nontrivial self-consistency check. In subsection 5.3 we 
compare our proposals (together with the computations of Yang-Mills 
correlators in section 3 and 4) with the predictions of string field
theory, and find substantial agreement.

\subsection{Three-string light-cone interactions from Yang-Mills
three-point functions}

Let $O_i$, $O_j$, and $O_k$ represent three single-trace BMN operators, 
of $U(1)$ charges $J_i$, $J_j$, and $J_k$, and normalized so that 
\begin{equation}\label{norm}
\langle \bar O_i(0)  O_j(x) \rangle = \frac{\delta_{ij}}{(2\pi x)^{2 
\Delta_i}}
\end{equation}
Let $|i'\rangle$  represent the  free single string states that correspond to 
these operators at zero bulk string coupling, normalized such that 
\begin{equation}\label{xyu}
\langle i' | j'\rangle = \delta_{ij}
\end{equation} 
Let
\begin{equation}\label{planarth}
\langle  \bar O_i(x_i)  O_j(x_j) O_k(x_k) \rangle_\te{free planar}
=\frac{\d_{J_i,J_j+J_k}C_{ijk}}
{(2\pi x_{ij})^{\Delta_i+\Delta_j-\Delta_k} 
(2\pi x_{ik})^{\Delta_i+\Delta_k-\Delta_j}
(2\pi x_{jk})^{\Delta_j+\Delta_k-\Delta_i}}
\end{equation}
The coefficients $C_{ijk}$ have been evaluated in \eqref{tpff} and 
\eqref{threebps2}.
At small $\lambda'$, we propose the following formula for the matrix 
element of the string field theory light-cone Hamiltonian
\begin{equation}\label{lcme}
\langle i'| P^- |j'k' \rangle
=\mu(\Delta_i-\Delta_j-\Delta_k)C_{ijk}.
\end{equation}
\eqref{lcme} is expected only to apply 
to leading order in $\lambda'$; we leave its generalization
to finite $\lambda'$ to future work. 

Equation \eqref{lcme} is one of the central proposals of our paper. 
In sections 5.2 and 5.3 below we will provide rather strong evidence for
its validity. Before proceeding to do so, however, we provide initial
motivation for the proposal \eqref{lcme}. Inner products of Yang-Mills 
states
on $S^3$ (and so, presumably, states of the dual string theory) are related
to correlation functions of the Euclidean Yang-Mills theory by the state
operator map. Thus it is plausible that matrix elements of the string
theory light-cone Hamiltonian are  given by Yang-Mills correlators,  
dressed by a factor of linear homogeneity in $p^-$. 

We now motivate the specific form of the dressing in \eqref{lcme}.
Yang-Mills correlators, correctly normalized (see below), are of order 
$g_2^2$. On the other hand, from section 4, torus mass renormalizations 
occur at order $\mu g_2^2 \lambda'$, and so go to zero when 
$\lambda'$ is taken to zero at fixed $g_2$. Consequently, the dressing
factor must go to zero as $\lambda'$ is taken to zero; this suggests the 
specific form of the formula \eqref{lcme}.

In the rest of this subsection we elaborate on the consequences of
\eqref{lcme}.

\subsubsection{Scaling with $N$ and $J$} 

Note that $C_{ijk}$ scales like $J^{3/2}/N$ for the BMN operators under 
consideration.\footnote{This is easiest to verify in a simple 
example. The normalized chiral operators $O^J=\Tr Z^J/\sqrt{N^J J}$
have planar three-point functions 
\begin{equation}\label{planarthh}
\langle \bar O^J(0)O^{J_1}(x_1)O^{J_2}(x_2) \rangle_\te{planar}
=
\frac{\d_{J_1+J_2, J}C_{J_1, J_2, J}}{(2\pi 
x_1)^{J_1+J} 
(2\pi x_2)^{J_2+J}}
\end{equation}
where $C_{J_1, J_2, J}=\sqrt{J_1J_2J}/N$.}
Further, in the large $\mu$ limit the energy splittings 
$(p_1^- +p_2^--p_3^-)$ are of order $\mu \lambda'$. Consequently, the 
right-hand side of \eqref{lcme} scales like  $\mu g_2 \lambda' /
\sqrt{J}$. As $J$ is taken to infinity in the BMN limit, these matrix 
elements scale to zero, which is puzzling at first sight. Note, however,
that the number of intermediate states (or final states) in any process
scales like $J$; consequently (see subsection 5.2)  the scaling of 
matrix elements is precisely correct to yield finite 
contributions to physical processes. Stated differently, the 
scaling of matrix elements like $\mu g_2 \lambda' /\sqrt{J}$
is merely a consequence of dealing with string states that are unit 
normalized. Switching to the more conventional delta function normalization 
 for states
\begin{equation}\label{cono}
\langle i | j \rangle = p^+_i \delta(p^+_i -p^{+}_j)=J_i \delta_{J_i, J_j}
\end{equation} 
requires a rescaling of states  
\begin{equation}\label{ts}
{|i\rangle}=\sqrt{J_i}{ |i'\rangle}.
\end{equation}
Light-cone Hamiltonian matrix elements may then be written as 
\begin{equation}\label{transnorm}
\langle i| P^- | jk \rangle
= \left[ (p_i^- - p_j^--p_k^-)\sqrt{\frac{J_j J_k}{J_i}}C_{ijk} \right]
p_i^+\delta(p_i^+ - p_j^+ - p^+_k)
\end{equation}
The term in the square bracket on the RHS of \eqref{transnorm} is
finite in the BMN  limit and is of order $\mu g_2 \lambda'$. 

\subsubsection{Effective Coupling}

It is instructive to perform the following exercise. Consider an effective 
two dimensional field theory with scalar fields $\phi_i$, interacting 
through a $p^+$ dependent cubic interaction  
\begin{equation}\label{modlelag}\propto  \int dx^+ dp_i^+ dp_j^+ dp_k^+ 
g(p_i^+, p_j^+, p_k^+) \delta(p_i^+ +p_j^+-p_k^+) 
\phi_i(x^+, p_i^+)\phi_j(x^+, p_j^+)\phi_k(x^+, p_k^+)
\end{equation}
Canonically quantizing this theory in the light-cone, it is not difficult 
to verify (for example, by adapting equation (23) of \cite{Bigatti:1997gm} 
to our
normalization) that the matrix elements for the light-cone Hamiltonian of
this system are 
\begin{equation}\label{modlaglch}\langle i| P^-|jk\rangle
\propto \int dp_i^+ dp_j^+ dp_k^+  g(p_i^+, p_j^+, p_k^+) \delta(p_i^+ +p_j^+-p_k^+) 
\end{equation}
Consequently we conclude that \eqref{transnorm} would be reproduced from a 
two dimensional cubic effective field theory with coupling (of dimension 
squared mass) given by $g(p_i^+, p_j^+, p_k^+)=(p_i^- - 
p_j^--p_k^-)\sqrt{J_j
J_k/J_i}\,C_{ijk}p_i^+$, i.e. 
\begin{equation}\label{stc} 
g(p_1^+, p_2^+, p_3^+) \approx \frac{1}{\alpha'}(\Delta^1+\Delta^2-\Delta^3)
\frac{\sqrt{J_1J_2J_3}}{\sqrt{gN}}C_{123}  \sim 
\frac{\CO(g_2\sqrt{\lambda'})}{\alpha'}
\end{equation}
leading to the identification of $g_2\sqrt{\lambda'}$ as the effective 
string coupling for these processes.

\subsubsection{Vanishing of on-shell amplitudes}

Recall that, in field theory, the decay of a particle is the result 
of the mixing between single particle states and multi particle states 
of the same energy. This mixing invalidates the use of non-degenerate 
perturbation theory in following the `evolution' of the unperturbed 
single particle state upon turning on an interaction. It fuzzes
out the very notion of a particle; in particular the mixing broadens out
delta function peaks in spectral functions, endowing the `particle' 
with a finite lifetime. 
 
It is striking that \eqref{lcme} prescribes the vanishing of 
matrix elements of the light-cone Hamiltonian between states of equal 
unperturbed energy. This prescription implies the stability 
of excited string states in the large $\mu$ limit even upon turning on
interactions. As the notion of a single particle continues to be well
defined in the interacting theory, it is thus natural to identify the 
BMN operator \eqref{bnnop} with the stable one-particle state, at large 
$\mu$, even upon turning on interactions.\footnote{It may 
be possible to derive this identification, together with our 
proposal \eqref{lcme}, from a careful analysis of the state-operator map. 
We hope to return to these issues in the future.}
This feature (the vanishing of matrix elements between states of equal 
unperturbed energy) also 
permits the use of non-degenerate quantum mechanical perturbation theory
in an analysis of mass renormalization of excited string states at large 
$\mu$. We will utilize this observation in subsection 5.2 below.

\subsection{Unitarity check\label{unitarity}}

As we argued in the introduction, eight transverse coordinates in the
pp-wave background 
are effectively compactified. The light-like direction is also compact
at finite $N$  (its conjugate momentum, $J$, is quantized) 
and string theory on the pp-wave background reduces to quantum mechanics.
In this subsection, we apply standard 
quantum mechanical second order perturbation theory to perform a 
self-consistency check of the amplitudes calculated from the gauge theory.  
The Hamiltonian here is $\Delta = J + P^-/ \mu$.

Consider the string state corresponding to the BMN operator $O_n^J$ defined
in equation \eqref{bnnop}. We will use the well known formula for
non-degenerate second order perturbation theory
\eqn{seconden}
{E_n^{(2)} = \sum_{m\neq n}
\frac{\abs{V_{mn}}^2}{E_n^{(0)}-E_m^{(0)}}}
to compute its second order energy shift. 

In \eqref{seconden}, the sum over states $m$ includes two types of
intermediate states:
\begin{itemize}
\item {\bf (A)}
the two-string states with strings corresponding to $O_m^{J_1}$ and
$O^{J_2}$  (these must be summed over the worldsheet momentum $m$). 
Using \eqref{lcme} and \eqref{tpff}, the squared matrix element 
that connects $O_n^J$ to this two-particle state is 
\begin{equation}\label{sme}
|V_{mn}|^2 =
\frac{g_2^2 \lambda^{\prime 2}(1-y)(ny+m)^2\sin^4(\pi ny)}{\pi^4Jy(ny-m)^2} 
\end{equation}
where we have defined $y=J_1/J$. The difference in energies 
between our state and the two-particle state is 
\begin{equation}
E_n-E_m=\frac{ \lambda' (n^2y^2-m^2) }{y^2}.
\end{equation}
\item {\bf (B)}
the two-string states described by the two 
chiral primaries $O_0^{J_1}$ and $O_0^{J_2}$, where the impurities are
$\phi$ and $\psi$ in the two operators respectively.
The squared matrix element of the light-cone 
Hamiltonian connecting $O_n^J$ to this two-particle state is
easily computed from \eqref{lcme} and \eqref{threebps2}:
\begin{equation}\label{smee}
|V_{n0}|^2 =
\frac{g_2^2 \lambda^{\prime 2}\sin^4(\pi ny)}{\pi^4J}.
\end{equation} 
The difference in unperturbed energies between the initial and intermediate
states is $E_n= \lambda' n^2$. 
\end{itemize}
In both cases, we must 
sum over $J_1=J-J_2$, i.e. integrate over $y=J_1/J$ from $0$ to $1$.

The total torus correction to the dimension of $O^J_n$ is therefore
\eqn{torusdim}{
\Delta^{(2)}_n =
J\int_0^1 dy \left( B^{(n)}+\sum_{m\in\Z} A_m^{(n)}\right),
}
where
\eqn{acko}
{A_m^{(n)} \equiv
\frac{\abs{V_{mn}}^2}{E_n-E_m} =
\frac{ g_2^2\lambda'y(1-y)(ny+m)\sin^4(\pi n y)}{\pi^4J(ny-m)^3}
}
and
\eqn{becko}{
B^{(n)}\equiv \frac{\vert{V_{n0}}\vert^2}{E_n-E_m} = 
\frac{4g_2^2\lambda'\sin^4 \pi n y}{\pi^2Jn^2}
}

The sum over $m$ in \eqref{torusdim} may be performed using the identity
\eqn{sumide}{\sum_{m\in \Z} \frac{ny+m}{(ny-m)^3}
=-\pi^2 \csc^2(\pi ny) \left(1+2\pi ny \cotg(\pi ny)\right).}
Adding $B^{(n)}$ to the result and performing the integral 
over $y$, we find
\eqn{torusdi}{\Delta^{(2)}_n = 
- \frac{g_2^2\lambda'}{4\pi^2}
\left(\frac 13 +\frac {5}{2\pi^2 n^2}\right),}
which is precisely \eqref{toruanoresult}, the genus one contribution to the
anomalous dimension. 

In conclusion, we have computed the one loop mass renormalization from 
gauge theory in two different ways; from the genus one contribution to the 
anomalous dimension of the corresponding operator, and independently using
our proposal \eqref{lcme} and standard second order perturbation theory. 
These two computations agree exactly. In the next subsection we will
proceed to compare our prescription \eqref{lcme} with the predictions of
string field theory. 

\FIGURE[ht]{\qquad\epsfig{file=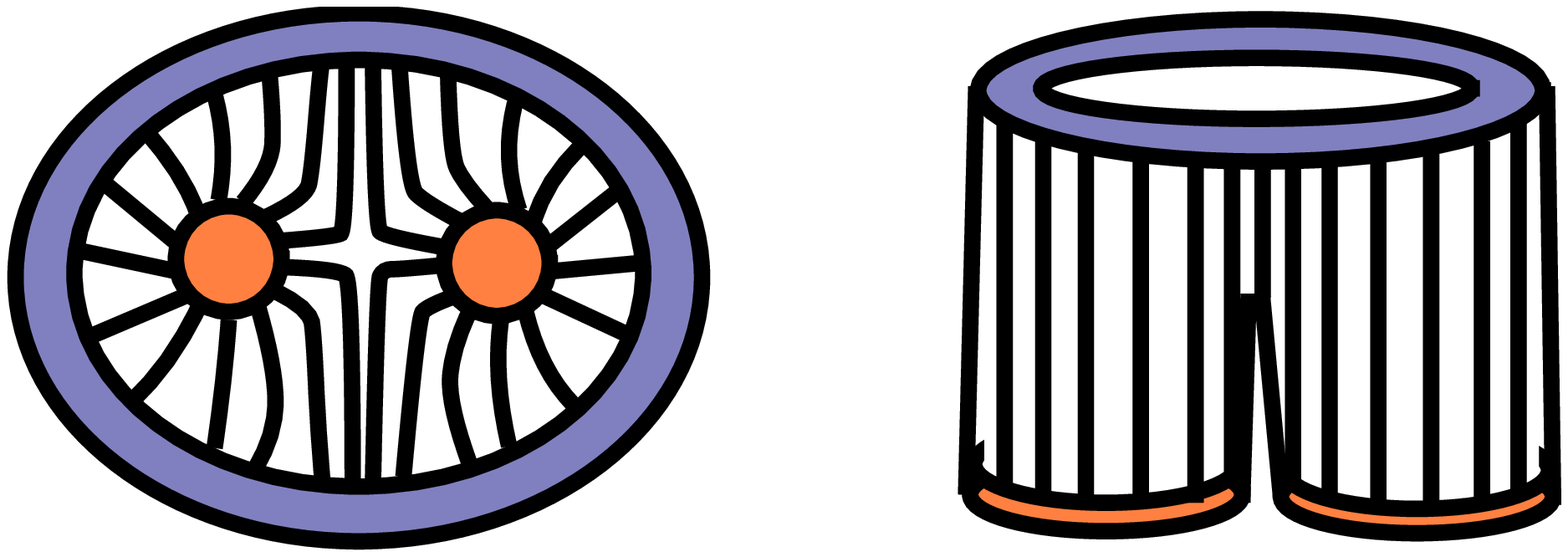,width=12cm}
\caption{A Feynman diagram for the planar three-string correlator
and the corresponding light-cone gauge history of joining strings.
\label{pants}}}

\subsection{Comparison with string field theory\label{sft}}

\subsubsection{The delta-functional overlap}

In this subsection we show that the free planar three-point function 
in Yang-Mills theory is identical to the delta-function overlap 
between string states in the large $\mu$ limit. 

In the light-cone gauge, the bosonic part of the 
worldsheet action of a string propagating in the pp-wave background is 
\begin{equation}\label{spp}
\frac{1}{4 \pi \ap} \int 
\left( \p_t X^i \p_t X^i - \p_{\sigma} X^i \p_{\sigma}X^i - 
\mu^2  X^i X^i \right).
\end{equation}
In the limit $\lambda' \to 0$, at finite $n$ the sigma derivative piece in
\eqref{spp} ($\p_{\sigma} X^i \p_{\sigma}X^i$) is negligible 
compared to the mass term, and may simply be ignored, to 
zeroth approximation. It is convenient to discretize the worldsheet 
\eqref{spp} into $J$ bits. On neglecting $\p_{\sigma} X^i \p_{\sigma}X^i$, 
the bits decouple from each other, and the string disintegrates into $J$ 
independent harmonic oscillators. 

Now consider an interaction process involving three of these strings. 
String splitting/joining interactions in light-cone gauge are local, and
the most important piece of the three-string interaction Hamiltonian 
is the overlap delta functional 
\begin{equation}\label{odf}
\int {\cal D}X^1(\sigma) {\cal D}X^2(\sigma) {\cal D}X^3(\sigma)\,
\Delta(X^1(\sigma)+X^2(\sigma)-X^3(\sigma)) \psi(X^1) \psi(X^2) 
\psi^*(X^3).
\end{equation}

\EPSFIGURE[r]{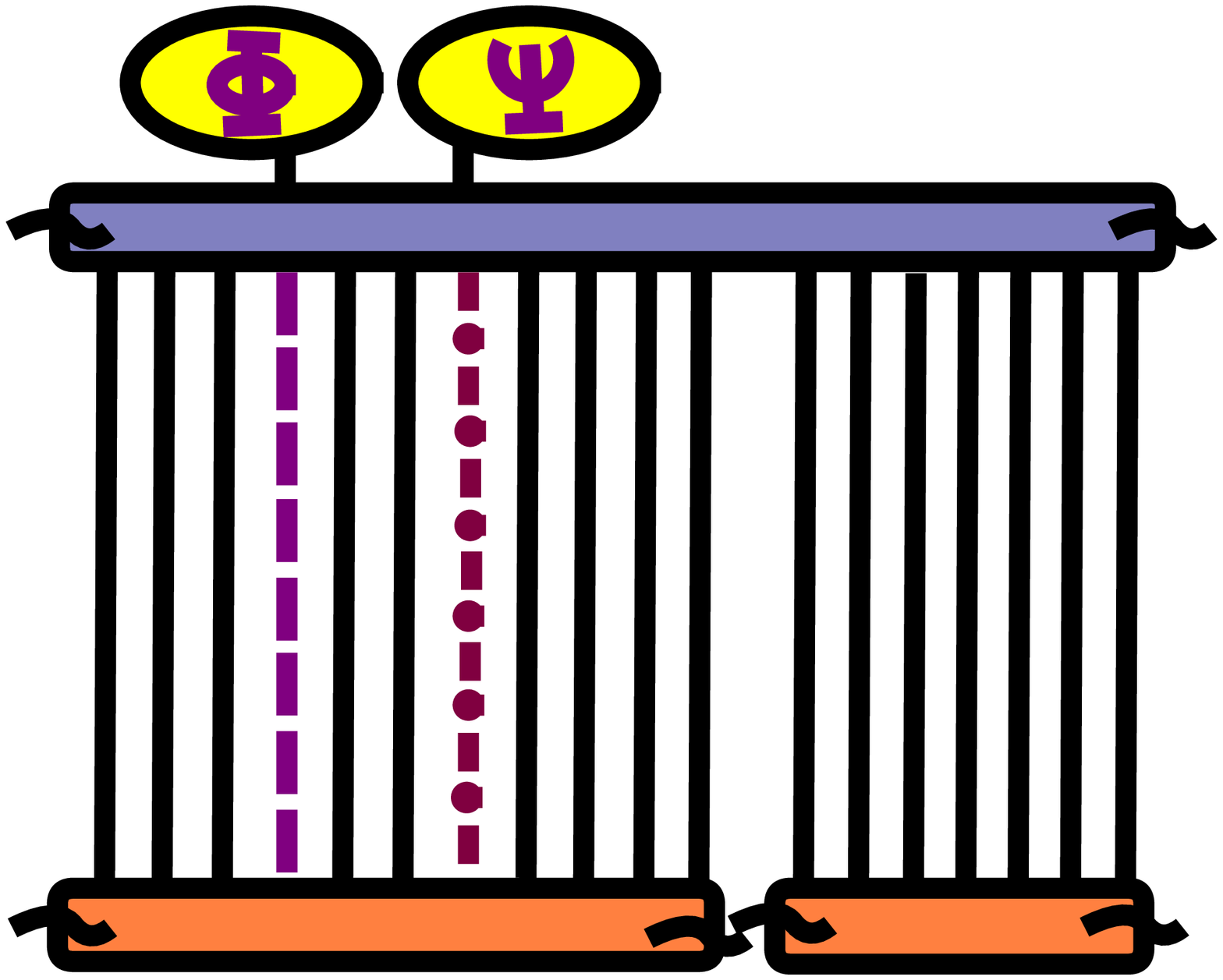,width=7cm}{Three-point function
in free theory as the delta functional overlap of the initial and
final state.\label{threedelta}}

On discretizing the three strings, every bit on the 
third string (the biggest of the three) is put in correspondence with 
a bit on one of the two smaller strings.  Specifically, two bits are in 
correspondence if they share the same value of $\sigma$. 
The wave functions $\psi(X)$ factor into wave functions for each bit. 
Any bit is either in its harmonic oscillator ground state (we could denote
that, in a diagram, by writing the letter $Z$ in the appropriate slot) 
or in the first excited state of $i^{\rm th}$ harmonic oscillator
(denoted by $\phi^i$ in the appropriate slot).  
The interaction \eqref{odf} is simply 
\begin{equation}\label{overlap} 
\!\!\!\!\!\!\!\!\!\!
\!\!\!\!\!\!\!\!\!\!
\!\!\!\!\!\!\!\!\!\!
\!\!\!\!\!\!\!\!\!\!
\!\!\!\!\!\!\!\!\!\!
\!\!\!\!\!\!\!\!\!\!
V=\prod_l \langle \chi_l' | \chi_l \rangle,
\end{equation}
\noindent where the index $l$ runs over the bits of the larger string, 
$|\chi_l \rangle$ is the harmonic oscillator wave function of the $l^{\rm th}$ 
bit on the larger string, and $|\chi_l' \rangle$ is the harmonic oscillator 
wave function on the corresponding bit on the corresponding smaller string
(either string one or string two, depending on the value of $\sigma$). 
Diagrammatically, $V$ is unity when $Z$'s sit on top of $Z$'s 
and $\phi^i$s on top of $\phi^i$s. It is zero otherwise 
(see figure \ref{threedelta}). 

This is precisely the rule to compute the free planar contribution to the
three-point function term by term in the series for the operators (2.5).
The sum over phases in (2.5) is just the discrete Fourier series which
may be taken simultaneously in Yang-Mills theory and on the discretized
string world sheet. It is thus guaranteed that the results (3.10) and
(3.11) from free planar gauge theory diagrams precisely reproduce the
delta-functional overlap of strings in the $\mu \rightarrow \infty$ limit.

\subsubsection{The prefactor}

Our prescription \eqref{lcme} for the matrix elements of the light-cone 
Hamiltonian involved two elements. 
\begin{itemize} 
\item{\bf (a)} The free Yang-Mills correlator.
\item{\bf (b)} The dressing factor $(p^-_1-p^-_2 - p^-_3)$.
\end{itemize}
Similarly, the one/two string light-cone Hamiltonian involves two
elements, a delta function overlap, and a prefactor acting on that 
delta function overlap. In the previous subsection we have demonstrated
that the delta function overlap is precisely equal to the free Yang-Mills
correlator. In this subsection we conjecture that  
the action of the prefactor on this delta function precisely produces 
the dressing factor $(p^-_1-p^-_2 - p^-_3)$ multiplying the delta function 
overlap in \eqref{lcme}.

The prefactor of string field theory involves two derivatives, 
and might naively have been estimated to be of order $p^+ \mu$ at large 
$\mu$. In fact, as $\mu \to \infty$, the prefactor vanishes when acting on 
the delta functional overlap!\footnote{
This result has been derived by M. Spradlin and  A. Volovich, and will 
be presented elsewhere.} This striking result is consistent 
with our proposal that the prefactor, acting on the delta functional 
produces a factor of
\begin{equation}\label{prefactordim}
p^-_1-p^-_2 -p^-_3 =\CO(1/ \mu)
\end{equation}
An honest verification of \eqref{prefactordim} appears to be algebraically 
involved, though it is straightforward in principle. 

A more detailed understanding of the structure of the string field theory 
prefactor in the large $\mu$ limit would
permit several predictions for gauge theory correlators. For example, 
a preliminary analysis appears to indicate that the normalization of the
prefactor \eqref{prefactordim} is proportional to the number of 
scalar impurities minus the number of $D_\mu Z$ impurities 
(with no contributions from fermionic impurities), and 
would seem to imply that the operators with one scalar and one 
$D_\mu Z$ excitation, as well as operators with fermionic excitations 
only, should have vanishing amplitudes.

\section{Conclusions and outlook}

In this paper we have begun an investigation into the  relationship between 
Yang-Mills correlators in the BMN limit and string interactions in the 
pp-wave background. Our analysis is valid at large $\apm p^+\mu
\sim(\lambda')^{-1/2}$ 
where the gauge theory appears effectively weakly coupled. We have employed 
perturbative Yang-Mills theory to make verifiable predictions for
interaction amplitudes and mass renormalizations of weakly coupled strings
on the pp-wave background.

The principal observations and proposals of our paper are 

\begin{itemize}

\item
For the appropriate class of questions, it appears that 
Yang-Mills perturbation theory in the BMN limit may be organized 
as a double expansion in an effective loop counting parameter 
$\lambda'=\gYM N/J^2$ \cite{ias} and an effective genus counting parameter
$g_2^2=J^4/N^2$. In particular, graphs of all genus continue to contribute,
even in the strict $\N \to \infty$ BMN limit. The effective genus 
counting parameter $g_2^2$ may independently be identified as the two 
dimensional Newton's constant for the dual string theory. 

\item
Mixing effects between single and multi trace Yang-Mills operators 
are also controlled by the parameter $g_2$. This implies a modification 
of the dictionary between Yang-Mills operators and perturbative
string states, proposed in \cite{ias} at the same order.

\item
We have proposed a relationship between Yang-Mills three point functions 
and three string interactions at large $\mu$. Our proposal, \eqref{lcme}, 
implies that Light-cone Hamiltonian matrix elements between a wide class 
of single and double string states are of order $\mu g_2 \lambda'$,
corresponding to invariant effective string coupling  of order 
$g_2 \sqrt{\lambda' }$. At large values of this coupling 
string perturbation theory breaks down and string states blow up into
giant gravitons. The detailed form of \eqref{lcme} also implies that, 
at large $\mu$, excited string states are stable, even at nonzero values 
of the effective string coupling.

\item
We have computed the one loop correction to the dispersion relation 
of these string states in two different ways: first by a direct
computation of the anomalous dimension associated with this operator at 
order $g_2^2$ (i.e. on the torus), and second from quantum mechanical 
perturbation theory, using matrix elements obtained from three-point
functions, as in the previous proposal. These computations agree exactly;
this constitutes a non-trivial check on our proposals. They also confirm
our identification of $g_2^2 \lambda'$ as the effective theoretic
genus counting parameter.

\end{itemize}

This paper suggests several directions for future investigation. To begin
with, it would be useful to check the proposals of this paper, and to 
better understand some its assumptions.
It is very important to check our proposal for the dictionary between 
Yang-Mills correlators and string interactions; to this end  
an explicit expression for the first term in an expansion in powers of 
$1/\mu$ of the string field theory interaction vertex formally derived in 
\cite{Spradlin:2002ar} is required. It is certainly important to thoroughly
understand when and why perturbative Yang-Mills can be employed in the 
study of this strongly interacting gauge theory (see \cite{grosscompete}). 
Finally, in the unitarity 
check of section 5.2 we did not account for intermediate states with 
different numbers of impurities from those in the  initial state. 
Contributions from such intermediate states are suppressed by large energy 
denominators. However, they could also be enhanced by parametrically large 
couplings. It would be useful to understand precisely when and why it is 
justified to omit such contributions.

Several generalizations of our work immediately suggest themselves. 
It should be straightforward to generalize our calculations and proposals 
to BMN operators involving $D_\mu Z$ and fermionic impurities. More
generally, it would be very interesting to extend the  dictionary between 
Yang-Mills and string theory proposed in this paper, 
beyond the large $\mu$ limit, and to wider classes of observables. 
It may also be possible to derive \eqref{lcme} and its generalizations 
from the more usual $AdS$/CFT prescriptions.

In conclusion, the BMN limit appears to allow us to  
re-interpret a sector of an effectively {\it perturbative}
Yang-Mills theory as an {\it  interacting}
string theory! This remarkable idea certainly merits further study.

\vspace{5mm}

\acknowledgments

We are grateful to A. Adams, N. Arkani-Hamed, D. Berenstein, 
M. Bianchi,
A. Dabholkar,
J. Gomis, R. Gopakumar, J. Maldacena, H. Ooguri, S. Shenker,
E. Silverstein,  M. Spradlin, A. Strominger, N. Toumbas, C. Vafa, M. van
Raamsdonk, D. Tong, H. Verlinde, A. Volovich, and S. Wadia for extremely
useful discussions.
The work of N. Constable, D. Freedman, A. Postnikov, and W. Skiba was
supported in parts by 
NSF PHY 00-96515, DOE~DF-FC02-94ER40818 and NSERC of Canada.
The work of A. Postnikov was 
also partially supported by NSF grant DMS-0201494.
The work of M. Headrick was supported in part by DOE grant DE-FG01-91ER40654
The work of S. Minwalla and L.~Motl was supported by
Harvard Junior Fellowships, and in part by  DOE grant DE-FG01-91ER40654.

\appendix

\section{Specification of operators}

The operators of the $\N =~4~SYM$ theory which map into
low-energy string states in the $pp$-wave background involve a large number
$J$ of scalar fields $Z$ together with several 
impurity fields $\phi$ and $\psi$. It simplifies things to take
these fields to be holomorphic combinations of the six real
elementary scalars $X^i$ of the theory, e.g. $Z =
(X^5+iX^6)/\sqrt{2}, \phi = (X^1+iX^2)/\sqrt{2}, \psi =
(X^3+iX^4)/\sqrt{2}$. We take $Z,\phi,\psi$ to be the lowest
states of the three chiral multiplets $\Phi^1,\Phi^2,\Phi^3$ which
appear in the $\N=1$ description of the $\N=4$ theory. BMN have
outlined the rules for the construction of these operators, but there
are some subtleties, and we thus briefly describe a clear
specification here.

The principles of the construction are the following:
\begin{itemize}
\item
{\bf (1)} For the case of $I$ impurity fields there are $I$ initially 
 independent phases 
   $q_j = \exp(2\pi i n_j/J)$. One writes a formal
   definition of the operators which reduces to a BPS operator
   when all $q_i=1$, specifically a ``level $I$'' $SU(4)$
   descendent of the chiral primary operator $\Tr(Z^{J+I})$.
   Many of the operators so defined vanish; specifically they
   vanish unless the product $q_1q_2\cdots q_I =1$ which is
   the level-matching condition on string states.
\item
{\bf (2)}
On the ``level-matching shell'' the non-vanishing operators
   still satisfy the BPS property when all $q_i=1$. They reduce
   to the familiar BMN operator (with a minor but necessary
   change) for $k=2$ and extend the construction to general $k$.
   Planar diagrams for the 2-point correlation functions of these
   operators vanish (both for free fields and order $g_{YM}^2$
   interactions) unless the momenta $n_j$ are conserved. This
   planar orthogonality property is approximate, holding to accuracy
   $1/J$ in the BMN limit $N \rarrow \infty, J \sim
   \sqrt{N}$. Since the momenta do not have any clear meaning in
   the field theory, one should not expect strict momentum
   conservation. Indeed non-conservation becomes a leading effect
   for diagrams of genus $\ge 1$, again both for order $g_{YM}^0$
   and order $g_{YM}^2$.
\end{itemize}
We start the discussion with the $I=2$ BMN operator
\begin{equation} 
 O_J = \sum_{l=0}^{J}q^l\Tr\left(\phi Z^l\psi Z^{J-l}\right),
\label{op}
\end{equation}
modified so that the sum begins at $l=0$ rather than $l=1$ as in
\cite{ias}. With
this definition the BPS property is exact when $q=1$. This minor
change is significant when one computes the planar correlator
including interactions. For the operator defined above one finds
by techniques described elsewhere in this paper that
\begin{equation}
\< O_J(x) \bar{ O}_J(0)\> = f_J(x) [1 +
cg_{YM}^2 N (1-q) (1-\bar{r}) \ln|x|] \sum_{l=0}^{J} (q\bar{r})^l,
\label{planarcorr}
\end{equation}
where $q=\exp(2\pi i n/J)$ and $r=\exp(2\pi i n'/J)$ are the phases
of the operators $ O_J(x)$ and $\bar{ O}_J(0)$, respectively,
and $f_J(x) =(N/4\pi^2x^2)^{J+2}$ involves the product of free scalar
propagators. The factor $(1-q)(1-\bar{r})$ is the discrete second
derivative of the phases coming from the stepping effect of the
interactions noted in \cite{ias}. This factor contributes a $1/J^2$
suppression in the BMN limit. With the sum in the operator beginning
at $l=1$ one would obtain a similar expression with the 
changes: a) the final sum in \eqref{planarcorr} starts at $l=1$,
and b) there is an additional term proportional to $g_{YM}^2 N 
(1+q\bar{r})\ln|x|$.
The last term arises because the construction of discrete second
derivative is incomplete at one site. It is not suppressed in
the BMN limit. If present the physical interpretation would be
spoiled, which is why is we chose the definition \eqref{op}.

\EPSFIGURE[l]{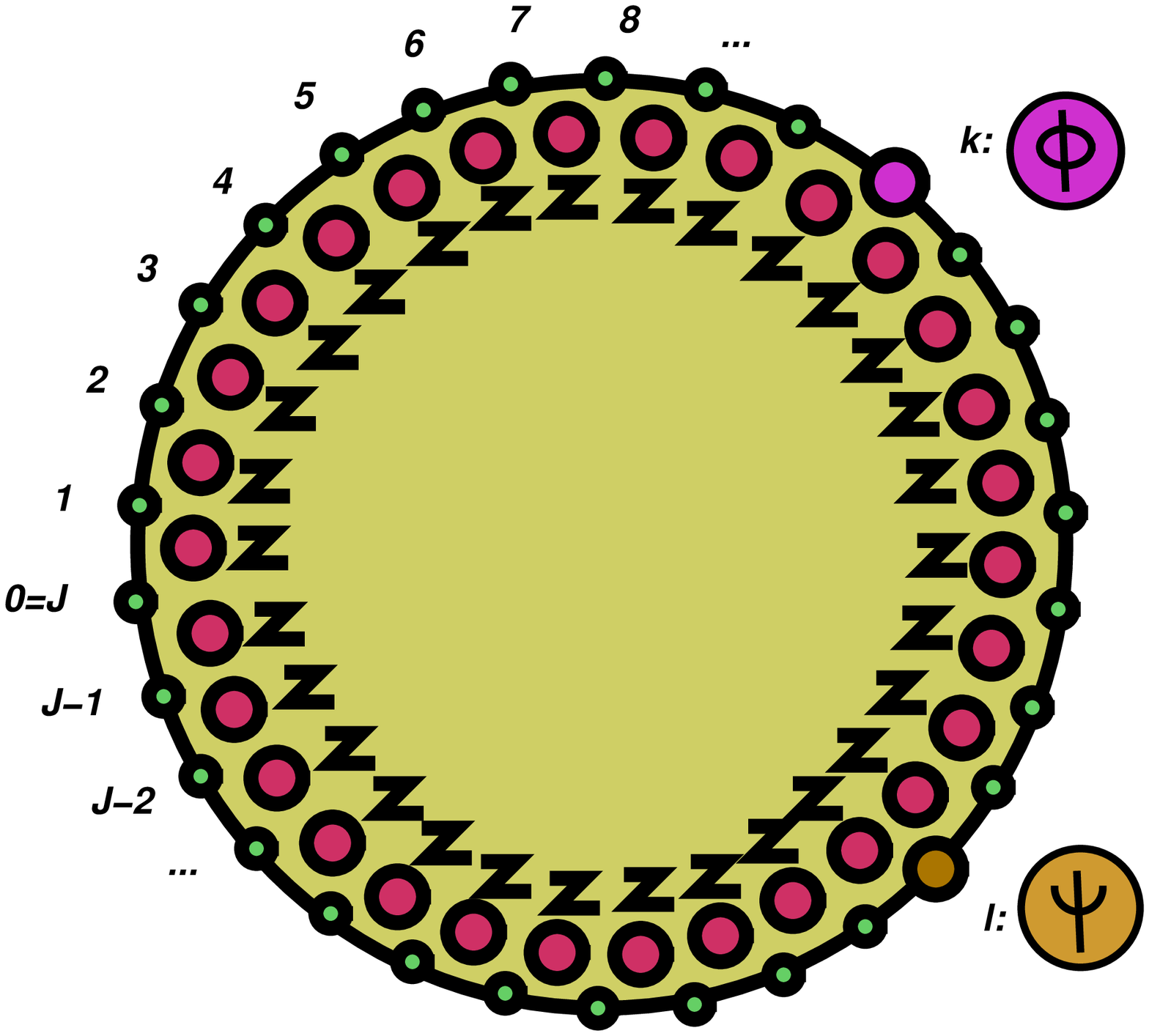,width=8cm}{A circular
array to indicate the $J$ $Z$-fields and the impurities
$\phi,\psi$ at positions $k,l$ from a chosen origin.\label{defopfig}}

The final sum in \eqref{op} has the value $J +1$ if $q\bar{r}=1$
and the value $1$ otherwise. In the BMN limit (to accuracy $1/J$)
one thus obtains 
the momentum conserving result $J\delta_{n,n'}$ as expected for a
string in the state $a^*(n)b^*(-n)|0\>$ in the state dual to the
operator $O_J$ in the limit $(\mu p^+ \rarrow \infty)$ when string
interactions vanish. 

The operator for two identical
$\phi$ impurities is obtained simply by replacing $\psi$ by $\phi$
in \eqref{op}. There are additional cross terms in the diagrams
for the 2-point function. The planar result is just
\eqref{planarcorr} with the final sum replaced by
\begin{equation}
\!\!\!\!\!\!\!\!\!\!\!\!\!\!
\!\!\!\!\!\!\!
\sum_{l=0}^{J} \left[(q\bar{r})^l + (qr)^l\right] = J\left(\d_{n,n'}
+\d_{n,-n'}\right)
\end{equation}
in the BMN limit. This is the correct orthogonality condition for
(non-interacting) strings in the level-matched state $a^*(n)a^*(-n)|0\>$.  

We now turn to the general program of defining operators off the
``level-matching shell''. The general method is embodied in figure 
\ref{defopfig}
which shows a circular array of $J$ points, the $J ~Z$'s in the
trace, with $\phi$ and $\psi$ in interstitial positions
at distances $k$ and $l$ from an arbitrary origin. The associated
phase is $q_1^kq_2^l$. It is clear from the figure that if we fix
the relative distance, say $l$ between $\psi$ and $\phi$, and sum
over rigid displacements of the positions of $\phi$ from
$k=0$ to $k=J-1$, we accumulate the phase polynomial
$q_2^l[1+q_1q_2+ \cdots (q_1q_2)^{J-1}]$ which vanishes unless
the level-matching condition $n_1+n_2=0$ is obeyed.

\vspace{1mm}

The analytic expression for the operator depicted in figure \ref{defopfig}
involves a sum over the two relative orders of $\phi,\psi$ within
the trace. It is
\bea
 O &=& O_1 +O_2;
\nonumber \\
 O_1 &=& \sum_{0\le k \le l \le J-1} q_1^kq_2^l \Tr(Z^k\phi
Z^{l-k}\psi Z^{J-l});
\nonumber \\
O_2&=& \sum_{0\le l' \le k' \le J-1} q_1^{k'}q_2^{l'} \Tr(Z^{l'}\psi
Z^{k'-l'}\phi Z^{J-k'}).  
\label{opos}
\eea
It is straightforward although a bit awkward to carry out
analytically the sum over rigid displacements of a configuration
of fixed relative phase $l$ pictured above. Cyclicity of the sum is 
vital, of course. Starting from $k=0$ in $O_1$ of \eqref{opos}
and moving to position $k=J-l-1$, one accumulates the phase
polynomial $q_2^l[1+q_1q_2+ \cdots (q_1q_2)^{J-l-1}]$. The next
step takes us into $O_2$ with $l'=0,~k'=J-l$, and we sum over
$l'$ in $l-1$ rigid steps until $\phi$ is in next-to-last
position in the trace. The phase polynomial from this traversal 
of $O_2$ is $q_1^{J-l}[1+ q_1q_2 + \cdots  (q_1q_2)^{l-1}]$.
The sum of these two polynomials is equal to the full polynomial
in the previous paragraph, thus giving the level-matching
condition exactly.

\vspace{1mm}

One may now impose the condition $q_1q_2=1$ and show that in
non-vanishing cases the operator in \eqref{opos} is just a factor
of $J$ times that of \eqref{op}. The first step is to substitute
$q_2=1/q_1$ in $O_1$, and use the relative position index $a=l-k$ to
rewrite the double-summed expression in \eqref{opos} as
\begin{equation}
O_1 =\sum_{a=0}^{J} (J-a) q_2^a \Tr(\phi Z^a \psi Z^{J-a}).
\label{op1}
\end{equation}
We have used the fact that, for fixed $a$, there are $J-a$
values of the original index $k$ which make identical
contributions to $O_1$. The sum over $a$ in \eqref{op1} actually
stops at $a=J-1$, but it is useful to add the vanishing entry as
we will see. The operator $O_2$ is handled similarly using
$b=k'-l'$ as the relative phase index. By symmetry one finds
\bea
O_2&=&\sum_{b=0}^{J} (J-b) q_1^b \Tr(\psi Z^b \phi Z^{J-b})
\nonumber \\
   &=& \sum_{a=0}^{J} a q_2^a \Tr(\phi Z^a \psi Z^{J-a}),
\label{op2}
\eea
where we have redefined $a=J-b$, used cyclicity and $q_1=1/q_2$
in the last step. We see that the sum of \eqref{op1} and
\eqref{op2} is equal to $J$ times the original on-shell operator
\eqref{op} as claimed.

We have gone into considerable detail in the simple case of 2
impurities in order to avoid an impossibly awkward discussion
in the general case which we now outline. In a set of $I \ge 3$
impurities, repetition of the fields $\phi,\psi$ occurs. However
these fields are effectively distinguished in the construction
of the operators because they carry
different phases $q_j$. Wick contractions in the correlators will
then impose Bose symmetry.

We therefore conceive of a set of $I$ independent impurity fields
$\phi_i,~i=1,\cdots I$, placed at arbitrary interstitial sites in
the circle of figure \ref{defopfig}, with assigned phase $q_i^{k_i}$. The
analytic expression for the corresponding
``off-level-matching-shell'' operator is the sum $I!$ terms, one for each
permutation of the impurities. The non-vanishing on-shell
operator contains $(I-1)!$ terms including all non-cyclic permutations.

For $I=3$, the first of six terms
can be written as
\begin{equation}
 O_1 = \sum_{0\le k\le l\le m \le J} q_1^k q_2^l q_3^m \Tr(Z^k
\phi_1 Z^{l-k} \phi_2 Z^{m-l} \phi_3 Z^{J-m}).
\end{equation}
There are two similar terms, $ O_2, O_3$ for the cyclic permutations  
$(\phi_2,\phi_3,\phi_1)$ and $(\phi_3,\phi_1,\phi_2)$ of the
impurity fields, and three more for anticyclic permutations. With
due diligence one may repeat the argument above for the case $I=2$
and show that the sum $ O_1 + O_2 + O_3$ vanishes unless the
level-matching condition $q_1q_2q_3 =1$ holds. The same property
holds separately for the sum of the three operators for
anti-cyclic permutations. The on-shell operator is the sum 
$ O =  O_c +O_a$ with the cyclic term 
\begin{equation}
 O_c = \sum_{0\le a,\,\, 0 \le b}^{a+b\le J} q_2^a q_3^{a+b} 
\Tr(\phi_1
Z^a \phi_2 Z^b \phi_3 Z^{J-a-b})
\end{equation}
and an analogous expression $ O_a$ for the anti-cyclic permutation
$(\phi_1,\phi_3,\phi_2)$, namely
\begin{equation}
 O_a = \sum_{0\le a,\,\, 0 \le b}^{a+b\le J} q_3^a q_2^{a+b} 
\Tr(\phi_1 Z^a \phi_3 Z^b \phi_ Z^{J-a-b}).
\end{equation}

With care, and with the $I=2$ case as a
model, one can insert the
on-shell condition $q_1 =1/(q_2q_3)$ in the operator $ O_1$,
introduce relative position indices $a=l-k,b=m-l$, and rewrite
$ O_1$ as
\begin{equation}
 O_1 = \sum_{0\le a,\,\, 0 \le b}^{a+b\le J}(J-a-b) q_2^a 
q_3^{a+b} \Tr(\phi_1
Z^a \phi_2 Z^b \phi_3 Z^{J-a-b}).
\end{equation}
Cyclic symmetry implies similar expressions for $ O_2, O_3$,
namely
\bea
 O_2 &=& \sum_{0\le c,\,\,0 \le d}^{c+d\le J}(J-c-d) q_3^d 
q_1^{c+d} \Tr(\phi_2
Z^c \phi_3 Z^d \phi_1 Z^{J-c-d}) ;
\nonumber \\ 
 O_3 &=& \sum_{0\le e,\,\,0 \le f}^{e+f\le J}(J-e-f) q_1^e 
q_2^{e+f} \Tr(\phi_3
Z^e \phi_1 Z^f \phi_2 Z^{J-e-f}).
\eea
With the redefinitions $c=b, d=J-a-b, f=a,e=J-b-a$, creative use
of the relation $q_1q_2q_3 =1$, and cyclic symmetry, one may show
that the sum $ O_1+ O_2+ O_3=J O_c$. The analogous result relating the
on-shell sum of three anti-cyclic permutations to $J$ copies of
the anti-cyclic $ O_a$ may be derived in the same way. This
discussion shows that when the level matching condition holds,
the off-shell operator for 3 impurities is just $J$ times the
on-shell operator $ O = O_a+ O_b.$

The 2-point correlation function of the operator with $I=3$
impurities may be denoted by
$\<( O_c(x)+ O_a(x))(\bar{ O}_c(0)+\bar{O}_a(0))\>$. Planar
diagrams come only from the diagonal terms
$\< O_c(x)\bar{O}_c(0)\>$ and
$\< O_a(x)\bar{O}_a(0)\>$. Contributions from both terms are
required for planar orthogonality in the two independent momenta.

To order $\CO(g_{YM}^2)$ we obtain
\bea
\label{planarcorr3}
\< O(x) \bar{ O}(0)\> &=& F(x) \left[1 +
g_{YM}^2 N P(q_i,r_i) \ln|x| \right] \sum_{l=0}^{J} (q_2\bar{r}_2)^l
   \sum_{m=0}^{J} (q_3\bar{r}_3)^m, \\
P(q_i,r_i)&=& (1-q_1^{-1}) (1-\bar{r}_1^{-1}) + (1-q_2) (1-\bar{r}_2)
             + (1-q_3) (1-\bar{r}_3). \nonumber
\eea
$F(x)=(N/4 \pi^2 x^2)^{J+3}$ is the product of $J+3$ scalar propagators
and the corresponding color factor. $P(q_i,r_i)$ is of order $1/J^2$
exactly like the $ O(g_{YM}^2)$ expression
for the two-point function of operators with two insertions
\eqref{planarcorr}. In the expression for $P(q_i,r_i)$ we
have neglected higher order terms in $1/J$ that multiply terms
of order $1/J^2$.

It is easy to understand how the structure of the interaction term
arises. We are interested in planar contributions so we consider
the nearest neighbor interactions only. The interactions can take
place between any of the marked fields $\phi_i$ and the neighboring fields
$Z$. The interaction in the conjugate operator $\bar{ O}$ has to take
place between the same marked field and neighboring fields $\bar{Z}$.
Thus, we get the sum of three contributions with different phase dependence
for each of the marked fields.

We hope that the discussion for $I=2,3$ impurities in this section makes the
construction of the case of arbitrary $I$ clear. 

\FIGURE[ht]{\qquad\qquad\epsfig{file=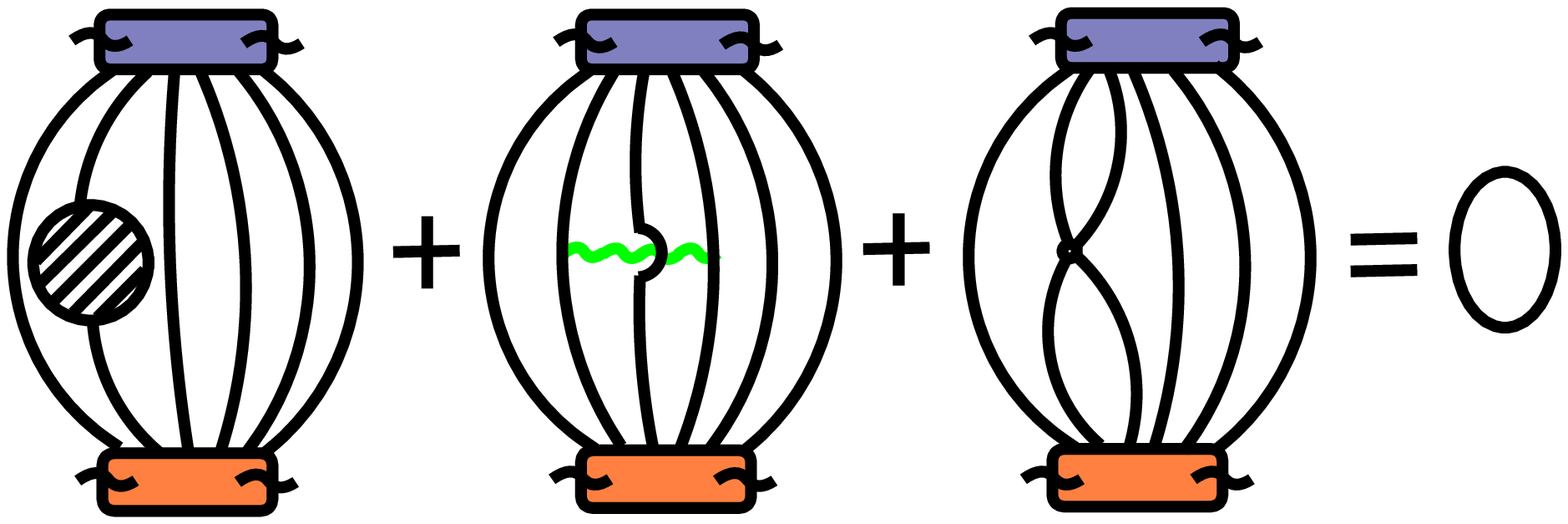,width=12cm}
\caption{Cancellations at order $g_{YM}^2$.\label{cancellationsbps}}}

\section{Irrelevance of D-terms}

In this appendix we will show that the F-term interactions studied in the
main text are the {\it only} interactions which need to be considered at 
order $g^2_{YM}$ in Yang-Mills perturbation theory. The
sum of the other contributions, from D-terms, gluon exchange and 
self-energy insertions, precisely vanishes at this order in both
2- and 3-point functions (for an example see figures 
\ref{cancellationsbps} and \ref{threepointgluons}). 
This simple but useful fact can be proved
by minor modifications of the techniques used for the same purpose
in the studies of BPS operators in either \cite{9907098}
or \cite{dhoker2}. We use the technique of \cite{9907098}.

We are concerned with the operators 
\begin{equation} 
 O_J = \sum_{l=0}^{J}q^l \Tr\left( \phi Z^l\psi Z^{J-l}\right),
\label{opp}
\end{equation}
and we will show that the sum all non-F-term contributions vanishes
term-by-term in the expansion in phases of $\< O_J \bar{O}_J\>$
and $\< O_J \bar{O}_{J_1} \Tr(\bar{Z}^{J_2})\>$.

\FIGURE[ht]{\quad\epsfig{file=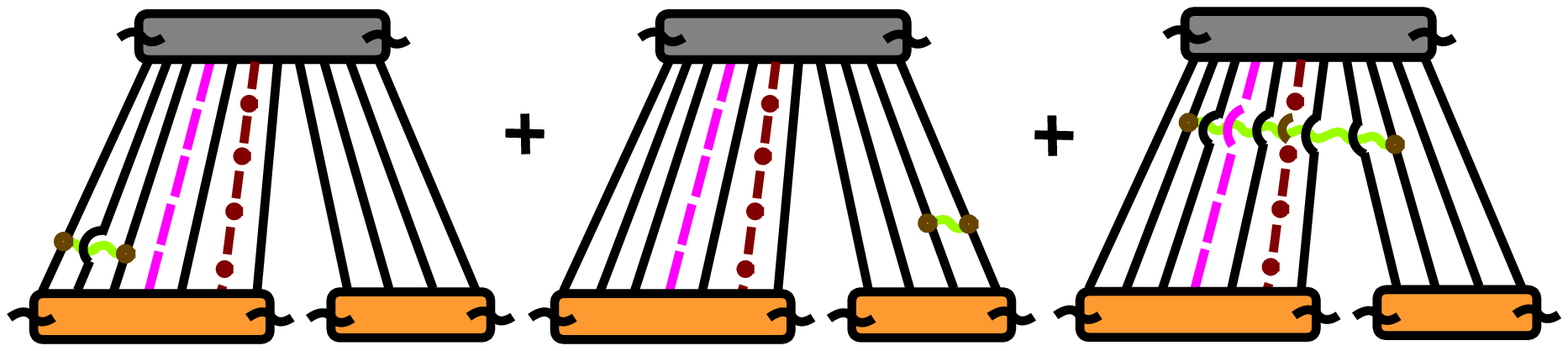,width=14cm} 
\caption{Gluon exchange contributions to a three-point
function.}\label{threepointgluons}}

The first relevant observation comes from inspection of the form
of the D-term potential in the $\N=4$ Lagrangian 
which is
\begin{equation}
V_D=g_{YM}^2\Tr\left([Z,\bar{Z}][Z,\bar{Z}]+2[\phi,\bar{\phi}][Z,\bar{Z}] 
+ 2[\psi,\bar{\psi}][Z,\bar{Z}]+ 2[\phi,\phi][\psi,\bar{\psi}]\right),
\label{alld}
\end{equation}
where we have dropped similar $\phi^4$ and $\psi^4$ terms which
do not contribute to the correlators listed above. One now sees
that all quartic vertices contribute to Feynman diagrams with the
same combinatorial weight, independent of $SU(3)$ flavor. Similar
remarks hold for gluon exchange diagrams. The 1-loop self-energy
insertion is also flavor blind. Thus for the purposes of this
Appendix, each summand in \eqref{op} can be replaced by $\Tr(Z^{J+2})$. 
One can now simply use the result of \cite{9907098} which shows
that all order $g_{YM}^2$ radiative corrections to  
$\<\Tr(Z^{J+2}) \bar{Z}^{J+2}\>$ cancel.  Nevertheless, we will
repeat the argument of \cite{9907098} briefly because we will
make a somewhat new application to 3-point functions below.

The first step is to observe that gluon exchange diagrams
and those from $V_D \sim \Tr([Z,\bar{Z}]^2)$ have the same color
structure, and must be summed over all pairs of lines in the second
Feynman diagram of figure~\ref{cancellationsbps}. Self-energy insertions on
each line must also be summed. The following identity holds for any
set of matrices $M_i,N$:
\begin{equation}
\sum_{i=1}^{n}\Tr\left(M_1\cdots[M_i,N]\cdots M_n\right) = 0.
\label{identity}
\end{equation}
Let $k=J+2$. Each diagram in figure~\ref{cancellationsbps} includes a sum 
over $k!$ 
permutations of the
fields in $\Tr(\bar{Z}^k)$ relative to a fixed permutation of the
fields of $\Tr(Z^{k})$. Let $T^{a_1} \cdots T^{a_k}$ denote the
fixed permutation of color generators of the fields in
$\Tr(Z^{k})$, and let $T^{b_1} \cdots T^{b_k}$ denote one of
the permutations of fields in  $\Tr(\bar{Z}^{k})$. For each pair
of fields $i \ne j$ the gluon exchange or D-term has a color
structure which may be expressed as a commutator with the
generators $T^i$ and $T^j$ in the product $T^{b_1} \cdots T^{b_k}$.
Summing over all pairs, we obtain the net contribution
\begin{equation}
-B(x)\Tr(T^{a_1}\cdots 
T^{a_{k}})\sum_{i\neq j=1}^{k}
\Tr(T^{b_1}\cdots[T^{b_i},T^e]\cdots[T^{b_j},T^e]
\cdots T^{b_k}),
\label{zint2}
\end{equation}    
where $B(x) = b_0 + b_1\ln(x^2M^2)$ is the space-time factor
associated with the interaction. We now use \eqref{identity} on
one of the commutators to rewrite \eqref{zint2} as
\bea
B(x)\Tr(T^{a_1}\cdots T^{a_{k}})\sum_{i=1}^{k}
\Tr(T^{b_1}\cdots[[T^{b_i},T^e],T^e]
\cdots T^{b_k}) 
\nonumber \\
= NB(x)\Tr(T^{a_1}\cdots T^{a_k})\sum_{i=1}^{k}
\Tr(T^{b_1}\cdots T^{b_i}\cdots T^{b_k}).
\label{zint3}
\eea

In the last step we recognize $[[~,T^e],T^e]$ as the $SU(N)$
Casimir operator in the adjoint representation which gives
$[[T^a,T^e],T^e] = N T^a$ for any generator. The final sum thus has
$k$ identical terms. Each self energy insertion also contains the
adjoint Casimir and has the form $NA(x)=N(a_0 +a_1\ln(x^2M^2))$,
and there are $k$ such terms. The sum of all diagrams in figure
\ref{cancellationsbps} is thus 
\begin{equation}
kN(B(x)+A(x))\Tr(T^{a_1}\cdots T^{a_{k}})\Tr(T^{b_1} \cdots T^{b_k}),
\end{equation}
which must be summed over all permutations 
$\{ b_i=\sigma(a_i) : ~i=1 \cdots k\}$ and
finally contracted on pairwise identical color indices. All
manipulations above are valid for the case $k=2$ which
is known to satisfy a non-renormalization theorem. Hence $B+A=0$,
and radiative corrections (other than from F-terms) cancel for
all $k$. Figure \ref{nonnearestwrong} shows a D-term diagram
which cancels with others despite the intuition that a gauge theory
vertex at the string interaction point, i.e.
the saddle point of the toroidal stringy diagram, should be significant.

\FIGURE[ht]{\quad\epsfig{file=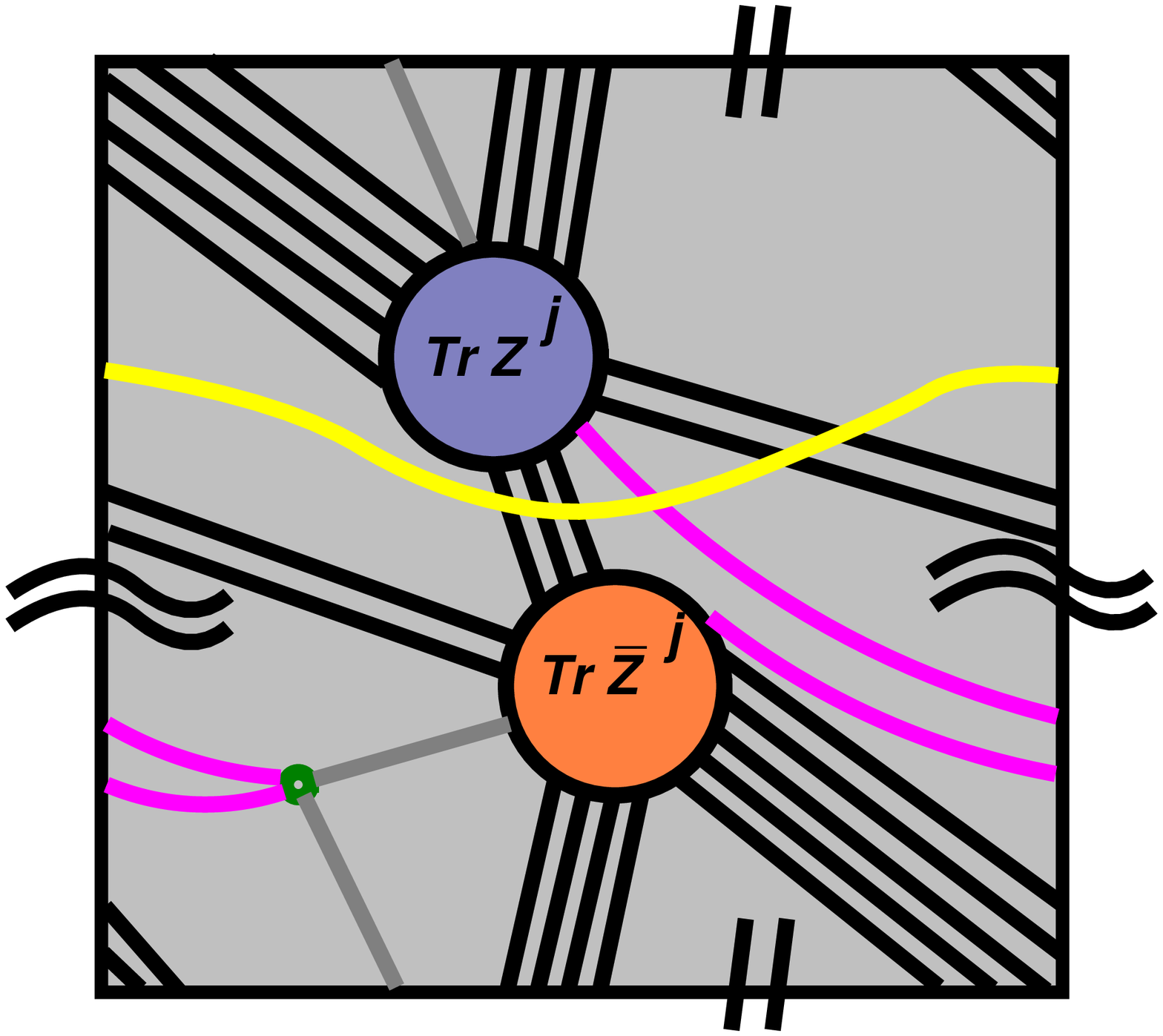,width=7cm}\qquad
\epsfig{file=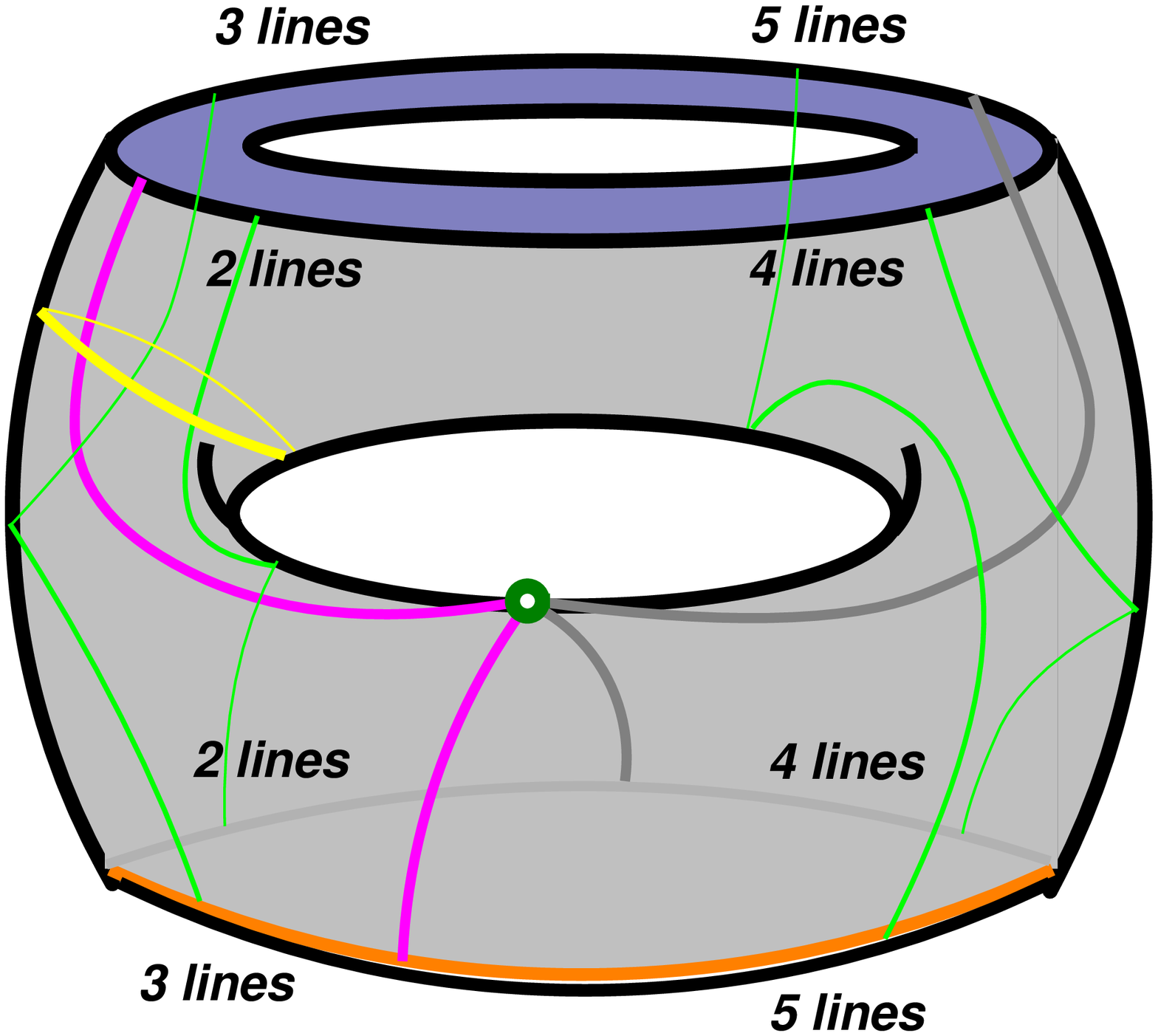,width=7cm}
\caption{Two representations of a diagram which cancels with 
others.}\label{nonnearestwrong}}

Next we study the 3-point function $\< O_J { \bar{O}_{J_1}}
\Tr(\bar{Z}^{J_2})\>$ with $J=J_1+J_2$. The flavor blind property
again means that the $JJ_1$ summands are all identical.
Gluon exchange interactions among pairs of lines are indicated in
figure \ref{threepointgluons}. Quartic vertices from $V_D$ are similarly 
summed. We consider a fixed permutation of fields in  $ O_{J_1}$ and in
$\Tr(Z^{J_2})$ and sum over permutations
(i.e. orderings of generators
$T^{c_1} \cdots T^{c_{J+2}}$) in the central operator $ O_J$ and sum 
over permutations of fields from $ O_{J_1}$ and from $\Tr(Z^{J_2})$. 
For interacting pairs which are connected to $ O_{J_1}$ the
previous argument applies mutatis mutandis. Radiative corrections
cancel when self-energy insertions on $J_1$ lines are included.
Idem for interacting pairs connected to $\Tr(Z^{J_2})$. The
remaining pair interactions include one line connected to each
operator. For these we use the color structure to place one
commutator in each position in $\Tr(T^{a_1} \cdots T^{a_{J_1}})$
and one commutator in each position within
$T^{b_1} \cdots T^{b_{J_2}}$. The resulting structure is then
\begin{equation} 
C(x,y,z) \sum_{i=1,j=1}^{J_1,J_2}\!\Tr(T^{a_1} \cdots [T^i,T^e]\cdots
T^{a_{J_1}})\,\Tr(T^{b_1} \cdots [T^j,T^e] \cdots T^{b_{J_2}}) 
\,\Tr(T^{c_1} \cdots T^{c_{J+2}}),
\end{equation}
where $C(x,y,z)$ is a spacetime factor which need not be
specified. However, for each fixed $i$, the sum on $j$ vanishes by
\eqref{identity}, and our task is complete.


\def\rbar{{\bar r}}
\def\proof{\medskip\noindent{\it Proof --- \ }}
\def\endproof{\hfill$\square$\medskip}
\def\I{i}

\section{Feynman diagrams and combinatorics}

In this appendix we give a self-contained approach to the two-point
correlation function discussed in the text. The purpose is to provide
the detailed basis of results for planar and genus one contributions
and to outline an algorithm to calculate genus $h > 1$ results. We hope
that the treatment below is readable both by physicists and
mathematicians.

Let us summarize the results of this appendix. 
We show that genus $h$ two-point function in the free case 
is given by a sum of ${(4h-1)!!} / (2h+1)$ terms that
correspond to the types of genus $h$ Feynman diagrams with 
$4h$ nonempty groups of edges. 
The two-point function with a single interaction equals 
$(2\pi)^2(nn' + (n-n')^2)\times\textrm{(free case)}$
plus sum of $(4h-1)(4h-1)!!/[3(2h+1)]$ terms
given by explicit formulas.  
The expression $(2\pi)^2nn'\times\textrm{(free case)}$ comes 
from nearest neighbor interactions, 
the expression $(2\pi)^2(n-n')^2\times\textrm{(free case)}$ 
comes from semi-nearest neighbor 
interactions, and the remaining expression comes from non-nearest
neighbor interactions.  The latter correspond to the 
types of genus $h$ diagrams with $4h-1$ nonempty groups of edges.  
For genus $h=1$, there is exactly one type of diagrams with 
4 groups of edges, and there is one type of diagrams with 
3 groups of edges. 
For genus $h=2$, there are 21 type of diagrams with 8 blocks
and 49 types of diagrams with 7 blocks, etc.

It is natural to assume that, for 2 interactions, we need to go one
level lower, i.e., the two-point function should be given by terms that
depend on the free case and the single interaction case plus new
additional part given by diagrams with $4h-2$ groups of edges, and so
on for higher $h$.  This gives a natural hierarchy of genus $h$ Feynman
diagrams according to their numbers of blocks.

It is important to note that the notation used in this Appendix differs
from that used in the main body of the paper in some respects. 
In subsections 3.3 and 4.2 of the main paper we have used the symbols 
$U=2\pi m$ and $V=2 \pi n$ to parameterize the phases $e^{iU}$ and 
$^{iV}$ in BMN operators. 
In this appendix, we will usually use the symbols $u,v$ for these
quantities. The indices $n,m$ will be denoted $n,n'$.
The symbol $A_{n, n'}$ used repeatedly in sections 3 and 4 of the
main body of the paper is identical to the symbol $A^1(n, n')$ in this 
appendix. 
Finally, the contribution of non-nearest interactions denoted
$B_{nm}$ in 
\eqref{generaltwopoint}
is identical to $G^1_{\textrm{nn}}(n,n')$ in the appendix.

\subsection{Correlation functions}

For a positive integer $J$ and two integers $n$ and $n'$, 
the operators $O^J_n$ and $\bar O^J_{n'}$ are given by 
\begin{equation}
O_n^J = \sum_{l=0}^J q^l \Tr(\phi Z^l \psi Z^{J-l})
\quad\textrm{and}\quad
\bar{O}_{n'}^J = \sum_{l'=0}^J \rbar^{l'} \Tr(\bar Z^{l'} \bar \psi 
\bar Z^{J-l'}\bar \phi)\,,
\label{eq:O-Obar}
\end{equation}
where $q=\exp(2\pi\I n/J)$ and $r=\exp(2\pi\I n'/J)$.
(Here $i=\sqrt{-1}$).
All fields $Z$, $\bar Z$,  $\phi$, $\bar\phi$, $\psi$, $\bar\psi$
are given by $N\times N$ Hermitian matrices.
We will discuss the free two-point correlation function:
\begin{equation}
\<O_n^J(x)\, \bar{O}_{n'}^J(y)\> = (4\pi^2(x-y)^2)^{-J-2} \,A_{J,N}(n,n')
\end{equation}
and the two-point function with one interaction:
\begin{equation}\label{defg}
\<O_n^J(x)\,[Z,\psi] \, [\bar Z, \bar\psi ] \,\bar{O}_{n'}^J(y)\> = 
-\frac{g^2N \ln|x-y| }{ 2\pi^2}
\,G_{J,N} (n,n').
\end{equation}
The functions $A_{J,N}$ and $G_{J,N}$ can be written as series in powers of $N$:
\begin{equation}
A_{J,N}=N^{J+2}\sum_{h\geq 0} N^{-2h}\, A_J^h 
\qquad\textrm{and}\qquad
G_{J,N}=N^{J+3}\sum_{h\geq 0} N^{-2h}\, G_J^h. 
\end{equation}
This is called the genus expansion of correlation functions,
because $A_J^h$ and $G_J^h$ are given by sums over Feynman diagrams
of some type drawn on an oriented  genus $h$ surface.

As we will see, $A_J^h$ is of order $J^{4h+1}$ and $G_J^h$ 
is of order $J^{4h-1}$ as $J\to\infty$.  Let
\begin{equation}
A^h(n,n')=\lim_{J\to\infty} A_J^h/ J^{4h+1}
\qquad\textrm{and}\qquad
G^h(n,n')=\lim_{J\to\infty} G_J^h/ J^{4h-1}\,.
\end{equation}
The limits $A(g_2;n,n')=\lim  A_{J,N}/(J\cdot N^{J+2})$
and $B(g_2;n,n')=\lim  G_{J,N}/(J^{-1}\cdot N^{J+3})$,
where $J,N\to\infty$ so that $J^2\sim g_2 N$ for a fixed constant $g_2$,  
can be written as
\begin{equation}
A(g_2;n,n')= \sum_{h\geq 0} (g_2)^{2h}\, A^h(n,n') 
\qquad\textrm{and}\qquad
G(g_2;n,n') =\sum_{h\geq 0} (g_2)^{2h}\, G^h(n,n'). 
\end{equation}
For any integer values of $n$ and $n'$, $A(g_2;n,n')$ and $B(g_2;n,n')$
are analytic functions of $g_2$.  In particular, 
\begin{equation}
\begin{array}{l}
\ds
A(g_2;0,0)=\frac{2 \sinh(g_2/2)}{ g_2}, \\[.1in] 
A(g_2;n,0)= A(g_2;0,n')=
G(g_2;0,0)=G(g_2;n,0)= G(g_2;0,n')=0,
\end{array}
\end{equation}
for $n,n'\ne 0$.
We will see that 
\begin{equation}
\begin{array}{l}
A^h(n,n') = \textrm{ sum of \  } (4h+1)\cdot \frac{(4h-1)!!}{2h+1} 
\textrm{ \  integrals},\\[.15in]
G^h(n,n') = (2\pi)^2\cdot (-nn'+(n-n')^2)\cdot A^h(n,n') + 
G^h_\mathrm{nn}(n,n'), \\[.15in]
\textrm{where \ }G^h_\mathrm{nn}(n,n') = \textrm{ sum of \  } 
4\cdot (4h-1)\cdot \frac{(4h-1)!!}{2h+1}
\textrm{ \  integrals}.
\end{array}
\end{equation}
Each of these integrals is given by an explicit formula.
As an example, we will present closed formulas 
for $A^h(n,n')$ and $G^h(n,n')$ for small values of genus $h$.
In general, the expressions for $A^h(n,n')$ and $G^h(n,n')$
have the form
\begin{equation}
\frac{\textrm{polynomial in $n$ and $n'$}}
{(n\cdot n')^a\cdot (n-n')^b \cdot (n+n')^c}.
\end{equation}


\subsection{Free two-point function via permutations}
\label{sec:2-point-permutations}

Feynman diagrams for the free two-point function are basically 
given by permutations.

Let us recall a few basic facts about permutations.
A {\it permutation} of order $m$ is a bijective map
$w:\{1,\dots,m\}\to\{1,\dots,m\}$.  
Multiplication of permutations is given by composition
of maps.  All permutations of order $m$ form the 
{\it symmetric group} $S_m$.
A {\it cycle} in a permutation $w$ is a subset of the form
$\{w(i),w^2(i),\dots,w^r(i)=i\}$.  In particular, each fixed point 
$w(i)=i$ is a cycle of size 1.  Thus each permutation gives a 
decomposition of $\{1,\dots,m\}$ into a disjoint union of cycles.  
The {\it number of cycles} of $w$ is the total number of cycles 
in this decomposition.
Let the {\it long cycle} $c\in S_m$ be the 
permutation that consists of a single $m$-cycle  given 
by $c:i\mapsto i+1\pmod m$. 
\medskip




Feynman diagrams that describe Wick couplings in the free case 
$\<O_n^J(x)\,\bar O_{n'}^J(y)\>$ are given by permutations
$w$ of order $m=J+2$ corresponding to mappings between the fields 
in $O_n^J(x)$ and the fields in $\bar O_{n'}^J(y)$.  
The diagram with permutation $w$ produces a term with some 
power $N^{C(w)}$ of the rank $N$ of the gauge group $U(N)$.
The exponent $C(w)$ is related to the genus $h(w)$ of the corresponding 
diagram by Euler's formula: 
\begin{equation}
C(w) = m - 2\,h(w).
\end{equation}
In physics language, $C(w)$ is the number of closed quark loops in the
ribbonized diagram (a.k.a.\ fat graph) corresponding 
to the Feynman diagram with $U(N)$ adjoint fields.
Combinatorially, $C(w)$ is the number of cycles in the twisted permutation
$c^{-1}w^{-1} c w\in S_m$, where $c$ is the long cycle in $S_m$.  
The expression $A_{J,N}$, which gives the free two-point function,
can be written as the following polynomial in the variables 
$q$, $\rbar$, and $N$
\begin{equation}
A_{J,N}=\sum_{w\in S_{m},\, w(1)=1}  N^{C(w)}
\, \sum_{i\geq 2}^m q^{i-2} \rbar^{w(i)-2}\,,
\label{eq:A}
\end{equation}
where the sum is over all permutations $w$ such that $w(1)=1$ (the first 
marked field $\phi$ is always contracted with $\bar\phi$), and the product 
is over all $i\ne 1$ (choice of position of the second marked field $\psi$).

Let us say that the number $h(w) = (m-C(w))/2$ is 
the {\it genus} of a permutation $w\in S_m$.
It is always a nonnegative integer because the parity of the number 
$C(w)$ of cycles in the twisted permutation $c^{-1}w^{-1}cw$ 
is the same as the parity of $m$.
Clearly, the genus of any cyclic shift $wc^r$
of $w$ is the same as the genus of $w$.  
Thus without loss of generality we will assume that $w(1)=1$.
The $h$-th term $A^h_J$ in the genus expansion of $A_{J,N}$
is given by the terms in~\eqref{eq:A} 
with $w$ of given genus $h$.
Since the only genus $0$ permutations are the identity permutation and 
its cyclic shifts, we have
\begin{equation}
A_{J}^0 = (q\rbar)^0+(q\rbar)^1+\cdots+(q\rbar)^J.
\end{equation}
In the next section we will show that $A_{J}^h$ is a polynomial 
in $q$ and $\rbar$ given by a sum of order $J^{4h+1}$ 
terms $q^i\rbar^j$.

\subsection{Block-reduction of permutations}
\label{sec:block-reduction}

For large values of $J$, the expression~\eqref{eq:A} 
involves a summation over permutations of large orders.
This makes it difficult to calculate the limit $J\to\infty$
of this expression.   
Nevertheless it is possible reduce a permutation $w$ of arbitrary 
large order $m$ and small genus $h$ to a permutation $\sigma$ of order 
$\leq 4h$.  The permutation $\sigma$ is 
the block-reduction of $w$ and is formally defined below.


For a permutation $w\in S_m$ with $w(1)=1$, let us subdivide 
the set $\{1,\dots,m\}$ into blocks as follows.
If there are consecutive indices $i,i+1,\dots,i+k$ 
in the cyclic order such that $w(i)=j, w(i+1)=j+1, \dots, w(i+k)= j+k$ 
then we combine the indices $i,i+1,\dots,i+k$ into a single block.  
Here all indices $i$ and values $w(i)$ are understood modulo $m$.
We will choose the blocks to be as maximal as possible.
Thus the permutation $w$ gives a subdivision of the set 
$\{1,\dots,m\}$ into disjoint union of blocks $B_1,\dots,B_k$ (where $1\in B_1$)
formed by cyclically consecutive elements and a permutation $\sigma\in S_k$ of 
blocks.  We call the permutation $\sigma$ the {\it block-reduction} of $w$.
A permutation $\sigma\in S_k$, $k\geq 2$, is a block-reduction for some $w$ 
if and only 
if 
\begin{equation}
\sigma(1)=1;\qquad
\sigma(i+1)\ne \sigma(i)+1, \quad\textrm{for }i = 1,\dots,k-1;
\quad\textrm{and}\quad \sigma(k)\ne k.
\end{equation}
Let us say that a permutation $\sigma\in S_k$ is {\it block-reduced} if it 
satisfies these conditions; and let $\BR_k^h$ be the set of block-reduced 
permutations of order $k$ and genus $h$.  
We will assume that $\BR_0^0$ contains one element---the empty permutation 
$\emptyset$ of order 0---which is the block-reduction of the identity 
permutation of any order.

Figure~\ref{fig:21and22} shows two genus $h=1$ Feynman diagrams given 
by permutations $1\,2\,5\,6\,7\,3\,4\,8\,9\,10$ and 
$1\,2\,9\,10\,6\,7\,8\,3\,4\,5\,11\,12$.  
The first permutation has 
3 blocks $B_1=\{8,9,10,1,2\}$, $B_2=\{3,4,5\}$, $B_3=\{6,7\}$
and its block-reduction is $1\,3\,2$.
The second permutation has 4 blocks $B_1=\{11,12,1,2\}$,
$B_2=\{3,4\}$, $B_3=\{5,6,7\}$, $B_4=\{8,9,10\}$ and its 
block-reduction is $1\,4\,3\,2$.


\FIGURE[ht]{\quad\epsfig{file=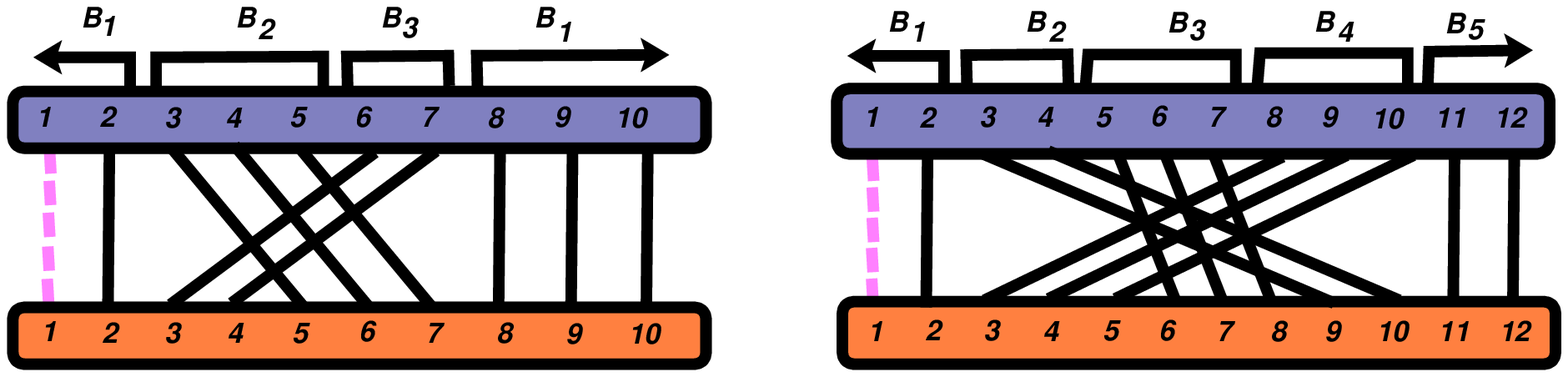,width=14cm}
\caption{Two genus 1 permutations with block-reductions (a) $1\,3\,2$ and
(b) $1\,4\,3\,2$.}
\label{fig:21and22}}

\begin{lemma}
A permutation $w$ and its block-reduction $\sigma$ have exactly the same genera
$h(w)=h(\sigma)$.  For each block-reduced permutation $\sigma\in S_k$ 
of genus $h$,
we have $k\leq 4h$.
\end{lemma}

\proof
Recall that genus of $w$ was defined in terms of the twisted permutation
$\tilde w=c^{-1}w^{-1}cw$.  The similar twisted permutation $\tilde\sigma$
for the block-reduction $\sigma$ of $w$ is obtained from $\tilde w$
by removing all its fixed points.  Thus $\sigma$ and $w$ have same
genera.  Also, $\tilde\sigma\in S_k$ is a fixed-point free permutation.
Thus its number of cycles is $C(\sigma)\leq k/2$ and the genus
of $\sigma$ is $h(\sigma)=(k-C(\sigma))/2\geq k/4$.
\endproof

There is only one genus 0 block-reduced permutation $\emptyset$ and
two genus 1 block-reduced permutations 132 and 1432.
For genus 2, there are 21 elements in $\BR_8^2$
and 49 elements in $\BR_7^2$.  
In general, the following statement holds.

\setcounter{footnote}{0}

\begin{proposition}  For $h\geq 1$, the numbers of elements in $\BR_{4h}^h$ 
and in $\BR_{4h-1}^h$ are given by\footnote{Recall that 
$(4h-1)!!=1\cdot 3\cdot 5\cdots (4h-1)$.}
\begin{equation}
|\BR_{4h}^h| = \frac{(4h-1)!!}{2h+1}
\qquad\textrm{and}\qquad
|\BR_{4h-1}^h| = \frac{4h-1}{3}\cdot |\BR_{4h}^h|.
\end{equation}
\end{proposition}

\proof
Elements of the set $\BR_{4h}^h$ are in one-to-one correspondence
with gluing of the $4h$-gon into a genus $h$ surface.   An element
$\sigma\in\BR_{4h}^h$ is determined by its twist 
$\tilde\sigma=c^{-1}\sigma^{-1}c\sigma\in S_{4h}$, where $c$ is the long 
cycle in $S_{4h}$.   The permutation $\tilde \sigma$ is a fixed-point
free permutation with $2h$ cycles.  Thus $\tilde \sigma$
is given by the 
product of $2h$ commuting transpositions.  
Let us label the sides of the $4h$-gon by the numbers $1,\dots,4h$
(in the clockwise order) and glue the side labelled $i$ with the side
labelled $\tilde\sigma(i)$, for $i=1,\dots,4h$.
This will produce a genus $h$ surface.

The numbers of gluings of any $2k$-gon into a genus $h$ surface 
were calculated by Harer and Zagier~\cite{HZ}.  In the case $k=2h$, 
their result implies the number of gluings is $(4h-1)!!/(2h+1)$.

For $\sigma'\in\BR_{4h-1}^h$, the twist $\tilde\sigma'\in S_{4h-1}$
should  be a permutation given by the product of $2h-2$ commuting
transpositions and a single 3-cycle.   Such $\tilde\sigma'$ can be
obtained from $\tilde\sigma$, for $\sigma\in\BR_{4h}^h$, by 
merging the first vertex $1$ with the last vertex $4h$ and replacing 
2 transpositions $(1,\tilde\sigma(1))$ and $(4h,\tilde\sigma(4h))$ 
with a single 3-cycle
$(1,\tilde\sigma(4h),\tilde\sigma(1))$.  We will get all 
$\tilde\sigma'$ such that the vertex 1 belongs to the 3-cycle.
In order to get all possible $\tilde\sigma'$ we need to take all cyclic 
shifts, which gives the factor $4h-1$, and then divide by 3, because 
we counted all elements 3 times.
Thus $|\BR_{4h-1}^h| = (4h-1)/3 \cdot |\BR_{4h}^h|$.
\endproof

In order to recover a permutation $w\in S_m$ from its block-reduction
$\sigma\in S_k$ we need to know the sizes of the blocks $B_i$ and the position 
of 1 in the first block, which is the placement of the marked field $\phi$ 
in~\eqref{eq:O-Obar}.  This information can be encoded as the 
sequence $(b_1,\dots,b_{k+1})$ of integers, where
$b_{i} = |B_i|$ for $i=2,\dots,k$; $b_1$ is the number of elements
of $B_1$ after 1 and $b_{k+1}$ is the number of elements of $B_1$
before 1 (in the cyclic order). So $b_1+b_{k+1}+1=|B_1|$.
This sequence satisfies the following conditions:
\begin{equation}
b_1+\cdots+b_{k+1}=m-1=J+1;\ b_1,b_{k+1}\geq 0;\ b_2,\dots,b_{k}>0.
\label{eq:bbb}
\end{equation}

For a block-reduced permutation $\sigma\in S_k$ and a sequence 
$b_1,\dots,b_{k+1}$ as above, 
let $b_1',b_2',\dots,b_{k+1}'$ be the sequence given
by $b_i'=b_{\sigma^{-1}}(i)$ for $i=1,\dots,k$ and $b_{k+1}'=b_{k+1}$.
Then the $h$-th term in the genus expansion of the
free two-point function can be written as
\begin{equation}
\begin{array}{l}
\displaystyle
A_{J}^h = \sum_{k\geq 1}\sum_{\sigma\in\BR_k^h} \sum_{i=1}^{k+1} 
\\[.2in]
\displaystyle
\left(\sum_{b_1+\cdots+b_{k+1}=J+1}
q^{b_1+\cdots+ b_{i-1}}\rbar^{b_1'+\cdots+b_{w(i)-1}'}
((q\rbar)^0+(q\rbar)^1+\cdots + (q\rbar)^{b_i})\right),
\end{array}
\end{equation}
where the sum is over all block-reduced permutations $\sigma$ of arbitrary
orders and fixed genus $h$ and the internal sum in the parenthesis
is over sequences $b_1,\dots,b_{k+1}$ that satisfy 
conditions~\eqref{eq:bbb}.

The number of terms $q^i\rbar^j$ (each of absolute value 1) 
in the internal
sum is of order $J^{k+1}$.  Thus only the terms with maximal possible
value $k=4h$ survive in the limit $J\to\infty$ and the whole expression
is of order $J^{4h+1}$.  Let 
\begin{equation}
A^h(n,n') = \lim_{J\to\infty} \frac{A_J^h}{J^{4h+1}}.
\end{equation}
Then $A_h(n,n')$ is given by the sum over $\sigma\in\BR^h_{4h}$ and 
$i=1,\dots,4h+1$.
In the limit $J\to\infty$, each sum over $b_1,\dots,b_{k+1}$ turns 
into a $(k+1)$-dimensional integral.
In the next section we show how to compute integrals of this type.

\subsection{Calculation of integrals}
\label{sec:integral}

Let us fix $r$ numbers $u_1,\dots,u_{r}\in\C$.
For a positive integer $J$, let $q_j=e^{u_j/J}$, $j=1,\dots,r$,
and let
\begin{equation}
S_J(q_1,\dots,q_{r})=\sum_{b_1+\cdots+b_{r}=J+1} q_1^{b_1}\cdots 
q_{r}^{b_{r}},
\end{equation}
where $b_1,\dots,b_{r}> 0$ run over all decomposition of $J+1$ into 
a sum of positive integers.
We are interested in the asymptotics of $S_J$ as $J\to\infty$.
Clearly, $S_J$ is a sum of order of $J^{r-1}$ monomials in the $q_j$.
The expression $S_J/J^{r-1}$ is just the $J$-th Riemann sum for 
an $(r-1)$-dimensional integral.  Thus
\begin{equation}
F_r(u_1,\dots,u_r)\stackrel{\mathrm{def}}=
\lim_{J\to\infty} S_J(q_1,\dots,q_r)/J^{r-1} =
\int_{\Delta_{r-1}} e^{u_1 x_1+\cdots+ u_r x_r}
\,d x_1 \cdots dx_{r-1},
\end{equation}
there the integration is over the $(r-1)$-dimensional simplex 
\begin{equation}
\Delta_{r-1}=\{(x_1,\dots,x_r)\mid x_1+\cdots+x_r=1,\, x_1,\dots,x_r\geq 0\}.
\end{equation}

\begin{proposition}  
The functions $F_r(u_1,\dots,u_r)$ are recursively
determined by the following relations.
If $u_1=\dots=u_r=u$ then
\begin{equation}
F_r(u,u,\dots,u) = e^u/(r-1)!.
\end{equation}
If $u_i\ne u_j$ for some $i$ and $j$ then $F_r$ is obtained
though $F_{r-1}$ by the divided difference operator:
\begin{equation}
F_r(u_1,\dots,u_r) = 
\frac{F_{r-1}(\dots,u_i,\dots,\widehat{u_j},\dots)- 
F_{r-1}(\dots,\widehat{u_i},\dots,u_j,\dots)}
{u_i-u_j}
\,,
\end{equation}
where $\widehat{u_i}$ means that the variable $u_j$ is omitted.
\label{prop:integral}
\end{proposition}  

%

\proof
In the first case $u_1=\cdots=u_r=u$ we just integrate the constant $e^u$
over the simplex $\Delta_{r-1}$, whose volume is $1/(r-1)!$.
In the second case we may assume that $i=r-1$ and $j=r$. The integral 
can be written as
\begin{equation}
\int_{\Delta_{r-2}} e^{u_1x_1+\cdots +u_{r-2} x_{r-2}}\,
\left(\int_0^{\alpha} 
e^{u_{r-1} x_{r-1} + u_r (\alpha - x_r)} d x_{r-1}\right) 
\,dx_1\cdots dx_{r-2}, 
\end{equation}
where $\alpha$ stands for $1-x_1-\cdots-x_{r-2}$.
The 1-dimensional integral in the parenthesis is equal to 
$(u_{r-1}-u_r)^{-1}(e^{u_r\alpha}-e^{u_{r-1}\alpha})$,
which gives the right-hand side of the recurrence relation.
\endproof

For nonnegative integers $a_1,\dots,a_r$, let
\begin{equation}
F_{(a_1,\dots,a_r)}(u_1,\dots,u_r) = F_k(u_1,\dots,u_1,u_2,\dots,u_2,\dots,
u_r,\dots,u_r),
\end{equation}
where $k=a_1+\cdots+a_r$ and we have $a_i$ copies of $u_i$, $i=1,\dots,r$,
in the right-hand side.
The function $F_{(a_1,\dots,a_r)}(u_1,\dots,u_r)$ is invariant under
simultaneous permutation of $a_i$'s and $u_i$'s.
Proposition~\ref{prop:integral} gives the following recurrence relations 
for these functions:
\begin{equation}
\begin{array}{l}
\displaystyle
F_{(a_1,\dots,a_r)} = 
\frac{F_{(\dots,a_i,\dots,a_j-1,\dots)}- 
F_{(\dots,a_i-1,\dots,a_j,\dots)}}{u_i-u_j}
\quad\textrm{if }u_i\ne u_j,\ a_i,a_j\geq 1, \\[.1in]
F_{(a_1,\dots,a_r)}(u_1,\dots,u_r) = 
F_{(\dots,a_i+a_j,\dots,\hat a_j,\dots)}(\dots,u_i,\dots,\hat u_j,\dots)
\quad\textrm{if }u_i=u_j, \\[.1in]
F_{(a_1,\dots,a_r)}(u_1,\dots,u_r) = 
F_{(\dots,\hat a_i,\dots)}(\dots,\hat u_i,\dots)
\quad\textrm{if }a_i=0,  \\[.1in]
F_{(a)}(u)=e^u/(a-1)!.
\end{array}
\label{eq:recurrence-relations}
\end{equation}
where $\hat a_i$ and $\hat u_i$ means that the corresponding terms
are omitted.

The next theorem presents an explicit expression for all these
functions.


\begin{theorem}
The function $F_r(u_1,\dots,u_r)$ is a continuous function 
of $u_1,\dots,u_r$ defined on $\C^r$.  If $u_i\ne u_j$ for all $i$
and $j$ then $F_r$ is given by
\begin{equation}
F_r(u_1,\dots,u_r) = \sum_{i=1}^r e^{u_i}\prod_{j\ne i} (u_i-u_j)^{-1}.
\label{eq:th:1}
\end{equation}
For arbitrary $u_1,\dots,u_r$, the function $F_r$ is obtained from 
this expression by continuity.  Also, $F_r$ with 
repeated arguments can be obtained by differentiation of the above
expression as follow.
For distinct $u_1,\dots,u_r$ and $a_1,\dots,a_r\geq 0$, we have
\begin{equation}
F_{(a_1+1,\dots,a_r+1)}(u_1,\dots,u_r) =
\frac{(\pa/\pa u_1)^{a_1}}{a_1!} \cdots
\frac{(\pa/\pa u_r)^{a_r}}{a_r!} F_r(u_1,\dots,u_r).
\label{eq:th:2}
\end{equation}
\label{th:F}
\end{theorem}

\proof
The first claim follows from Proposition~\ref{prop:integral} by 
induction on $r$.
In order to prove the second claim, remark that the $(k-1)$-dimensional
integral for $F_{(a_1+1,\dots,a_r+1)}$, $k=r+a_1+\cdots+a_r$, can be 
reduced to the following $(r-1)$-dimensional integral:
\begin{equation}
F_{(a_1+1,\dots,a_r+1)}(u_1,\dots,u_r)=
\int_{\Delta_{r-1}} \left(\prod_{i=1}^r\frac{x_i^{a_i}}{a_i!}\,
\right)
 e^{u_1 x_1+\cdots+ u_r x_r}
\,d x_1 \cdots dx_{r-1}\,.
\end{equation}
Now this integral for $F_{(a_1+1,\dots,a_r+1)}$ 
is obtained from the integral for $F_r = F_{(1,\dots,1)}$
by applying the differential operator 
$\prod_{i=1}^r (a_i!)^{-1}
(\pa / \pa u_i)^{a_i}$
to the integrand.
\endproof

Formula~\eqref{eq:th:1} says that 
$F_r(u_1,\dots,u_r)$ is the top coefficient of the Lagrange
interpolation of the exponent $f(x)=e^x$ at points $u_1,\dots,u_r$.
The second claim~\eqref{eq:th:2} 
can be reformulated in terms of the generating function, as follows:
\begin{equation}
\sum_{a_1,\dots,a_r\geq 0} 
z_1^{a_1}\cdots z_r^{a_r}\, F_{(a_1+1,\dots,a_r+1)}(u_1,\dots,u_r)
= F_r(u_1+z_1,\dots,u_r+z_r)\,.
\end{equation}

For example, according to Theorem~\ref{th:F}, we have,
for distinct $u,v,w$, 
\begin{equation}
\begin{array}{l}
F_{(1)}(u)=F_{(2)}(u)=e^u\,,
\quad
F_{(3)}(u)=e^u/2\,,\\[.1in]
F_{(1,1)}(u,v)=\frac{e^u-e^v}{u-v}\,,
\quad
F_{(2,1)}(u,v)=\frac{e^u}{u-v}-\frac{e^u-e^v}{(u-v)^2}\,,\\[.1in]
F_{(1,1,1)}(u,v,w)=\frac{e^u}{(u-v)(u-w)}-\frac{e^v}{(u-v)(v-w)} +
\frac{e^w}{(u-w)(v-w)}\,.
\end{array}
\end{equation}


\subsection{Formula for free two-point function}

In this section we put everything together and give a formula for
free two-point function.

For a permutation $\sigma\in S_k$ 
and $1\leq i\leq k+1$, let us define the
numbers $\ll_i(\sigma)$, $\lr_i(\sigma)$, $\rl_i(\sigma)$, and
$\rr_i(\sigma)$ as
\begin{equation}
\begin{array}{ll}
\ll_i(\sigma)=\#\{j\mid j<i,\ \sigma'(j)<\sigma'(i)\},\quad&
\lr_i(\sigma)=\#\{j\mid j<i,\, \sigma'(j)>\sigma'(i)\},\\[.05in]
\rl_i(\sigma)=\#\{j\mid j>i,\, \sigma'(j)<\sigma'(i)\},&
\rr_i(\sigma)=\#\{j\mid j>i,\, \sigma'(j)>\sigma'(i)\},
\end{array}
\end{equation}
where $\sigma'\in S_{k+1}$ is the permutation obtained from $\sigma$ 
by adding a fixed point $k+1$.

\begin{theorem} The $h$-th term of the genus expansion of 
the free two-point function is given by
\begin{equation}
A^h(n,n')=\sum_{\sigma\in\BR_{4h}^h}\sum_{i=1}^{4h+1}
F_{(\ll_i(\sigma)+1,\lr_i(\sigma),\rl_i(\sigma),\rr_i(\sigma)+1)},
\label{eq:AF}
\end{equation}
where the arguments of all $F$'s are 
$(2\pi \sqrt{-1} (n-n'), 2\pi\sqrt{-1}n,-2\pi\sqrt{-1}n',0)$.
\label{th:AF}
\end{theorem}
The functions $F$ are explicitly given by Theorem~\ref{th:F}.
This formula is valid for any complex values of $n$ and $n'$.
It involves rational expressions in $n$ and $n'$ and in the exponents
of $2\pi \sqrt{-1} n$, $-2\pi \sqrt{-1} n'$, and $2\pi \sqrt{-1}(n-n')$.
If $n$ and $n'$ are integers then all exponents are equal to 1.


\subsection{Example: free case, genus = 1, 2}
\label{sec:free,genus=1}

Assume that $u=2\pi n$ and $v=2\pi n'$.

In the case of genus $h=1$, there exists only one permutation 
$\sigma=1432$ in $\BR_{4}^1$.
Theorem~\ref{th:AF} gives the following expression
for $A^1(n,n')$: 
\begin{equation}
A^1(n,n') = F_{(1,0,0,5)}+F_{(2,0,2,2)}+F_{(2,1,1,2)}+F_{(2,2,0,2)}+
F_{(5,0,0,1)}\,,
\label{eq:A1nn'}
\end{equation}
all functions $F$ are in the variables 
$(i\,(u-v),i\,u,-i\,v,0)$.

The above formula is valid for arbitrary complex $n$ and $n'$.
Assume now that $n$ and $n'$ are integers.  Then 
$e^{i\, u}=e^{i\,v}=1$ and $A^1(n,n')$
does not involve any exponents.
In the cases when some of the arguments $i(u-v)$, $i\,u$, $-i\,v$, 
and $0$ coincide, 
we can use the reduction relations~\eqref{eq:recurrence-relations}
to simplify the expression~\eqref{eq:A1nn'}:
\begin{equation}
\begin{array}{l}
A^1(0,0) = 5F_{(6)}, \\[.1in]    
A^1(n,0) = A^1(0,n') = 
F_{(5,1)} + F_{(4,2)}+F_{(3,3)}+F_{(2,4)}+F_{(1,5)} = 0,\\[.1in]    
A^1(n,n) = 2 F_{(0,0,6)} + F_{(0,2,4)} + F_{(2,0,4)} + F_{(1,1,4)},\\[.1in]    
A^1(n,-n) = F_{(1,0,5)} + 3 F_{(2,2,2)} + F_{(5,0,1)},
\end{array}
\end{equation}
where $n$ is non-zero, the arguments of $F$'s in the third line
are $(i u,-i u,0)$, and $(2i u,i u,0)$ in the fourth line.

We can calculate the expression for $A^1(n,n')$ explicitly 
using Theorem~\ref{th:F}.  There are several cases that depend
on which pairs of arguments $i(u-v)$, $iu$, $-iv$, and $0$ coincide. 
The function $A^1(n,n')$ is given by
\begin{equation}
A^1(n,n')= 
\ \left\{
\begin{array}{cl}
\frac{1}{24}  & \textrm{if }  n = n' = 0;\\[.1in]
0   
& \textrm{if exactly one $n$ or $n'$ is 0}; 
\\[.1in]
\frac{1}{60} + \frac{1}{6\,u^2} + \frac{7}{u^{4}} \ 
& \textrm{if }  n = n' \textrm{ are non-zero}; 
\\[.1in]
-\frac{1}{12\,u^2} + \frac{35}{8\,u^4}
     & \textrm{if }  n = -n' \textrm{ are non-zero}; 
\\[.1in]
\frac {1}{(u-v)^2}\left(- \frac {1}{3}  + \frac {4}{u^2} 
+ \frac {4}{v^2}  - \frac {6}{uv}  - \frac{2}{(u-v)^2}\right) \ 
           & \textrm{otherwise.}
\end{array}
\right.
\end{equation}

%
%
%
%
%
%
%
%
%
%

\medskip
The genus $h=2$ free two-point function $A^2(n,n')$ can be written as a sum 
of $9\cdot |\BR_8^2|= 189$ integrals.
We can calculate all these integrals using {\tt Maple}. 
Explicitly, $A^2(n,n')$ is given by 
\begin{equation}
A^2(n,n')=\left\{
\begin{array}{cl}
\frac{21}{8!}=\frac{1}{2^4 5!}  & \textrm{if }  n = n' = 0;\\[.1in]
0          & \textrm{if exactly one $n$ or $n'$ is 0}; \\[.1in]
\frac{43}{72\cdot 7!} + \frac{1}{504\,u^2} + \frac{3}{10\, u^4} + 
\frac{107}{12\, u^{6}} + \frac{143}{8\, u^{8}}  \ 
           & \textrm{if }  n = n' \textrm{ are non-zero}; \\[.1in]
- \frac{1}{4\cdot 5!\, u^2}  + \frac{2^6}{3^2\, u^{4}} + 
\frac{2\cdot 53}{3\, u^{6}}
           & \textrm{if }  n = -n' \textrm{ are non-zero}; \\[.1in]
\frac{\textrm{some polynomial of degree 18 in $u$ and $v$}}
{ u^4 v^4 (u-v)^8 (u+v)^4 }
           & \textrm{otherwise}.
\end{array}
\right.
\end{equation}
We have skipped the numerator in the last case.  

\medskip
For genus $h=3$, there are $11!! / 7=1485$  elements in $\BR_{12}^3$
and each gives 13 terms.  In total we need to calculate 19305 terms.
This can also be done on a computer.

\subsection{Two-point function with an interaction}

The Feynman diagrams that correspond to Wick couplings in the case of the 
two-point function with a single interaction 
$\<O(x)\,[Z\psi]\, [\bar Z \bar\psi ] \,\bar O(y)\>$
can also be easily described in terms of permutations.
A coupling is given by a permutation $w\in S_n$ of order $m=J+2$ with
$w(1)=1$ together with a choice of two indices $i,j=2,\dots,m$,  
$i\ne j$, that correspond to the interacting fields.
The corresponding diagram is obtained from the free Feynman diagram, 
given by permutation $w$, by adding an ``interaction edge'' 
$i\rightsquigarrow j$ between the $i$th and $j$th edges of the free diagram.
Each such diagram given by a triple $(w,i,j)$ produces four terms in $G_{J,N}$ 
that correspond to four terms in the expansion of the product of commutators 
$[Z\psi]\,[\bar Z \bar\psi ]$.  All these terms come with 
a factor $N^{C(w,i,j)}$, where the exponent $C(w,i,j)$ counts the
number of closed loops in the ribbonized Feynman diagram.
It is not hard to express this number combinatorially in terms of 
cycles in permutations:
\begin{equation}
C(w,i,j) = \#\{\textrm{cycles in } \tilde w\} +
\delta_{w(i),\,j-1},
\end{equation}
where
$\tilde w= t_{w(i),\,j-1} c^{-1} w^{-1} c w$ and 
$t_{w(i),\,j-1}$ is the transposition of $w(i)$ and $j-1$.
Notice that the number $C(w,i,j)$ depend on the order of $i$ and $j$.
The {\it genus} of the Feynman diagram associated with triple $(w,i,j)$ 
is combinatorially determined by Euler's formula
\begin{equation}
h(w,i,j)=(n+1-C(w,i,j))/2.
\end{equation}

The two-point function with an interaction $G_{J,N}$ can now be written as
\begin{equation}
\begin{array}{l}
\displaystyle
G_{J,N}=\sum_{w\in S_{m},\, w(1)=1} \quad 
\sum_{i,\,j=2,\dots,n;\, i\ne j}
N^{J+3-2 h(w,i,j)}\,\, \times \\[.25in]
\displaystyle
\qquad\times
\left(q^{i} \rbar^{w(i)} - q^{i} \rbar^{w(j)} - q^{j} \rbar^{w(i)}
+q^{j} \rbar^{w(j)}\right) (q\rbar)^{-2}.
\end{array}
\label{eq:B}
\end{equation}
The $h$-th term of the genus expansion is given by the sum
\begin{equation}
G^h_J = \sum
\left(q^{i} \rbar^{w(i)} - q^{i} \rbar^{w(j)} - q^{j} \rbar^{w(i)}
+q^{j} \rbar^{w(j)}\right) (q\rbar)^{-2}
\end{equation}
over all triples $(w, i, j)$ with fixed genus $h(w,i,j)=h$.

We will see that $G^h_J$ is of order $J^{4h-1}$.  Let 
\begin{equation}
G^h(n,n')=\lim_{J\to\infty} G^h_J/J^{4h-1}.
\end{equation}

We have $h(w,i,j)\geq h(w)$, where $h(w)$ is the 
genus of the free Feynman diagram as defined in
Section~\ref{sec:2-point-permutations}.
According to Section~\ref{sec:block-reduction},
the total number of permutation $w\in S_m$ of genus $h$ is of 
order $J^{4h}$.  This implies that the total number of all triples $(w,i,j)$ 
with $h(w)<h(w,i,j)=h$ is of order $J^{4(h-1)}J^2=J^{4h-2}$
and each of these triples give 4 terms of absolute value 1.
Thus, in the limit of expression~\eqref{eq:B},
the pairs $(i,j)$ such that $h(w,i,j)>h(w)$ will not make any
contribution to $G^h(n,n')$.

It is natural to subdivide all possible choices for the interaction 
$(i,j)$ that does not increase the genus $h(w,i,j)=h(w)$ 
into the following there classes:
{\it nearest-neighbor interactions\/} ($j=i+1$ and $w(j)=w(i)+1$),
{\it semi-nearest neighbor interactions\/} (exactly one of the conditions
$j=i+1$ or $w(j)=w(i)+1$ holds), and 
{\it non-nearest neighbor interactions\/}
($j\ne i+1$ and $w(j)\ne w(i)+1$).
Let $G_{\textrm{ne}}^h(n,n')$, $G_{\textrm{se}}^h(n,n')$,
and $G_{\textrm{nn}}^h(n,n')$ be the contributions to $G^h(n,n')$
of these tree case, respectively.  Thus
\begin{equation}
G^h(n,n') = 
G_{\mathrm{ne}}^h(n,n') +
G_{\mathrm{sn}}^h(n,n')+
G_{\mathrm{nn}}^h(n,n')
\end{equation}

Let us show how to calculate these three expressions.
First, for a given permutation $w\in S_m$, we describe all pairs
$(i,j)$ such that  $h(w,i,j)=h(w)$.  
Suppose that $B_1,\dots,B_k$ are the blocks of 
a permutation $w\in S_m$.
Let us connect all blocks $B_1,\dots,B_k$ by directed edges 
of two types ``$\longrightarrow$'' and ``$\dashrightarrow$'' as follows:
\begin{equation}
\begin{array}{ccccccccc}
B_1 &\longleftarrow &B_2 &\longleftarrow &\cdots &\longleftarrow &
B_k &\longleftarrow&  B_1 \\[.05in]
B_{\sigma(1)} &\dashrightarrow &B_{\sigma(2)} & \dashrightarrow&\cdots &
 \dashrightarrow& B_{\sigma(k)} & \dashrightarrow & B_{\sigma(1)} \\[.05in]
\end{array}
\end{equation}
An {\it alternating chain} of blocks of length $l$ is a chain of the 
following type:
\begin{equation}
B_{s_1}\dashrightarrow B_{s_2} \longrightarrow B_{s_3} \dashrightarrow
B_{s_4} \longrightarrow B_{s_5}\dashrightarrow\cdots
\end{equation}
The {\it length} of a chain is its number of edges.
An 
{\it alternating cycle} of blocks is a closed chain 
$B_{s_1}\dashrightarrow\cdots \longrightarrow B_{s_r}=B_{s_1}$.

\begin{lemma}
Let $w\in S_m$ and  let $i,j\in\{2,\dots,m\}$, $i\ne j$.   
Then $h(w,i,j)=h(w)$ 
if and only if one of the following two conditions
is satisfied:
\begin{enumerate}
\item $j=i+1$ and $i,j$ belongs to the same block $B_s$.
\item $i$ the last element in some block $B_s$; $j$ is the first
element in some block $B_t$; and $B_s$ is connected with $B_t$
by an alternating chain of odd length:
\begin{equation}
B_s\dashrightarrow \cdots \dashrightarrow B_t.
\end{equation}
\end{enumerate}
\label{lem:wij}
\end{lemma}

According to Section~\ref{sec:block-reduction},
the total number of permutations $w\in S_m$ of genus $h$ 
with $k$ blocks 
is of order $J^k$.  Since $G_J^h$ is of order $J^{4h-1}$,
only permutations with $k=4h$ or $k=4h-1$ blocks can give a nonzero 
contribution to $G^h(n,n')$.

Let us first consider the case $k=4h$.   We have two options:

\smallskip
\noindent{\bf I. Nearest neighbor interactions: }
Suppose that $i$ and $j$ are such as in case~(1) of Lemma~\ref{lem:wij}.
Then 
\begin{equation}
q^{i} \rbar^{w(i)} - q^{i} \rbar^{w(j)} - q^{j} \rbar^{w(i)} 
+q^{j} \rbar^{w(j)} 
= (1-q)(1-\rbar)\, q^i \rbar^{w(i)}.
\end{equation}
The contribution of these terms to $G_J^h$ is asymptotically
equal to $(1-q)(1-\rbar) J^{4h+1} A_J^h$.  Note that it of order
$J^{4h-1}$,  because $(1-q)(1-\rbar)$ is of order $J^{-2}$.
Thus the contribution of these terms to $G^h(n,n')$ is equal to
\begin{equation}
G_{\mathrm{ne}}^h(n,n') = u\,v\,A^h(n,n'),
\end{equation}
where $A^h(n,n')$ is the $h$-th term in the free two-point function
given by Theorem~\ref{th:AF}.
As before, $u=2\pi  n$ and $v=2\pi  n'$.

\smallskip
\noindent{\bf I{}I. Semi-nearest neighbor interactions: }  This is case~(2)
of Lemma~\ref{lem:wij}.  For $k=4h$ all possible
alternating cycles of blocks have length 4.
Thus the only possible pairs $i$ and $j$ are the following:
$i$ is the last element of $B_s$; $j$ is the first element of $B_t$;
and ($B_s\dashrightarrow B_t$ or $B_s\longleftarrow B_t$).
In this case we can recombine the terms as follows:
\begin{equation}
\begin{array}{l}
\displaystyle
\sum\left(q^{i} \rbar^{w(i)} - q^{i} \rbar^{w(j)} - 
q^{j} \rbar^{w(i)} +q^{j} \rbar^{w(j)}\right)=
\\[.2in]
\displaystyle
\quad
=(2-q-\rbar)\left(\sum_{i\in\mathrm{LAST}(w)} q^{i} \rbar^{w(i)}\right) 
+(2-q^{-1}-\rbar^{-1})\left(\sum_{j\in\mathrm{FIRST}(w)} 
q^{j} \rbar^{w(j)}\right), 
\end{array}
\label{eq:LAST-FIRST}
\end{equation}
where $\mathrm{LAST}(w)$ is the set of last elements in blocks of $w$
and $\mathrm{FIRST}(w)$ is the set of first elements in blocks of $w$. 
The right-hand side of this expression involves $8h$ terms, each 
come with a factor  $(2-q-\rbar)\sim -\sqrt{-1}(u-v)\,J^{-1}$ or 
$(2-q^{-1}-\rbar^{-1})=\sqrt{-1}(u-v)\,J^{-1}$.
Since the number of elements $w$ with $\sigma\in\BR_{4h}^h$ 
is of order $J^{4h}$, the total contribution of terms of this type is
of order $J^{4h} J^{-1} = J^{4h-1}$.
In the limit $J\to\infty$, the sum over $w$ of given genus $h$ 
becomes a finite sum integrals given by Theorem~\ref{th:F}.
Using the notation of Theorem~\ref{th:AF}, we can write two
terms in the right-hand side of~\eqref{eq:LAST-FIRST}
as the following sums over $\sigma$ and $i$ such as in~\eqref{eq:AF}: 
\begin{equation}
\begin{array}{l}
\displaystyle
(2-q-\rbar)\sum_{j\in\mathrm{LAST}(w)} q^j\rbar^{w(j)} = -\sqrt{-1}(u-v)
\sum_{\sigma,\,i} F_{(\ll_i+1,\lr_i,\rl_i,\rr_i)}, \\[.2in]
\displaystyle
(2-q^{-1}-\rbar^{-1})\sum_{j\in\mathrm{FIRST}(w)} q^j\rbar^{w(j)} = 
\sqrt{-1}(u-v)
\sum_{\sigma,\,i} F_{(\ll_i,\lr_i,\rl_i,\rr_i+1)},
\end{array}
\label{eq:L-F-sum}
\end{equation}
where the arguments of $F$'s are $(\sqrt{-1}(u-v),\sqrt{-1}u,-\sqrt{-1}v,0)$.
According to~\eqref{eq:recurrence-relations}, the sum of the 
two summands in the right-hand sides of~\eqref{eq:L-F-sum} is 
\begin{equation}
(u-v)^2\,F_{(\ll_i+1,\lr_i,\rl_i,\rr_i+1)},
\end{equation}
which is exactly $(u-v)^2$ times the  summand in~\eqref{eq:AF}. 
Thus the total contribution of semi-nearest interactions is equal to
\begin{equation}
G_{\mathrm{sn}}^h(n,n')= (u-v)^2 A^h(n,n'),
\end{equation}
where $A^h(n,n')$ is the free two-point function given by Theorem~\ref{th:AF}.
If $n=n'$ then $G_{\mathrm{sn}}^h(n,n)=0$.

\smallskip
Let us now consider the case $k=4h-1$.  The total number of 
permutations $w$ with $\sigma\in\BR_{4h-1}^h$ is of order $J^{4h-1}$.
In this case we may ignore the nearest neighbor interactions because their
contribution comes with a prefactor $(1-q)(1-\rbar)\sim J^{-2}$, which
makes the total order $J^{4h-1} J^{-2}$ subdominant to $J^{4h-1}$.
In this case all alternating cycles of blocks have size 4, except
a single cycle of size 6. 
The 4-cycles produce semi-nearest interactions that come with a prefactor 
$(2-q-\rbar)$ or $(2-q^{-1}-\rbar^{-1})$ of order $J^{-1}$, which again makes 
their contribution negligible in the limit $J\to\infty$.

\smallskip
\noindent{\bf I{}I{}I. Non-nearest neighbor interactions: }  
Suppose that $k=4h-1$ and $w$ is a genus $h$ permutation with 
$4h-1$ blocks.
Then there exists a unique alternating 6-cycle 
\begin{equation}
B_{s_1}\dashrightarrow B_{s_2}\longrightarrow
B_{s_3}\dashrightarrow B_{s_4}\longrightarrow
B_{s_5}\dashrightarrow B_{s_6}\longrightarrow B_{s_1}\,.
\end{equation}
Let $f_r$ and $l_r$ be the first and the last elements, respectively, 
in the block $B_{s_r}$. 
There are only 3 possible choices $(i,j)= (l_1,f_4), (l_3,f_6), (l_5,f_2)$
for the interaction edge, whose contribution survive in the limit $J\to\infty$. 
Each of these 3 pairs produces 4 terms,
which gives the following 12 terms:
\begin{equation}
\begin{array}{l}
q^{l_1}\rbar^{w(l_1)}+ q^{f_4}\rbar^{w(f_4)}-
q^{l_1}\rbar^{w(f_4)}- q^{f_4}\rbar^{w(l_1)}\\
q^{l_3}\rbar^{w(l_3)}+ q^{f_6}\rbar^{w(f_6)}-
q^{l_3}\rbar^{w(f_6)}- q^{f_6}\rbar^{w(l_3)}\\
q^{l_5}\rbar^{w(l_5)}+ q^{f_2}\rbar^{w(f_2)}-
q^{l_5}\rbar^{w(f_2)}- q^{f_2}\rbar^{w(l_5)}.
\end{array}
\label{eq:12-lf}
\end{equation}
Since the number of genus $h$ permutations $w$ with $4h-1$ blocks
is of order $J^{4h-1}$, the total contribution of these terms
to $G_J^h$ is again of order $J^{4h-1}$.
The sum of the 12 terms in~\eqref{eq:12-lf}
over all genus $h$ permutations $w$ with 
$4h-1$ blocks can be written as
the sum
\begin{equation}
\sum_{\hat w} \left(
q^{i-1}\rbar^{\hat w(i-1)} - 2 q^{i}\rbar^{\hat w(i)}
+ q^{i+1}\rbar^{\hat w(i+1)}\right),
\label{eq:NN-qr-4h}
\end{equation}
over all genus $h$ permutations $\hat w$ with $4h$ blocks
such that $\hat w$ contains a block $B_{\mathrm{sing}}$ with a single 
element, $|B_{\mathrm{sing}}|=1$,  and $i\ne 1$ is the position of the block 
$B_{\mathrm{sing}}$ in $\hat w$.

Indeed, there are the following 3 ways to transform $w$
with $4h-1$ blocks to a permutation $\hat w$ with $4h$ blocks 
by inserting a block with a single element and preserving the genus.
Let $w^{(1)}$ be the permutation obtained from $w$ by inserting 
a new edge to its Feynman diagram between the blocks $B_1$ and $B_6$
on the top and the blocks $B_3$ and $B_4$ on the bottom.  
Similarly, we construct the permutation $w^{(2)}$
by inserting a new block with a single element between $B_3$ and $B_2$
on the top and $B_5$ and $B_6$ on the bottom; and the permutation 
$w^{(3)}$ by inserting a new block between $B_5$ and $B_4$ on the top
and $B_1$ and $B_2$ on the bottom.
One can easily check that the sum of the summands
in~\eqref{eq:NN-qr-4h} for 3 permutations 
$\hat w=w^{(1)}, w^{(2)}, w^{(3)}$ produces
exactly the 12 terms in~\eqref{eq:12-lf}.

As before, in the limit $J\to\infty$, the sum~\eqref{eq:NN-qr-4h}
reduces to a finite sum of integrals 
over block-reduced $\sigma\in\BR_{4h}^h$.
Using the notation of Theorem~\ref{th:AF}, we can write 
the contribution $G^h_{\mathrm{nn}}(n,n')$ of non-nearest neighbor 
interactions as follows:
\begin{equation}
\begin{array}{l}
\displaystyle
G^h_{\mathrm{nn}}(n,n')= 
-2 \sum_{\sigma\in\BR_{4h}^h} \sum_{i=2}^{4h} F_{(\ll_i,\lr_i,\rl_i,\rr_i)} 
\ +
\\[.3in]
\displaystyle
\qquad +\sum_{\sigma\in\BR_{4h}^h}
\left( 
\sum_{i=1,\dots,4h-1}^{\sigma_i'<\sigma_{i+1}'}
F_{(\ll_i+1,\lr_i,\rl_i,\rr_i-1)} 
+\sum_{i=1,\dots,4h-1}^{\sigma_i'>\sigma_{i+1}'}
F_{(\ll_i+1,\lr_i,\rl_i-1,\rr_i)} 
\right)
+
\\[.3in]
\displaystyle
\qquad +\sum_{\sigma\in\BR_{4h}^h} 
\left(
\sum_{i=3,\dots,4h+1}^{\sigma_{i-1}'<\sigma_{i}'}
F_{(\ll_i-1,\lr_i,\rl_i,\rr_i+1)} 
+\sum_{i=3,\dots,4h+1}^{\sigma_{i-1}'>\sigma_{i}'}
F_{(\ll_i,\lr_i-1,\rl_i,\rr_i+1)} 
\right),
\end{array}
\label{eq:Gnn}
\end{equation}
where the arguments of all 
$F$'s are $(\sqrt{-1}(u-v),\sqrt{-1}u,-\sqrt{-1}v,0)$.

%

\subsection{Example: interaction case, genus = 1}

Then the contribution of  nearest and semi-nearest interactions 
is equal to
\begin{equation}
G_{\mathrm{ne}}^1(n,n') +
G_{\mathrm{sn}}^1(n,n')= (uv+(u-v)^2)\,A^1(n,n') =
(u^2+v^2-uv)\,A^1(n,n').
\end{equation}
This reduces to the free genus 1 case given in Section~\ref{sec:free,genus=1}.

Formula~\eqref{eq:Gnn} gives the following expression for
$G_{\mathrm{nn}}(n,n')$:
\begin{equation}
\begin{array}{ll}
\displaystyle
G_{\mathrm{nn}}^1(n,n')= 
&
-2(F_{(1,2,0,1)} + F_{(1,1,1,1)}+ F_{(1,0,2,1)}) + \\[.05in]
\displaystyle
&+ F_{(1,0,0,3)} + F_{(2,0,1,1)} + F_{(2,1,0,1)} + \\[.05in]
&+ F_{(1,0,1,2)} + F_{(1,1,0,2)} + F_{(3,0,0,1)},
\end{array}
\end{equation}
where, as usual, the arguments of all
$F$'s are $(i(u-v),i u,-i v,0)$.
Using relations~\eqref{eq:recurrence-relations}, we can simplify 
this expression:
\begin{equation}
\begin{array}{l}
G_{\mathrm{nn}}^1(0,0)= G_{\mathrm{nn}}^1(n,0)= 
G_{\mathrm{nn}}^1(0,n')=0,\\[.1in]
G_{\mathrm{nn}}^1(n,n)= 
2(F_{(0,0,4)}- F_{(0,2,2)}- F_{(2,0,2)}+ F_{(0,1,3)} + F_{(1,0,3)} - 
F_{(1,1,2)}),\\[.1in]
G_{\mathrm{nn}}^1(n,-n)= 
F_{(1,0,3)}+ F_{(3,0,1)}+ 2 F_{(1,1,2)}+ 2 F_{(2,1,1)} -6 F_{(1,2,1)}.
\end{array}
\end{equation}
where $n$ is non-zero, the  arguments in the second line are $(iu,-iu,0)$,
and the arguments in the third line are $(2iu,iu,0)$.

Finally, using Theorem~\ref{th:F}, we obtain
\begin{equation}
G^1(n,n')=(u^2+v^2-uv)\,A^1(n,n') + 
G_{\mathrm{nn}}^1(n,n'), 
\end{equation}
where
\begin{equation}
G^1_{\mathrm{nn}}(n,n')= 
\ \left\{
\begin{array}{cl}
0  & \textrm{if $n$ or $n'$ is 0}; 
\\[.1in]
\frac {1} {3} + \frac{10}{u^2} \ 
& \textrm{if }  n = n' \textrm{ are non-zero}; \\[.1in]
-\frac{15}{2\,u^2} 
     & \textrm{if }  n = -n' \textrm{ are non-zero}; \\[.1in]
\frac{6}{uv} + \frac{2}{(u-v)^2} \ 
           & \textrm{otherwise.}
\end{array}
\right.
\end{equation}

\section{Effective operator approach to Wick contractions}

In this appendix we outline an approach to handling the 
color combinatorics of planar and toroidal Feynman diagrams 
which makes use of perhaps more  familiar techniques based on 
Wick contractions.
To demonstrate these methods we consider the planar and genus one
contributions to the two-point function at $\CO(g^2_{YM})$. This provides
an independent check of many of the calculations presented
elsewhere in this paper.

Once again the BMN operators are,
\begin{equation} \label{oP}
 O = \sum_{l=0}^{J} q^l \Tr\left(\phi Z^l \psi Z^{J-l}\right) \,\,\,\,\,\,
{\rm and}\,\,\,\,\,\, 
\bar{ O}=\sum_{l'=0}^{J} \bar{r}^{l'} \Tr\left(\bar{Z}^{J-l'} \bar{\psi}
\bar{Z}^{l'} \bar{\phi}\right)
\end{equation}
with $q = \exp( 2\pi i n/J)$ and $r=\exp( 2\pi i n^{\prime}/J)$. 

\medskip

As discussed in Appendix B it is only necessary to 
consider the F-term interactions in the ${\N}=4$ SYM action. 
These can be written,
\begin{equation}
-4g^2_{YM} \Tr([\bar{Z},\bar{\phi}][Z,\phi])=
4g^2_{YM}f^{pab} \bar{Z}^a \bar{\phi}^b~f^{pcd} Z^c \phi^d
\label{fterm}
\end{equation}
where the $f^{pab}$ are the structure constants of $SU(N)$, and we
trivially extend them to $U(N)$ by adding the $N\times N$ matrix
$T^0 = I/\sqrt{N}$ to the standard set of $N^2-1$ $SU(N)$
generators $T^a$, $a=1, \dots, N^2-1$. The full basis of $U(N)$
generators is then normalized by\footnote{This normalization 
differs from that common in physics by a
factor of 2 so that the $f^{pab}$ differ by $\sqrt{2}$.}
\begin{equation}
  (ab) \equiv \Tr{T^a T^b} = \delta^{ab}.
\end{equation}

Using equation \eqref{fterm} we define an effective operator, ${ O}_{\rm 
eff}$, as the
sum of Wick contractions of $\phi$ and each $Z$ in
\eqref{oP} with the factor $f^{pab} \bar{Z}^a \bar{\phi}^b$ of the
interaction. After trivial manipulation this produces
\begin{multline} \label{effop}
O_{\rm eff} =  -i \sum_{l=0}^{J} q^l \\
\times\left(\sum_{m=0}^{l-1}
   \Tr([T^a,T^p]Z^m T^a Z^{l-m-1} \psi Z^{J-l}) +
  \sum_{m=0}^{J-l-1} \Tr([T^a,T^p]Z^l \psi Z^{J-l-m-1} T^a 
Z^m)\right)
\end{multline}

From now on we compactify the notation by replacing all explicit
generators by their index values, i.e. $T^a \rarrow a$ and
replace the explicit trace of an arbitrary $N\times N$ matrix $M$ by
$\Tr(M) \rarrow (M)$. The following `splitting/joining' rules can
then be used to evaluate traces and
products thereof which involve summed repeated color indices:
\eqn{rules}{\begin{array}{rclrcl}
(MaM'a) &=& (M)(M')\qquad&(ab) &=& \delta^{ab}\\
(Ma)(aM') &=& (MM')\qquad&(a) &=& \sqrt{N} \delta^{a0}\\
 aa &=& N I&(~) &=& N
\end{array}}
These follow from results derived in ref.~\cite{cvit}.

Finally it is useful to write $ O_{\rm eff} =  O_{\rm eff}^1 +  O_{\rm eff}^2$ where 
$  O_{\rm eff}^1$ includes only the `nearest neighbor interactions  
{\it i.e.,} the $m=0$ terms in
\eqref{effop}. The $m>0$ terms are contained in $ O_{\rm eff}^2$ and 
represent interactions between fields which are non-nearest neighbors. 
These operators may be expressed as
\begin{equation}
 O_{\rm eff}^1=-iN(q-1)\sum_{l=0}^{J-1}q^l(pZ^l\psi Z^{J-l-1});
\label{nearest}
\end{equation}
\bea
 O_{\rm eff}^2&=&-i\frac{q}{q-1}\sum_{m=1}^{J-1}(pZ^m)(Z^{J-m-1}\psi)
(1+q^{-1}-q^m-q^{-m-1})
\nonumber \\
&-&i\sum_{m=1}^{J-1}\sum_{l=m+1}^{J}q^l(1-q^{-m-1})
(Z^m)(pZ^{l-m-1}\psi Z^{J-l}).
\label{nonnear}
\eea
In obtaining equations \eqref{nearest} and \eqref{nonnear} we have 
applied 
the $U(N)$ trace identities in equation \eqref{rules} 
and explicitly summed several of the geometric progressions in $q^l$ by 
reversing the order of summations.

With equations \eqref{nearest} and \eqref{nonnear} at our disposal we may 
now 
discuss the order $g^2_{YM}$ contributions to the correlator
\begin{equation} \label{split}
\<{ O} (x) \bar{ O}(y)\> = \<{ O}_{\rm eff}^1 \bar{{ O}}_{\rm eff}^1\> 
+\<{ O}_{\rm eff}^1\bar{{ O}}_{\rm eff}^2\> +\<{ O}_{\rm eff}^2 \bar{{ O}}_{\rm eff}^1\> +
\<{ O}_{\rm eff}^2 \bar{{ O}}_{\rm eff}^2\>.
\end{equation}
From now on we omit reference to the space-time points $x,y$ since our
main concern is to capture the order $N^{J+3}$ planar
and order $N^{J+1}$ genus one contributions to this correlator. 
This is handled here by performing the Wick
contractions of the fields $Z,\bar{Z}$ and evaluating the ensuing color
contractions and traces.  Evaluating the various terms in 
equation \eqref{split} requires the following `contraction identities',
\bea
\Tr(Z^a\bar{Z}^a) &=& N^{a+1} + 
\binom{a+2}{4}
N^{a-1} + \O(N^{a-3})
\nonumber \\
\Tr(Z^a)\Tr(\bar{Z}^a)&=& aN^{a} + \O(N^{a-2})
\label{idents} \\
\Tr(Z^a)\Tr(\bar{Z}^{a+b}Z^{b}) &=& a(b+1)N^{a+b} + \O(N^{a+b-2})
\nonumber \\
\Tr(Z^a\bar{Z}^cZ^{b}\bar{Z}^d) &=& 
\left({\mathrm{min}}(a,b,c,d)+1\right)N^{a+b+1} 
+\O(N^{a+b-1}) \,\,\,\,{\rm with} \,\,\,\, a+b=c+d
\nonumber
\eea
which are derived by counting the number of ways one may perform the Wick 
contractions within each trace structure while obtaining a maximal power 
of $N$. The remaining $\O(N^{J+1})$ contributions are the semi-nearest and
   non-nearest diagrams of section 4 of the main text. In the effective
   operator formalism the semi-nearest diagrams are given by the second
   and third terms in (D.8), and computation shows that they vanish in
   the special case $q=r$ we are considering. The non-nearest diagrams
   are given by the last term in (D.8). It is straightforward to evaluate
   and sum the relevant Wick contractions and obtain
\begin{equation} 
\<{ O}_{\rm eff}^1 \bar{{ O}}_{\rm eff}^1\>= N^2(q-1)(\bar{q}-1)
\sum_{l=0}^{J-1}\sum_{\bar{l}=0}^{J-1}q^l\bar{q}^{\bar{l}}
(Z^l\bar{Z}^{\bar{l}})(Z^{J-l-1}\bar{Z}^{J-\bar{l}-1})
\label{near1}
\end{equation}
for which planar contributions are only possible for $l=\bar{l}$. 
It is interesting to note that up to the $(q-1)(\bar{q}-1)$ prefactor this is
exactly the same expression which generates the all-genus polynomial at order
$g^0_{YM}$.
Use of the identities in equation \eqref{idents} and explicit evaluation 
of the
resulting sums yields an expression which in the limit of large $J$ is given
by
\bea
\<{ O}_{\rm eff}^1 \bar{{ O}}_{\rm eff}^1\> &=& 
g_2^2A_{nn}\left(1-n^2\lambda^{\prime}\ln(\Lambda^2x^2)\right),
\nonumber \\
{\rm where} \,\,\,\, A_{nn} &=& 
\left(\frac{1}{60}-\frac{1}{6}\frac{1}{(2\pi n)^2}+
\frac{7}{(2\pi n)^4}\right).
\label{near2}
\eea
Notice that the non-planar corrections are of order $g_2^2$ as has been 
emphasized throughout this paper. In order to obtain the remaining 
$\CO(N^{J+1})$
contributions we need only consider the final term in equation 
\eqref{split} since
in the special case we are considering, {\it i.e.,} $q=r$, the other terms
conspire to precisely vanish at leading order in a large $J$ expansion. 
Evaluation of the final term in equation \eqref{split}
is a straightforward calculation which yields,
\begin{equation}
\<{ O}_{\rm eff}^2 \bar{{ O}}_{\rm eff}^2\> = 
\frac{g_2^2\lambda^{\prime}\ln(\Lambda^2 x^2)}{4\pi^2}\left(\frac{1}{3}+
\frac{5}{2\pi^2n^2}\right)
\label{nonnearA}
\end{equation}     
Putting the results of equations \eqref{near2} and \eqref{nonnearA} the 
complete 
planar and torus contributions to the $g^2_{YM}$ piece of the two-point
function is given by,
\begin{equation}
(1+g_2^2A_{nn})\left(1-n^2\lambda^{\prime}\ln(\Lambda^2x^2)\right)+
\frac{g_2^2\lambda^{\prime}\ln(\Lambda^2 x^2)}{4\pi^2}\left(\frac{1}{3}+
\frac{5}{2\pi^2n^2}\right)
\label{final}
\end{equation}
which is precisely the same answer as obtained in the main text.
Extending these techniques to arbitrary $q,r$ is straightforward
and in all cases is found to agree with the methods used elsewhere in 
this paper.


\begin{thebibliography}{99}
%
%

\bibitem{'tHooft:1973jz}
G.~'t Hooft,
{\it ``A Planar Diagram Theory For Strong Interactions,''}
Nucl.\ Phys.\ B {\bf 72}, 461 (1974).

\bibitem{ias}
{D.~Berenstein, J.~Maldacena and H.~Nastase,
{\it ``Strings in flat space and pp waves from $N = 4$ super 
Yang-Mills,''}
arXiv:hep-th/0202021.
}

\bibitem{Blau:2001ne}
M.~Blau, J.~Figueroa-O'Farrill, C.~Hull and G.~Papadopoulos,
{\it ``A new maximally supersymmetric background of IIB superstring 
theory,''}
JHEP {\bf 0201}, 047 (2002)
[arXiv:hep-th/0110242].

\bibitem{Blau:2002dy}
M.~Blau, J.~Figueroa-O'Farrill, C.~Hull and G.~Papadopoulos,
{\it ``Penrose limits and maximal supersymmetry,''}
Class.\ Quant.\ Grav.\  {\bf 19}, L87 (2002)
[arXiv:hep-th/0201081].

\bibitem{Metsaev:2001bj}
{R.~R.~Metsaev,
{\it ``Type IIB Green-Schwarz superstring in plane wave Ramond-Ramond  
background,''}
Nucl.\ Phys.\ B {\bf 625}, 70 (2002)
[arXiv:hep-th/0112044].
}

\bibitem{Lee:1998bx}
S.~M.~Lee, S.~Minwalla, M.~Rangamani and N.~Seiberg,
``Three-point functions of chiral operators in $D = 4$, $N = 4$ SYM at  
large $N$,'' Adv.\ Theor.\ Math.\ Phys.\  {\bf 2}, 697 (1998)
[arXiv:hep-th/9806074].


\bibitem{9907098}
{E.~D'Hoker, D.~Z.~Freedman and W.~Skiba,
{\it ``Field theory tests for correlators in the $AdS$/CFT 
correspondence,''}
Phys.\ Rev.\ D {\bf 59}, 045008 (1999)
[arXiv:hep-th/9807098].
}

\bibitem{Spradlin:2002ar}
M.~Spradlin and A.~Volovich,
{\it ``Superstring interactions in a pp-wave background,''}
arXiv:hep-th/0204146.


\bibitem{germanpaper}
C.~Kristjansen, J.~Plefka, G.~W.~Semenoff and M.~Staudacher,
{\it ``A New Double-Scaling Limit of $N=4$ Super Yang-Mills Theory and 
PP-Wave Strings,''}
arXiv:hep-th/0205033.

\bibitem{berencompete}
D.~Berenstein, H.~Nastase,
{\it ``On light-cone string field theory from Super Yang-Mills and 
holography,''}
arXiv:hep-th/0205048.

\bibitem{grosscompete}
D.J.~Gross, A.~Mikhailov, R.~Roiban,
{\it ``Operators with large R charge in $N=4$ Yang-Mills theory,''}
arXiv:hep-th/0205066.



\bibitem{giant}
J.~McGreevy, L.~Susskind and N.~Toumbas,
{\it ``Invasion of the giant gravitons from anti-de Sitter space,''}
JHEP {\bf 0006}, 008 (2000)
[arXiv:hep-th/0003075].


%
\bibitem{Metsaev:1999gz}
R.~R.~Metsaev,
{\it ``light-cone gauge formulation of IIB supergravity in $AdS_5 \times 
S^5$ background and $AdS$/CFT correspondence,''}
Phys.\ Lett.\ B {\bf 468}, 65 (1999)
[arXiv:hep-th/9908114].


\bibitem{Blau:2002rg}
M.~Blau, J.~Figueroa-O'Farrill and G.~Papadopoulos,
{\it ``Penrose limits, supergravity and brane dynamics,''}
arXiv:hep-th/0202111.


\bibitem{Berenstein:2002ke}
D.~Berenstein, C.~P.~Herzog and I.~R.~Klebanov,
{\it ``Baryon spectra and $AdS$/CFT correspondence,''}
arXiv:hep-th/0202150.


\bibitem{Itzhaki:2002kh}
N.~Itzhaki, I.~R.~Klebanov and S.~Mukhi,
{\it ``PP wave limit and enhanced supersymmetry in gauge theories,''}
JHEP {\bf 0203}, 048 (2002)
[arXiv:hep-th/0202153].

\bibitem{Gomis:2002km}
J.~Gomis and H.~Ooguri,
{\it ``Penrose limit of $N = 1$ gauge theories,''}
arXiv:hep-th/0202157.


\bibitem{Russo:2002rq}
J.~G.~Russo and A.~A.~Tseytlin,
{\it ``On solvable models of type IIB superstring in NS-NS and R-R plane 
wave backgrounds,''}
JHEP {\bf 0204}, 021 (2002)
[arXiv:hep-th/0202179].


\bibitem{Zayas:2002rx}
L.~A.~Zayas and J.~Sonnenschein,
{\it ``On Penrose limits and gauge theories,''}
arXiv:hep-th/0202186.

\bibitem{Hatsuda:2002kx}
M.~Hatsuda, K.~Kamimura and M.~Sakaguchi,
{\it ``Super-PP-wave algebra from super-$AdS \times  S$ algebras in
eleven-dimensions,''}
arXiv:hep-th/0204002.


\bibitem{Hatsuda:2002xp}
M.~Hatsuda, K.~Kamimura and M.~Sakaguchi,
{\it ``From super-$AdS_5 \times S^5$ algebra to super-pp-wave algebra,''}
arXiv:hep-th/0202190.

\bibitem{Alishahiha:2002jj}
M.~Alishahiha and M.~M.~Sheikh-Jabbari,
{\it ``Strings in PP-waves and worldsheet deconstruction,''}
arXiv:hep-th/0204174.

\bibitem{Alishahiha:2002ev}
M.~Alishahiha and M.~M.~Sheikh-Jabbari,
{\it ``The PP-wave limits of orbifolded $AdS_5 \times  S^5$,''}
arXiv:hep-th/0203018.


\bibitem{Billo:2002ff}
M.~Billo and I.~Pesando,
{\it ``Boundary states for GS superstrings in an Hpp wave background,''}
arXiv:hep-th/0203028.


\bibitem{David:2002wn}
J.~R.~David, G.~Mandal and S.~R.~Wadia,
{\it ``Microscopic formulation of black holes in string theory,''}
arXiv:hep-th/0203048.


\bibitem{Cvetic:2002si}
M.~Cveti\v{c}, H.~Lu and C.~N.~Pope,
{\it ``M-theory pp-waves, Penrose limits and supernumerary 
supersymmetries,''}
arXiv:hep-th/0203229.



\bibitem{Takayanagi:2002hv}
T.~Takayanagi and S.~Terashima,
{\it ``Strings on orbifolded pp-waves,''}
arXiv:hep-th/0203093.

\bibitem{Gursoy:2002tx}
U.~Gursoy, C.~Nunez and M.~Schvellinger,
{\it ``RG flows from Spin(7), CY 4-fold and HK manifolds to $AdS$, Penrose
limits and pp waves,''}
arXiv:hep-th/0203124.


\bibitem{Floratos:2002uh}
E.~Floratos and A.~Kehagias,
{\it ``Penrose limits of orbifolds and orientifolds,''}
arXiv:hep-th/0203134.



\bibitem{Michelson:2002wa}
J.~Michelson,
{\it ``(Twisted) toroidal compactification of pp-waves,''}
arXiv:hep-th/0203140.


\bibitem{Gueven:2002cj}
R.~Gueven,
{\it ``Randall-Sundrum zero mode as a Penrose limit,''}
arXiv:hep-th/0203153.

\bibitem{Chu:2002in}
C.~S.~Chu and P.~M.~Ho,
{\it ``Noncommutative D-brane and open string in pp-wave background with
B-field,''}
arXiv:hep-th/0203186.


\bibitem{Cvetic:2002hi}
M.~Cveti\v{c}, H.~Lu and C.~N.~Pope,
{\it ``Penrose limits, pp-waves and deformed M2-branes,''}
arXiv:hep-th/0203082.

\bibitem{Dabholkar:2002zc}
A.~Dabholkar and S.~Parvizi,
{\it ``Dp branes in pp-wave background,''}
arXiv:hep-th/0203231.

\bibitem{Berenstein:2002zw}
D.~Berenstein, E.~Gava, J.~M.~Maldacena, K.~S.~Narain and H.~Nastase,
{\it ``Open strings on plane waves and their Yang-Mills duals,''}
arXiv:hep-th/0203249.


\bibitem{Gauntlett:2002cs}
J.~P.~Gauntlett and C.~M.~Hull,
{\it ``pp-waves in 11-dimensions with extra supersymmetry,''}
arXiv:hep-th/0203255.

\bibitem{Lee:2002cu}
P.~Lee and J.~w.~Park,
{\it ``Open strings in PP-wave background from defect conformal field
theory,''}
arXiv:hep-th/0203257.

\bibitem{Das:2002cw}
S.~R.~Das, C.~Gomez and S.~J.~Rey,
{\it ``Penrose limit, spontaneous symmetry breaking and holography in 
pp-wave background,''}
arXiv:hep-th/0203164.

\bibitem{Kiritsis:2002kz}
E.~Kiritsis and B.~Pioline,
{\it ``Strings in homogeneous gravitational waves and null holography,''}
arXiv:hep-th/0204004.

\bibitem{Kumar:2002ps}
A.~Kumar, R.~R.~Nayak and Sanjay,
{\it ``D-brane solutions in pp-wave background,''}
arXiv:hep-th/0204025.


\bibitem{Gubser:2002tv}
S.~S.~Gubser, I.~R.~Klebanov and A.~M.~Polyakov,
{\it ``A semi-classical limit of the gauge/string correspondence,''}
arXiv:hep-th/0204051.


\bibitem{Skenderis:2002vf}
K.~Skenderis and M.~Taylor,
{\it ``Branes in $AdS$ and pp-wave spacetimes,''}
arXiv:hep-th/0204054.


\bibitem{Mukhi:2002ck}
S.~Mukhi, M.~Rangamani and E.~Verlinde,
{\it ``Strings from quivers, membranes from moose,''}
arXiv:hep-th/0204147.


\bibitem{Balasubramanian:2002sa}
V.~Balasubramanian, M.~x.~Huang, T.~S.~Levi and A.~Naqvi,
{\it ``Open strings from N = 4 super Yang-Mills,''}
arXiv:hep-th/0204196.


\bibitem{Imamura:2002xq}
Y.~Imamura,
{\it ``Large angular momentum closed strings colliding with D-branes,''}
arXiv:hep-th/0204200.

\bibitem{Frolov:2002av}
S.~Frolov and A.~A.~Tseytlin,
{\it ``Semiclassical quantization of rotating superstring in $AdS_5 \times 
S^5$,''}
arXiv:hep-th/0204226.


\bibitem{Takayanagi:2002je}
H.~Takayanagi and T.~Takayanagi,
{\it ``Open strings in exactly solvable model of curved space-time and
PP-wave limit,''}
arXiv:hep-th/0204234.


\bibitem{Bakas:2002qh}
I.~Bakas and K.~Sfetsos,
{\it ``PP-waves and logarithmic conformal field theories,''}
arXiv:hep-th/0205006.

\bibitem{Oh:2002sv}
K.~Oh and R.~Tatar,
{\it ``Orbifolds, Penrose limits and supersymmetry enhancement,''}
arXiv:hep-th/0205067.

\bibitem{Bigatti:1997gm}
D.~Bigatti and L.~Susskind,
{\it ``A note on discrete light cone quantization,''}
Phys.\ Lett.\ B {\bf 425}, 351 (1998)
[arXiv:hep-th/9711063].



\bibitem{HZ} 
{J.~Harer, D.~Zagier, 
{\it ``The Euler characteristic of the moduli space of curves,''}
{Invent.\ Math.\ \bf 85} (1986), no.~3, 457--485.
}


\bibitem{Balasubramanian:2001nh}
V.~Balasubramanian, M.~Berkooz, A.~Naqvi and M.~J.~Strassler,
{\it ``Giant gravitons in conformal field theory,''}
JHEP {\bf 0204}, 034 (2002)
[arXiv:hep-th/0107119].


\bibitem{Corley:2001zk}
S.~Corley, A.~Jevicki and S.~Ramgoolam,
{\it ``Exact correlators of giant gravitons from dual $N = 4$ SYM 
theory,''}
arXiv:hep-th/0111222.

\bibitem{dhoker2}
{E.~D'Hoker and A.~V.~Ryzhov,
{\it ``Three-point functions of quarter BPS operators in $N = 4$ SYM,''}
JHEP {\bf 0202}, 047 (2002)
[arXiv:hep-th/0109065].
}

\bibitem{cvit}
P.~Cvitanovic,
{\it ``Group Theory For Feynman Diagrams 
In Nonabelian Gauge Theories: Exceptional Groups,''}
Phys.\ Rev.\ D {\bf 14}, 1536 (1976).


\end{thebibliography}
\end{document}